# On membrane-based approaches for rare earths separation and extraction – recent developments

Joanna Kujawa[1*], Samer Al Gharabli[2], Anthony Szymczyk[3], Artur P. Terzyk[1], Sławomir Boncel[4,5], Katarzyna Knozowska[1], Guoqiang Li[1], Wojciech Kujawski[1]

[1)] Faculty of Chemistry, Nicolaus Copernicus University in Toruń, 7 Gagarina Street, 87-100 Toruń, Poland

[2)] Pharmaceutical and Chemical Engineering Department, German Jordanian University, Amman 11180, Jordan

[3)] Univ Rennes, CNRS, ISCR (Institut des Sciences Chimiques de Rennes) - UMR 6226, F-35000 Rennes, France

[4)] Department of Organic Chemistry, Bioorganic Chemistry and Biotechnology, Faculty of Chemistry, Silesian University of Technology, Krzywoustego 4, 44-100 Gliwice, Poland

[5)] Centre for Organic and Nanohybrid Electronics, Silesian University of Technology, Konarskiego 22B, 44-100 Gliwice, Poland

## Abstract

The current progress in the improvement of membrane materials enlarges the utilizations and necessities for innovative, more efficient, resistant, and highly specialized separation materials. One of the growing areas of the implementation of membrane-based strategies is the selective separation of rare earth elements (REEs). Owing to their high potential and essential meaning in modern industry, demands for raw REEs have increased significantly in the last years. Moreover, considering the fact that isolation of REEs has listened as one of "the seven chemical separation processes that, if improved, would rear great global benefits" their application is emerging. Membrane strategies are gaining popularity in metal separation and wastewater treatment as sustainable green approaches with a simple operation. Many membrane strategies for REEs separation have been developed, but only a few reviews have been found. The main message coming from the literature review is that more light should be shed on the progress and implementation of non-liquid membranes. The previously utilized liquid membranes faced many problems related mainly to the long-term utilization, stability, and extraction yield. As a consequence, the REEs separation processes have moved in the direction of much more durable materials, i.e., non-liquid membrane strategies, in which the carriers are physically or chemically connected with the membrane or porous structure of support. The most significant advantage is the lack of a problem of losing extractant into the aqueous phase.

This article aims to: i) review the current membrane-based methods in REEs separation focusing on non-liquid membranes (imprinted, polymer inclusion, nanocomposite, metal-/covalent organic framework membranes), ii) present the considerations of the essential scientific and technical issues, e.g. extraction performances, separation efficiency, REEs transport features, and iii) discuss their transport models and mechanisms as well as membrane stability.

## Table of content:

## Introduction

As awareness of climate change grows, some countries are developing green policies and swiftly transitioning to sustainable energy technology [1-4]. There is an increasing demand for resources required to make crucial components of such technologies. The European Commission has already pointed out some of these components as necessary raw materials, raising worries about the supply security [3, 4]. Such examples of these essential materials are rare earth elements (REEs), which are required to produce permanent magnets for electric vehicle motors and wind turbine generators, as well as other emerging utilizations. For the abovementioned application, there are four crucial elements, praseodymium (Pr), neodymium (Nd), dysprosium (Dy), and terbium (Tb). REE materials are now mostly supplied by China [3], and with demand growing rapidly, there are concerns about supply constraints in the midst of geopolitical tensions [5-8]. In 2020, the EU started a raw materials alliance focused primarily on REEs and permanent magnet production, as well as a detailed action plan to address possible risks associated with vital raw resources [9]. Notably, for these four elements (Pr, Nd, Dy, and Tb), the intensification in their usage is expected in both low-carbon technologies and other applications. Forthcoming REEs demand for wind turbines and e-mobility will be driven by both technological advancements and material optimization, as well as the political ambitions underlying their development. In contrast, requisition in other sectors, such as electronics and specialized equipment, will be primarily influenced by market dynamics. For that reason, there is a possibility for a broad range of probable future scenarios (Fig. 1).

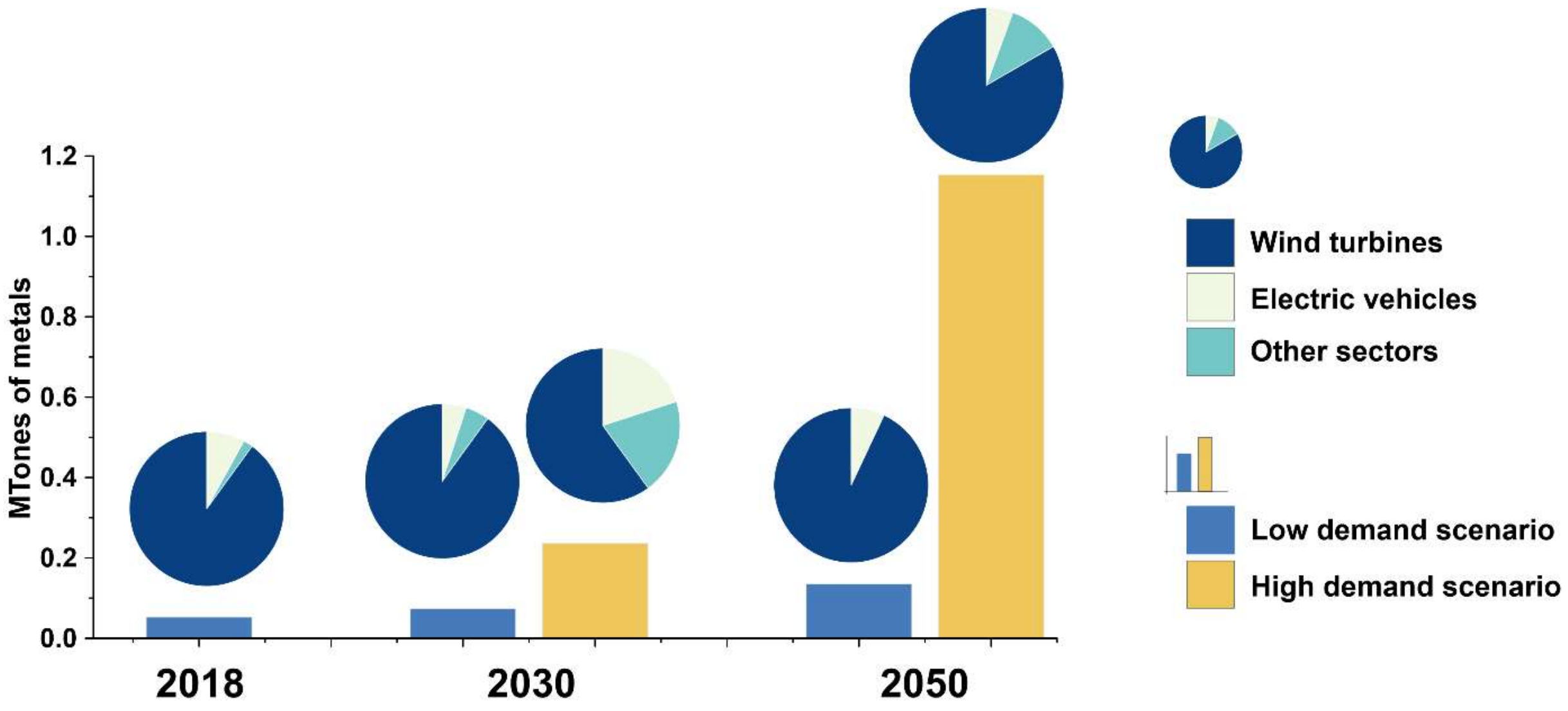

**Fig. 1**. Projected global requisition for Nd, Pr, Dy, and Tb metals for wind turbine generators, motors for electric vehicles and other sectors, based on the low and high demand scenarios [10-14].

Other critical applications of rare earths [15], which are considered as "the secret ingredients of modern industry" [16] were identified in the sectors and devices strictly associated with human life, e.g. magnets [17, 18], catalysts [19-22], inorganic materials including ceramics [23-27], petrochemicals [28, 29], electronics [30, 31], battery alloys [32-37] and additives/enhancers for more advanced materials [38, 39]. Paradoxically, industrial separation for REEs from mineral ores seemed less clean than their by-end uses [40].

Bearing in mind such tremendous needs for REEs, other pathways away from the exploration and mining of natural resources should be considered. The alternative options should focus on improving recycling facilities and waste collection [41] (permanent magnets (Nd), fluorescent powders, television tubes, flat panel displays) [42-49], developing novel solid or liquid extraction systems [50-55], and implementing separation and bio-separation technologies [56-62]. At the European level recycling process is an essential solution owing to limited options for growing the primary supply. For that reason, the mentioned source of REEs should be upgraded since current recycling input rates for REEs are only around 1 % (Fig. 2).

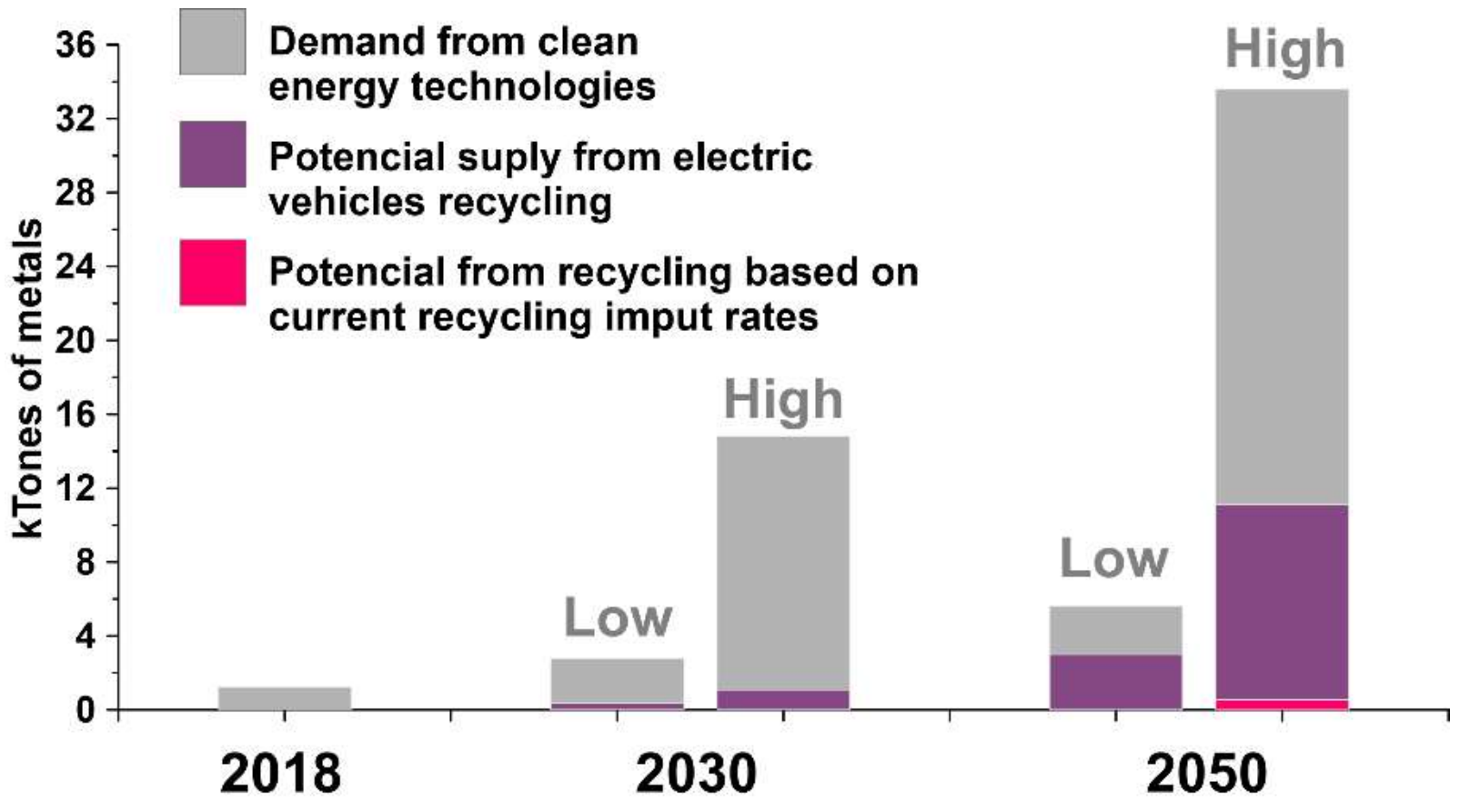

**Fig. 2**. Estimated demand for Nd, Pr, Dy, and Tb metals for clean energy technologies compared with the potential supply form the recycling of REEs from electric vehicles (EU-27). Data presented according to the low and high demand scenarios [10-14, 63, 64].

Considering the large similarities in the ionic radius of the REEs (Table 1), they are typically found in multi-element combinations. Nevertheless, the purity of a single REE is more important in industrial applications than the mixture. For that reason, a lot of work has been done to develop efficient methods for rare earth separation and purification. The presented review covers the topic of the extraction, separation, and purification of REEs with the implementation of membrane-based separation processes focusing mainly on non-liquid materials. Principally, the progress within the last five years (79 % of literature from the last 5 years and 54 % from the last 2 years) has been highlighted (Fig. 3). Starting from the in-depth characterization of the REEs and the requirement for membrane separation materials that need to meet to be effective in the REEs separation are presented, followed by extraction and separation mechanisms. Finally, the broad range of emerging applications in REEs separation with the implementation of non-liquid membrane-based separation materials is described. The most important milestones in the area have been collected on the timeline presented in Fig. 4. Moreover, the future perspective and new possible pathways are highlighted.

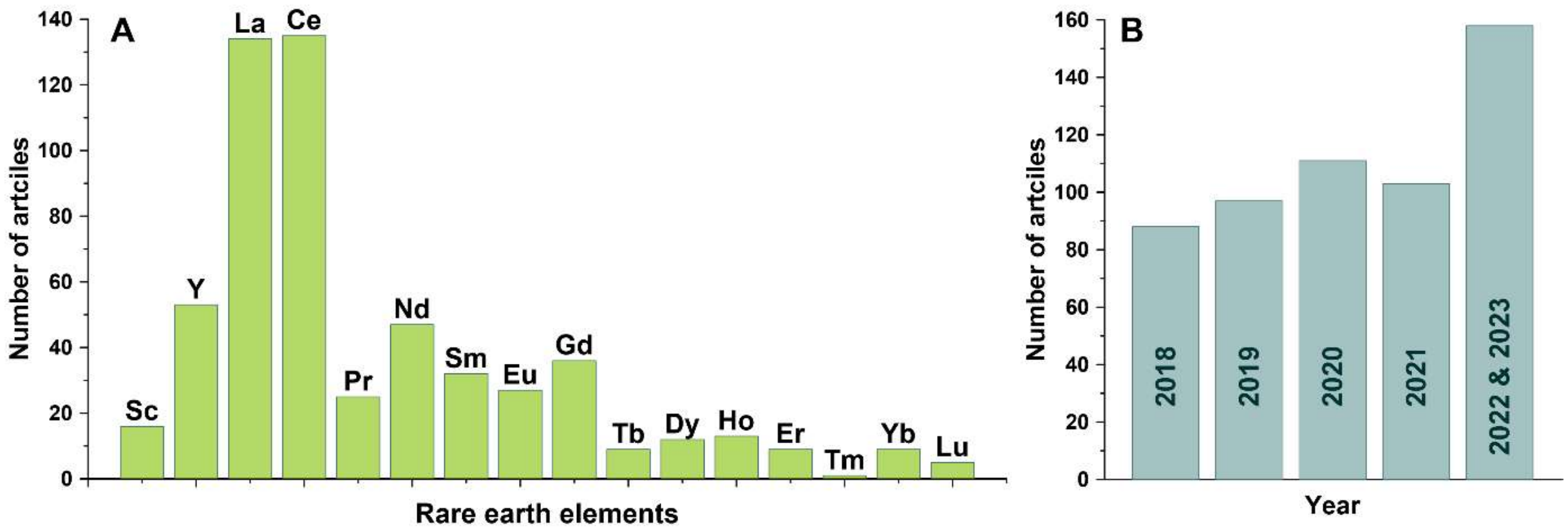

**Fig. 3**. Number of published articles on each REE with the implementation membrane-based separation technology (A). Total number of articles in a certain year. Keywords: each REE and "membrane separation" Scopus database (based on 26th July 2023 data).

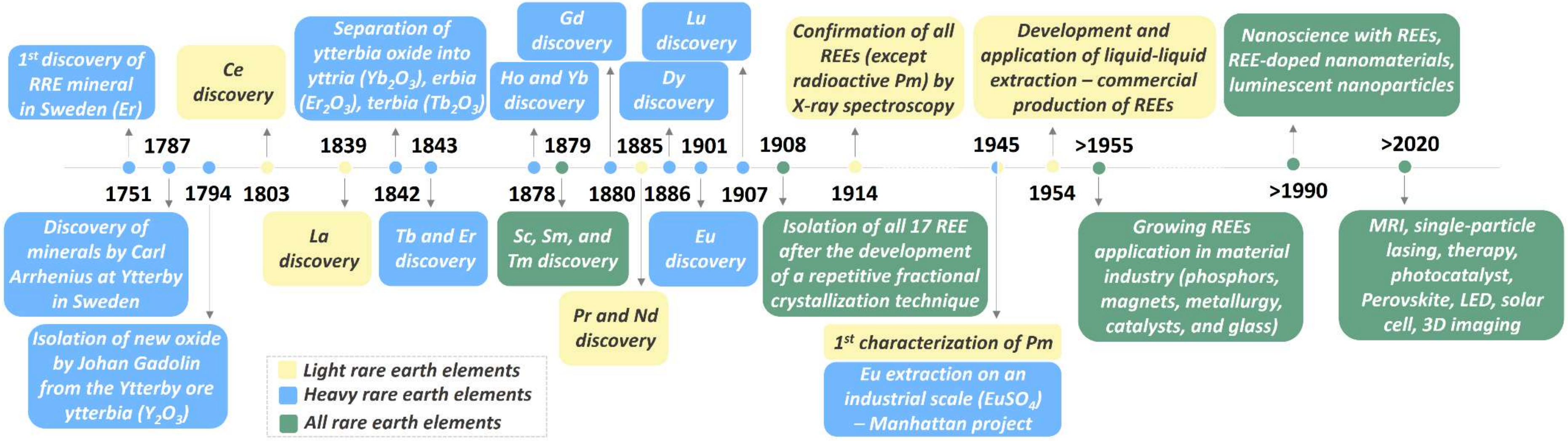

**Fig. 4.** The time-line with the most important milestones in the topic of the review.

## 1.1 Historical view of REE

Rare-earths is a group of 17 elements gathering the entire block of lanthanides (ranging from La to Lu), and additionally Sc and Y. These elements display many chemicals as well as physical similarities and their geological deposits frequently coincide. The beginning of REEs can be traced back to the middle of 18$^{th}$ century. Based on the sources, in 1754, the first rare-earth mineral was found in Sweden [65]. Unfortunately, this material was previously misidentified as calcium iron silicate. Carl Arrhenius discovered another comparable mineral at Ytterby, Sweden, in 1787. Johan Gadolin, a Finnish chemist, extracted new material and called it ytterbia from the Ytterby ore in 1794. Mosander later divided the ytterbia oxide into yttria, erbia, and terbia in 1842, while further research found that the mentioned oxides were likewise rare-earth mixtures [66]. Owing to the similar chemical features of rare-earth elements, the isolation of all seventeen REEs was not finalized till 1908-1909, when a repetitive fractional crystallization method was developed [67]. Later on, Henry Moseley applied X-ray spectroscopy in 1914 to prove the presence of all REEs, excluding radioactive promethium (Pm) [68] that it has not been characterized still 1945 at Oak Ridge National Laboratory, previously Clinton Laboratories.

Even though the entire group of REEs was characterized, the separation and scale-up production of single rare-earth compounds remained problematic. Before World War II, only Eu could be separated on a larger scale as europium(II) sulphate owing to that Eu ions can be simply reduced from $Eu^{3+}$ to $Eu^{2+}$ in comparison with other REEs [69]. As a consequence of dynamic scientific and technological progress during WWII, the primary efficient REE separation was accomplished with the implementation of ion exchange chromatography. The principle of the separation was the diversification in the stability of rare-earth citrate chelates [70-72]. Later on, around 1954, liquid−liquid extraction was developed and effectively applied in the commercial production [73]. The large-volume production of REEs brought a stable foundation for their practical utilization in the modern society. One example is the $Eu^{3+}$-doped $YVO_4$ which turns out to be the first red phosphor, allowing the progress of color television displays [74-76]. Starting from the 1950s, REEs have aided progress in the materials industry in a variety of fields, including metallurgy, phosphors, catalysts magnets, and glass. Rare earths are often employed as additions or dopants in products [38]. The advantage of employing rare earth is that they may substantially boost material features even in small concentrations. Consequently, REEs have been regarded as contemporary industry's "vitamins", and the manufacturing of rare earth-doped materials has become one of the most critical components of technological growth. With the evolution of nanotechnology and nanoscience, the meaning of REE-enriched nanomaterials for various usage has been progressively recognized. The early work was focused on REE-doped luminescent nanoparticles [77, 78]. The inspiration for the designing such materials was taken from the formation of inorganic nanophosphors [79, 80] and colloidal quantum dots [81-84].

It was not until the beginning of the twenty-first century that new synthetic pathways for generating high-quality crystals < 100 nm in size sparked growth in this field. Monodisperse REE-enriched nanomaterials with straightforwardly adjustable crystallinity, morphology, and size, can now be formed with precisely designed structures owing to recent advances in new synthetic strategies e.g., thermal decomposition [85, 86], hydro(solvo)thermal processing [87-89], and coprecipitation [90-92].

Nanomaterials based on rare-earth have evolved into a highly multidisciplinary field of study that bridges the gap between solid-state chemistry and materials science (Fig. 4). Owing to the unique features of REEs, a broad range of utilizations have been developed, including environmental protection (clean energy development and catalytic treatment of pollutants) [93, 94], healthcare (therapy and bioimaging) [95, 96] as well as innovative technologies including three-dimensional displays, luminescent inks [97] and optical communications [98, 99].

### 1.2 Characterization and classification of REEs

In the periodic table, the *4f* elements, according to IUPAC [100], are called the lanthanoids (previously and still frequently the 'lanthanides') or "rare earth elements". Although "the rare earth elements" is a broadly used terminology, they are not particularly rare, except for promethium, which has no stable isotope (product of natural nuclear fission in uranium ore [69, 101-103]). Generally, together with lanthanides, the yttrium (Y) and scandium (Sc) are considered rare earth elements owing to their chemical similarities. The chemistry of the entire group is principally ionic and is determined primarily by the size of the $M^{3+}$ ion (Fig. 5). Since Y, which is located above La in the Transition Group III and possesses a similar 3-ion with a noble-gas core, has both atomic and ionic radii lying close to the corresponding values for Tb and Dy that results from the lanthanide contraction (Table 1, Fig. 5).

The characteristic regularity of the *4f* elements reflects the fact that the *4f* orbitals filled across the series are core-like and overlap little with orbitals on donor atoms. The elements from lanthanum to lutetium possess the 3-oxidation state, and the chemistry is mostly that of the Ln(III) ion. The Ln(III) ions with incompletely filled *f* orbitals exhibit electronic, optical, and magnetic features, which are broadly exploited in different technologies [104]. The *4f* orbitals are deeply buried owing to the lanthanide contraction, and therefore, the *4f* electrons are not available for the covalent bonding [105, 106]. These features are essential from the practical point of view and possible pathways of modification and utilization. Compounds of Ln(II) and Ln(IV) can be generated when the relatively favorable energy of a particular $4f^n/5d^1$ electron configuration shows a possibility, and these uncommon species are frequently very reactive [107, 108]. Moreover, to describe the orbitals for the materials having *3d-*, *4f-,* and *5d*-electron systems, the linear augmented-Slater-type-orbital (LASTO) approach should be applied [109].

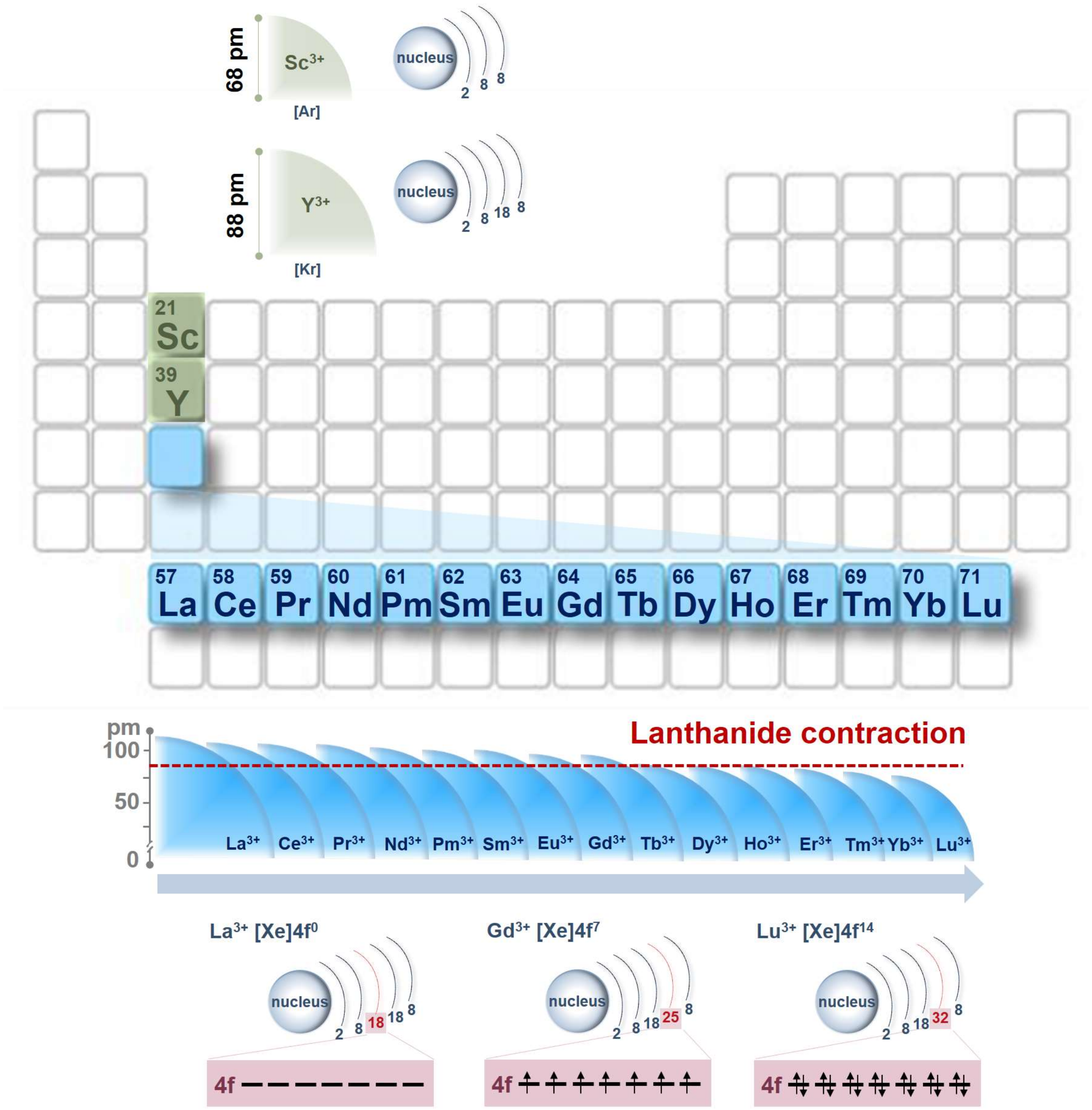

**Fig. 5.** Characterization of rare-earths (ionic radius and valence configuration). The number of electrons in the *4f* orbitals rises with increasing atomic number from $La^{3+}$ to $Lu^{3+}$. The electron configurations of $La^{3+}$, $Gd^{3+}$, and $Lu^{3+}$ reveal empty, half-filled, and entirely filled 4f orbitals, respectively.

**Table 1**. Selected properties of Lanthanides atoms and ions plus yttrium (Y) and scandium (Sc).

| Z | Name | Symbol | Electronic configuration Atom | $M^{2+}$ | $M^{3+}$ | $M^{4+}$ | Σ*I* [eV] | $E_0$ [V] | Radius [pm] Atom | $M^{3+}$ |
|---|---|---|---|---|---|---|---|---|---|---|
| | | | | **Light REE** | | | | | | |
| 57 | Lanthanum | La | $[Xe]6s^25d^1$ | $[Xe]5d^1$ | $[Xe]4f^0$ | - | 36.2 | -2.52 | 187 | 106 |
| 58 | Cerium | Ce | $[Xe]4f^16s^25d^1$ | $[Xe]4f^2$ | $[Xe]4f^1$ | $[Xe]4f^0$ | 36.4 | -2.48 | 183 | 103 |
| 59 | Praseodymium | Pr | $[Xe]4f^36s^2$ | $[Xe]4f^3$ | $[Xe]4f^2$ | $[Xe]4f^1$ | 37.55 | -2.47 | 182 | 101 |
| 60 | Neodymium | Nd | $[Xe]4f^46s^2$ | $[Xe]4f^4$ | $[Xe]4f^3$ | - | 38.4 | -2.44 | 181 | 99 |
| 61 | Promethium | Pm | $[Xe]4f^56s^2$ | $[Xe]4f^5$ | $[Xe]4f^4$ | - | - | -2.42 | 181 | 98 |

| | | | | | | | | | |
|---|---|---|---|---|---|---|---|---|---|
| 62 | Samarium | Sm | $[Xe]4f^{6}6s^{2}$ | $[Xe]4f^{6}$ | $[Xe]4f^{5}$ | - | 40.4 | -2.41 | 180 | 96 |
| | | | | **Heavy REE** | | | | | | |
| 63 | Europium | Eu | $[Xe]4f^{7}6s^{2}$ | $[Xe]4f^{7}$ | $[Xe]4f^{6}$ | | 41.8 | -2.41 | 199 | 95 |
| 64 | Gadolinium | Gd | $[Xe]4f^{7}6s^{2}5d^{1}$ | $[Xe]4f^{7}5d^{1}$ | $[Xe]4f^{7}$ | - | 38.8 | -2.40 | 180 | 93 |
| 65 | Terbium | Tb | $[Xe]4f^{9}6s^{2}$ | $[Xe]4f^{9}$ | $[Xe]4f^{8}$ | $[Xe]4f^{7}$ | 39.3 | -2.39 | 178 | 92 |
| 66 | Dysprosium | Dy | $[Xe]4f^{10}6s^{2}$ | $[Xe]4f^{10}$ | $[Xe]4f^{9}$ | $[Xe]4f^{8}$ | 40.4 | -2.35 | 177 | 91 |
| 67 | Holmium | Ho | $[Xe]4f^{11}6s^{2}$ | $[Xe]4f^{11}$ | $[Xe]4f^{10}$ | - | 40.8 | -2.32 | 176 | 89 |
| 68 | Erbium | Er | $[Xe]4f^{12}6s^{2}$ | $[Xe]4f^{12}$ | $[Xe]4f^{11}$ | - | 40.5 | -2.30 | 175 | 88 |
| 69 | Thulium | Tm | $[Xe]4f^{13}6s^{2}$ | $[Xe]4f^{13}$ | $[Xe]4f^{12}$ | - | 41.85 | -2.28 | 174 | 87 |
| 70 | Ytterbium | Yb | $[Xe]4f^{14}6s^{2}$ | $[Xe]4f^{14}$ | $[Xe]4f^{13}$ | - | 43.5 | -2.27 | 194 | 86 |
| 71 | Lutetium | Lu | $[Xe]4f^{14}6s^{2}5d^{1}$ | $[Xe]4f^{14}5d^{1}$ | $[Xe]4f^{14}$ | - | 40.4 | -2.25 | 173 | 85 |
| 21 | Scandium | Sc | $[Ar]4s^{2}3d^{1}$ | $[Ar]3d^{1}$ | [Ar] | - | 44.11 | -1.88 | 164 | 68 |
| 39 | Yttrium | Y | $[Kr]5s^{2}4d^{1}$ | $[Kr]3d^{1}$ | [Kr] | - | 38.98 | -2.37 | 180 | 88 |

**Σ*I*** – the sum of first three ionization potentials [69].

Basing the varied physical and chemical features of the REEs, they can be classified into two groups, the light rare earth elements (LREEs) from La to Eu and the heavy REEs (HREEs) from Gd to Lu along with Sc and Y [110] (Table 1). When compared, the heavy ones are considered more critical resources owing to their extensive utilization in photo-electromagnetism, e.g., radiation sources [111, 112], laser media [113, 114], scintillation crystals [115-117] and magnetic materials [118, 119], that are irreplaceable for the advanced technology and national defense [120, 121]. On the other hand, the light REE are broadly applied in superconductors [122, 123], catalysis [124, 125], ferromagnetism [126, 127], quantum cutting [128], and sensitizer [129, 130].

### 1.3 Membrane separation techniques for recovery of REEs

Membrane separation techniques (MSTs) represent the most technologically required 'all-in-one' processing which encompasses simultaneous extraction and stripping operation(s). This approach eliminates operating under elevated temperature regimes such as (fractional) distillation, evaporation or drying. Nowadays, MSTs are therefore considered as the most environmentally benign and the most economic, hence their advancement toward the highest technological readiness levels experiences a pressing need for tunability-driven creativity, both scientific and engineering. Those requirements aimed at the recovery of REEs, albeit competitive advanced solutions are continuously reported globally such as liquid or solid phase extraction employing novel solvents or biosorption (i.e., sorption *via* biomass passively concentrating and binding REE cations – physically and chemically – onto its porous structures), locate MSTs as the most rapidly developing separation techniques [131].

Among MSTs as particularly important, polymer inclusion membranes (PIMs) characterize membranes in which a 3D-polymer network (PVDF, PTFE, CTA, etc.) embeds an ion-complexing carrier. Simultaneously, the target flexibility of the membrane under operational conditions is achieved by a rather straightforward addition of a plasticizer or a selection of the well-defined neat polymer material (in terms of molecular weight, morphology as well as bulk and surface chemistry) [132]. PIMs, fundamentally advantageous over any other non-polymer competitive alternatives (only marginally present in the scientific literature), feature simple compositions, yet of unbeaten versatility (e.g., one type of a membrane dedicated for numbers of REE solute types), a straightforward and scalable synthesis/manufacturing toward tunable porosity, effective immobilization of the carrier upon manufacturing accompanied by its high concentration, high-performance mechanics, prolonged time and operational stability, convenient and fast installation in the facilities as well as, last but not least, low overall costs. Fundamentally, PIMs offer high transport flux, fast permeation, and a low consumption of yet expensive extracting/complexing carriers. The nature of PIMs easily allows their immediate physicochemical compatibilization with broad-scale diverse-property chemicals. This scenario allows considering them as the first-choice systems toward multi-functionality such as development of *in situ* chemistries, i.e., synthesis of nanoparticles (e.g. electro-/photocatalysts, optodes, etc.), speciation, or pretreatment/preconcentration, to list the few most important ones [38].

It should be emphasized that – from the central postulates of *green chemistry* perspective – PIMs constitute the most promising candidates for sustainable systems dedicated for the REE separation [133]. Nevertheless, and undoubtedly, an elaboration and development of the original solutions in the area of PIMs would require from the scientists and technologists broad and fresh views. Those would encompass multi-level 'properties-by-design' approaches employing, e.g., computational [134] and experimental design of efficient and selective carriers compatible with the novel material-based membrane and the media [135] – in this sequence or *vice versa*. Overall, the innovations concern polymer matrices, plasticizers and carriers while here the nomenclature of PIM components could be treated somewhat arbitrary. For instance, the carrier could play a role of a plasticizer or the polymer matrix macromolecules could be covalently functionalized and eventually bear all of the components as 'enchanted' in just one separating entity.

Although economically justifiable, neat polymer membranes display several drawbacks. Among them, fouling, low thermal and chemical stability, and a lack of specificity remain as the most troublesome ones [136]. Indeed, if operated under the highly acidic or otherwise harsh conditions in the REE hydrometallurgy processing, the membranes emerge as prone to damage and the consequent/immediate decline of the performance. Taking into account the separation nature of membranes, such as preferential adsorption feature and shape selectivity, together with the facility in preparation and thermal or chemical stabilities, it is believed that the following strategies could be the potential choices for REEs separation and purification in the near future.

Firstly, to overcome the shortcomings of neat polymeric membranes, numerous nanoparticles (NPs) such as Ag, Fe, Al, Ti, Zr, Mg, Pd, $SiO_2$ as well as different types of two-dimensional nanomaterials, e.g. functionalized carbon nanotubes (f-CNTs), graphene oxide (GO), and others were amalgamated with the polymer matrices to enhance their multiple characteristics [137]. Those metallic and inorganic NPs were intended and finally proved to augment permeability, selectivity, hydrophilicity, antifouling, tensile strength, and thermal stability. NP-embedded polymeric membranes were frequently applied for the enhanced gas or organic molecules permeability and separation. And although dispersibility of the above NPs is challenging and the fabrication of membranes may be multi-stage and/or high-energy demanding, the potential of the membrane characteristics improvement is undisputed. This trend, in our opinion, accompanied by a tunable morphology and surface physicochemistry of NPs, will be continued toward separation of REEs.

Apart from PIMs, molecular/ion imprinted membranes (MIMs/IIMs) are considered as promising alternative techniques of enhanced stability and selectivity. Molecular imprinting technique has become recently a hot topic in the separation science. Polymers of a dedicated compromise between rigidity and flexibility constitute the prospective matrices capable of preserving the cavities – programmable by functional groups from polymerization monomers – matching the size and shape of the template molecules or ions toward the specific separation. Despite a rapid progress in the field adsorption and separation of REEs by imprinted polymers in recent years, the studies on the behavior of REE ions in the fully functional imprinted films are rather scarce. The complexity of manufacturing MIMs/IIMs involving a limited number of polymerization methods, troublesome tailoring of the porosity, challenging permselectivity, and poor recyclability represent the most critical obstacles in the scalability [138]. The so-far applied functional monomers and cross-linking agents cannot satisfy the precise requirements for the molecular recognition. Additionally, as the majority of separation processes proceeds in the aqueous media, the hydrogen bonds formed between functional monomers and templates during the self-assembly-driven pre-polymerization would be significantly affected in polar protic solvents [139]. Consequently, the synthesis of imprinted polymers must be performed in non-polar solvents to exclude the interferences in water environments while many templates are insoluble therein, which severely hampers the full-scale application of the imprinting. Hence, the new routes of development of the imprinting technique cover: water-based synthesis of template-incorporated membranes, novel carriers with specific selectivity for the target template REE ions, and new functional monomers with this task specific recognition ability.

A new class of hybrid organic/inorganic material toward preparation of membranes is metal-organic frameworks (MOFs) [140]. In MOFs, the regularly porous structural lattice is composed from metal ions coordinated with organic ligands. They are contemporarily applied as gas and liquid storage and separation media, sensors, catalysts, proton conductors, etc. MOFs are considered as competitive against other adsorbents such as mesoporous silica, activated carbon or polymeric resins due to milder conditions of methods of synthesis, higher porosity and specific surface area. In gas and liquid separation, MOF layer are tethered to the surface of a substrate surface such as silica, alumina, (nano)carbons, polymers, etc. The main problem of the majority of MOFs is the degradability in the aqueous environments, hence development of water stable MOFs are considered as well-applicable in the REE separation [141] and the future of MOFs membranes in REEs separation emerges as bright.

PIMs can be confronted to the less stable liquid membranes (LMs). Generally, LMs are divided into: (1) unsupported liquid membrane such as (1a) bulk liquid membrane (BLM) and (1b) emulsion liquid membrane (ELM), and (2) supported liquid membrane (SLM). A supported liquid membrane contains a porous membrane support and a carrier enabling the separation process and has been applied as an effective and versatile – in light of the separation of a broad palette of metal ions from dilute solutions [142, 143].

Indeed, development of membrane separation of REEs witnessed a global contribution from LM techniques. Nevertheless, scale-up experiments were not satisfying mainly because of the problems with LM stability and efficiency while the efforts allow for a moderate optimism [144].

Unquestionably, there is a clear potential of applications in REE ions enrichment and separation. Yet, there are challenges which need to be addressed in the scaled-up facilities.

These include an involvement of cleaning as a time-consuming separate unit operation, replacement of the fouled membranes with the overall economy of processing. So far, these aspects were non-optimized. Similarly, a relative hydrophobicity, lowered fluxes resulting from (bio)fouling, low thermal and chemical stability are still incompletely addressed. Among which the fouling problem must be mitigated before the commercialization since it will decrease the transport flux and consequently rise the cost of membrane replacement and operation [145, 146]. That is the main obstacle of these membrane systems to be applied in the treatment of REE metallurgy industry.

The key output from the most recent state-of-art literature review is that more light should be shed on the development and application of non-liquid membranes, in which the carriers are physically or chemically bound to the membrane or the porous structure of the support. Indeed, the greatest advantage of PIMs is the absence of the problem of losing extractant into the aqueous phase and the 'all-in-one' implementation scheme. Primarily, low-cost PIMs display a high transport flux, fast permeation, low consumption of the high-priced coordinating carriers, versatility as well as time- and operation-stability than SLMs. On the other hand, the imprinting technique is emerging as an exclusively promising technique of REE separation, but the stability of the template cavity must be well-designed and proven in the long-term operation. PIMs and MIMs/IIMs – as a part of the non-liquid membrane family – bring hope to become the real-file industrial solutions. Nonetheless, innovative novel synthetic polymers and their nanocomposites of well-defined and tunable structures – designed for the specific REE(s) – should continue the avalanche of research. And all of the targeted industrial applications should focus on minimization of the energy, materials, and all resources input in a 'cradle-to-grave' scheme and reagent saving.

## 2. Extraction and separation mechanisms

### 2.1. Complexation/Diffusion mechanism

Transport of REEs in liquid membranes (LMs; bulk liquid membranes, emulsification liquid membranes or supported liquid membranes) and polymer inclusion membranes (PIMs; solid membranes) results from a complexion-diffusion process. Indeed, the hydrophobic nature of LMs and PIMs prevents the diffusion of REEs ions in the absence of a carrier (organic extractant) whose role is to form a complex with REEs ions to facilitate their diffusion across the membrane (Fig. 6A) [147].

The transport of REEs in those systems is commonly described as a five-step process (Fig. 7A). The REE first diffuses from the feed solution through the boundary layer in the vicinity of the membrane/feed solution interface. The second step consists of the REE extraction by complexation with the carrier. The REE-carrier complex then diffuses through the membrane towards the membrane/stripping solution interface due to the concentration gradient across the membrane phase.

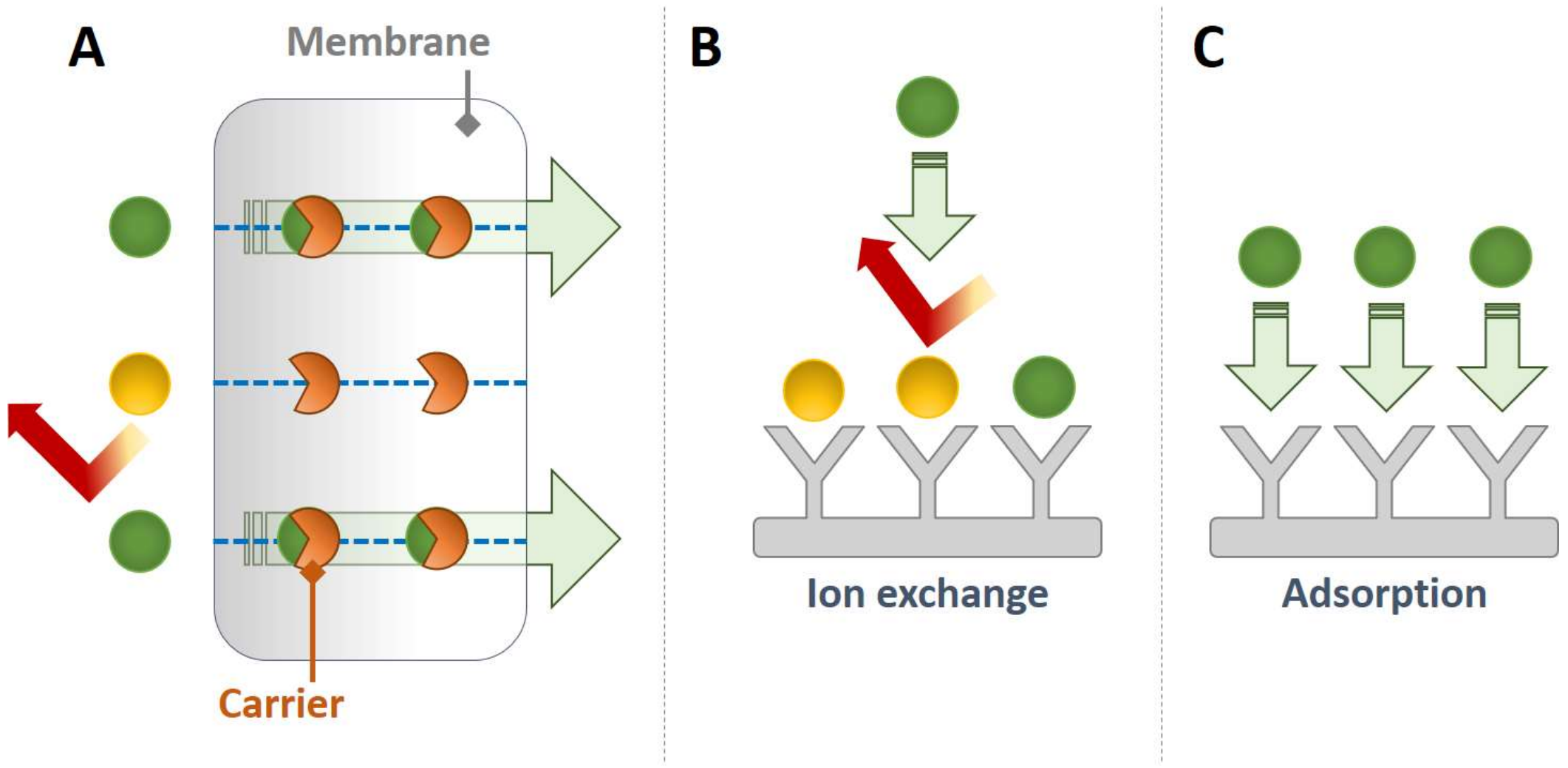

**Fig. 6**. Schematic representation of carrier-assisted transmembrane transport (A). Green sphere: species forming a complex with the organic carrier. Yellow sphere: species not complexed by the organic carrier. A REE selectively interacts with an ion exchange resin (B) or adsorbent (C).

A decomplexation reaction (stripping reaction) occurs at the membrane/extraction solution interface (fourth step), and the REE finally diffuses through the boundary layer between the membrane and the stripping solution.

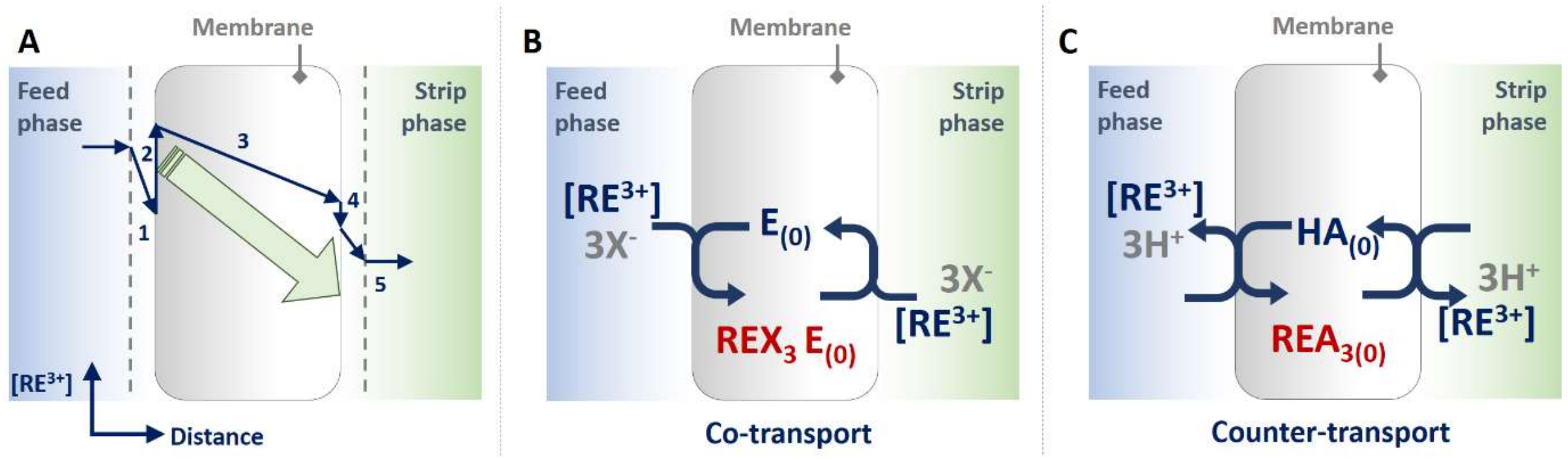

**Fig. 7**. Schematic description of REEs transport process through membranes (A). Rare earth cation ($RE^{3+}$) extraction by (B) co-transport and (C) counter-transport mechanisms. $X^{-}$: inorganic anions; $E_{(o)}$: carrier (in the organic phase); $HA_{(o)}$: acidic carrier (in the organic phase).

It may be noted that the literature on REEs separation sometimes refers to the terms co-transport and counter-transport [131]. In this context, co-transport means that the inorganic anions contained in the feed solution (i.e., the REEs counter-ions) are extracted together with the REE cations (Fig. 7B). On the other hand, counter-transport (occurring with acidic carriers such as mono-2-ethylhexyl (2-ethylhexyl)phosphonate) involves a cation-exchange mechanism between REE cations and hydrogen ions (Fig. 7C).

Let us consider for example the complexation reaction associated with Fig. 7B:

$$RE^{3+}_{(aq)} + 3X^{-}_{(aq)} + E_{(o)} \leftrightarrow REX_3E_{(o)} \quad (1)$$

The thermodynamic equilibrium constant of the complexation reaction ($K$) the constant can be written approximately as:

$$K = \frac{[REX_3E_{(o)}]}{\left[RE^{3+}_{(aq)}\right]\left[X^{-}_{(aq)}\right]^3\left[E_{(o)}\right]} \quad (2)$$

Considering transport at steady state through the membrane phase and the surrounding boundary layers, the following approximate expression of the REE molar flux ($j_{RE}$) can be derived [148].

$$j_{RE} = \left( \frac{1}{D_{RE,f}/\delta_f} + \frac{1}{K\left[X^-_{(aq)}\right]^3 \left[E_{(o)}\right] \frac{D_{REX_3E_{(o)}}}{\Delta x}} + \frac{1}{D_{RE,s}/\delta_s} \right)^{-1} \left( \left[RE^{3+}_{(aq)}\right]_f - \left[RE^{3+}_{(aq)}\right]_s \right) \quad (3)$$

where $D$ stands for the diffusion coefficient (note that for supported liquid membranes and PIMs, the diffusion coefficient of the REE-carrier complex has to be corrected for both the membrane porosity and tortuosity), $\delta$ is the thickness of the boundary layer, $\Delta x$ is the effective thickness of the membrane phase, and the subscripts $f$ and $s$ stand for feed solution and stripping solution, respectively.

According to Eq. (3), the transport of REEs through the membrane phase increases with the extraction constant ($K$). For instance, Croft et al. [149] performed the sequential separation of $La^{3+}$, $Gd^{3+}$ and $Yb^{3+}$ using a PIM composed of 55 wt% poly(vinyl chloride) and 45 wt% di-(2-ethylhexyl) phosphoric acid (D2EHPA), the thermodynamic extraction constants of these three REEs by D2EHPA having being estimated at 0.776, 81.2 and $7.45 \times 10^4$, respectively [149]. This behavior can be rationalized by the Hard and Soft Acids and Bases (HSAB) theory [150]. Indeed, D2EHPA can be considered as a hard (Lewis) base owing to its chemical structure with localized partial charges and is therefore prone to interact more strongly with heavy REEs that are harder Lewis acids than light REEs. Soukeur et al. [151] reported selectivity factor up to 5,600 between ytterbium and cerium using D2EHPA in chloroform.

Eq. (3) also indicates that extraction performance should increase by increasing the concentration of the organic extractant in the membrane phase (due to the shift of the complexation equilibrium towards the formation of the REE-carrier complex). However, an optimal concentration of organic extractant is expected because the increase in its concentration leads to an increase in the viscosity of the membrane phase and therefore to a decrease in the diffusion coefficient of the REE-carrier complex in the membrane phase ($D_{REX_3E_{(o)}}$). Tehrani and Rahbar-Kelishami [148] indeed reported increasing extraction of gadolinium by supported (nano)liquid membranes for carrier (Aliquat 336 ionic liquid) concentrations up to 2-3 mol $L^{-1}$ followed by a decreasing permeability for higher carrier concentrations. Qualitatively similar results were observed by Wannachod et al. [152] in the case of (counter)transport of neodymium using 2-ethylhexyl-2-ethylhexyl phosphoric acid (HEHEPA) carrier dissolved in octane.

### 2.2. Facilitated/retarded permeation mechanism

Another strategy to increase the selectivity of the membranes towards species with very similar structures (REEs or, for instance, enantiomers) is based on molecular imprinting approaches [153, 154]. They consist in elaborating a polymer membrane in the presence of a template species and subsequently removing this template in order to obtain a material with “memory” sites, which have the ability to selectively rebind the original template from a mixture [155].

In recent years, various ion-imprinted membranes (IIMs) for selective separation of REEs were reported [139, 156, 157] (Fig. 8).

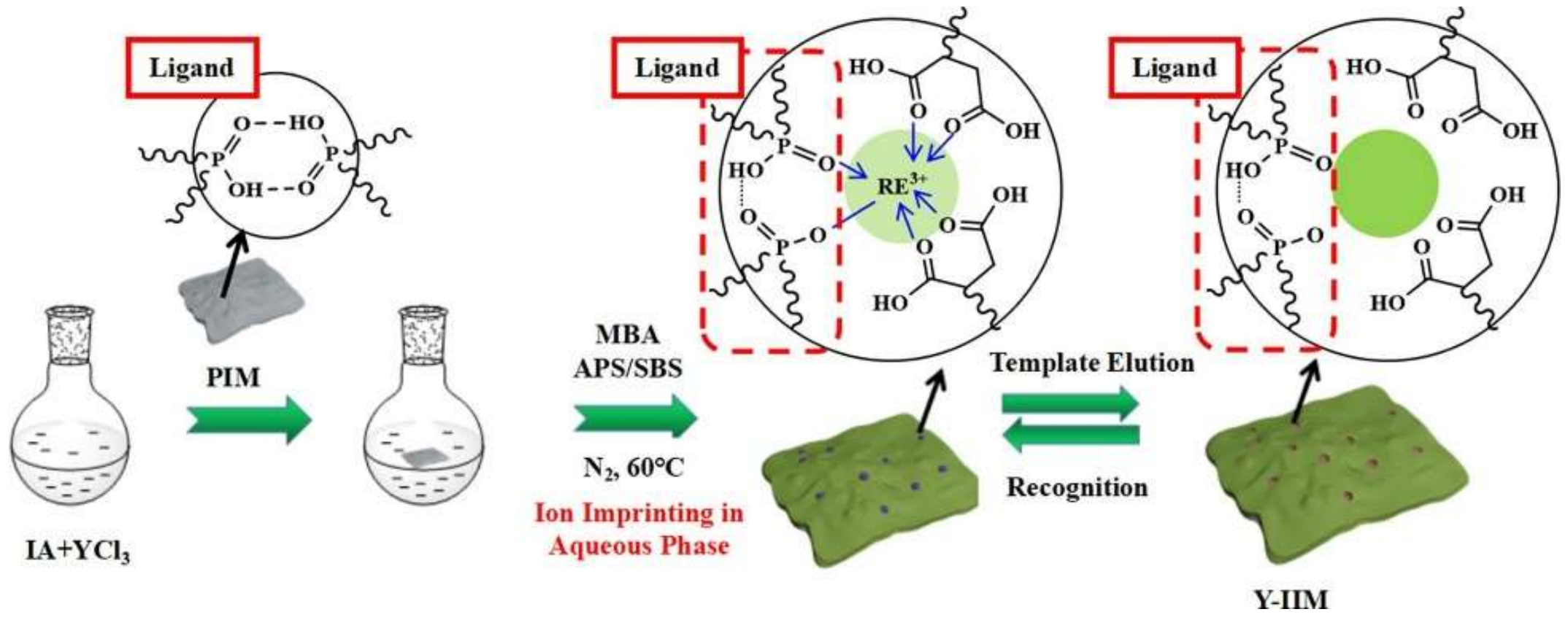

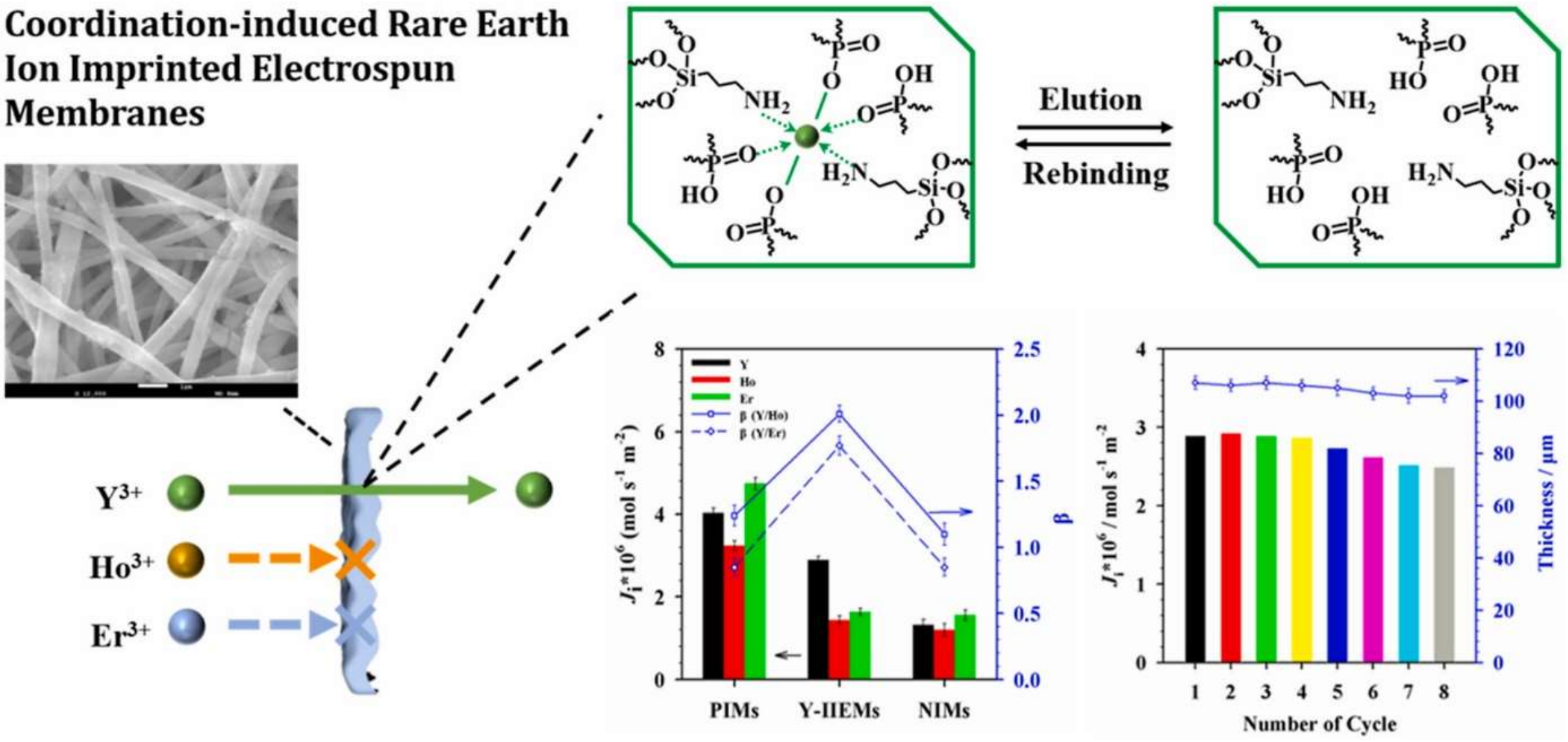

**Fig. 8**. Schematic illustration of IIM preparation for specific recognition of REEs. Reproduced with permission from [139] Copyright © 2021, Elsevier. (upper scheme). Reproduced with permission from [157] Copyright © 2022, Elsevier. (bottom scheme)

Depending on the binding affinity and structural features of the membrane, the so-called facilitated or retarded permeation mechanisms can be observed with imprinted membranes [158].

Membranes based on facilitated transport mechanism are characteristically defined as diffusion-selective membranes [154]. As shown in Fig. 9A, the facilitated transport mechanism is based on the enhanced permeation of the target compound through the membrane *via* binding/desorption cycles between neighboring recognition sites, while diffusion of the other species (non-interacting with the binding sites) is impeded by the microporous structure of the membrane. Efficient separations can then only be achieved with relatively dense membranes with an appropriate density and distribution of recognition sites [159].

The facilitated transport mechanism therefore involves a sufficiently high binding affinity between the target compound and the specific recognition sites. However, if the binding affinity becomes so strong that it makes desorption of the preferentially adsorbed species difficult, then the latter may be transported more slowly across the membrane than other compounds with no specific interactions with the membrane sites (Fig. 9B). Membranes based on such a retarded transport mechanism are typically described as adsorption-selective membranes and separation efficiency is mainly controlled by the membrane binding capacity. Separation performance of membranes based on facilitated transport mechanism is constrained by selectivity/permeability trade-off while membranes based on the retarded transport mechanism might improve both permeability and selectivity simultaneously [154].

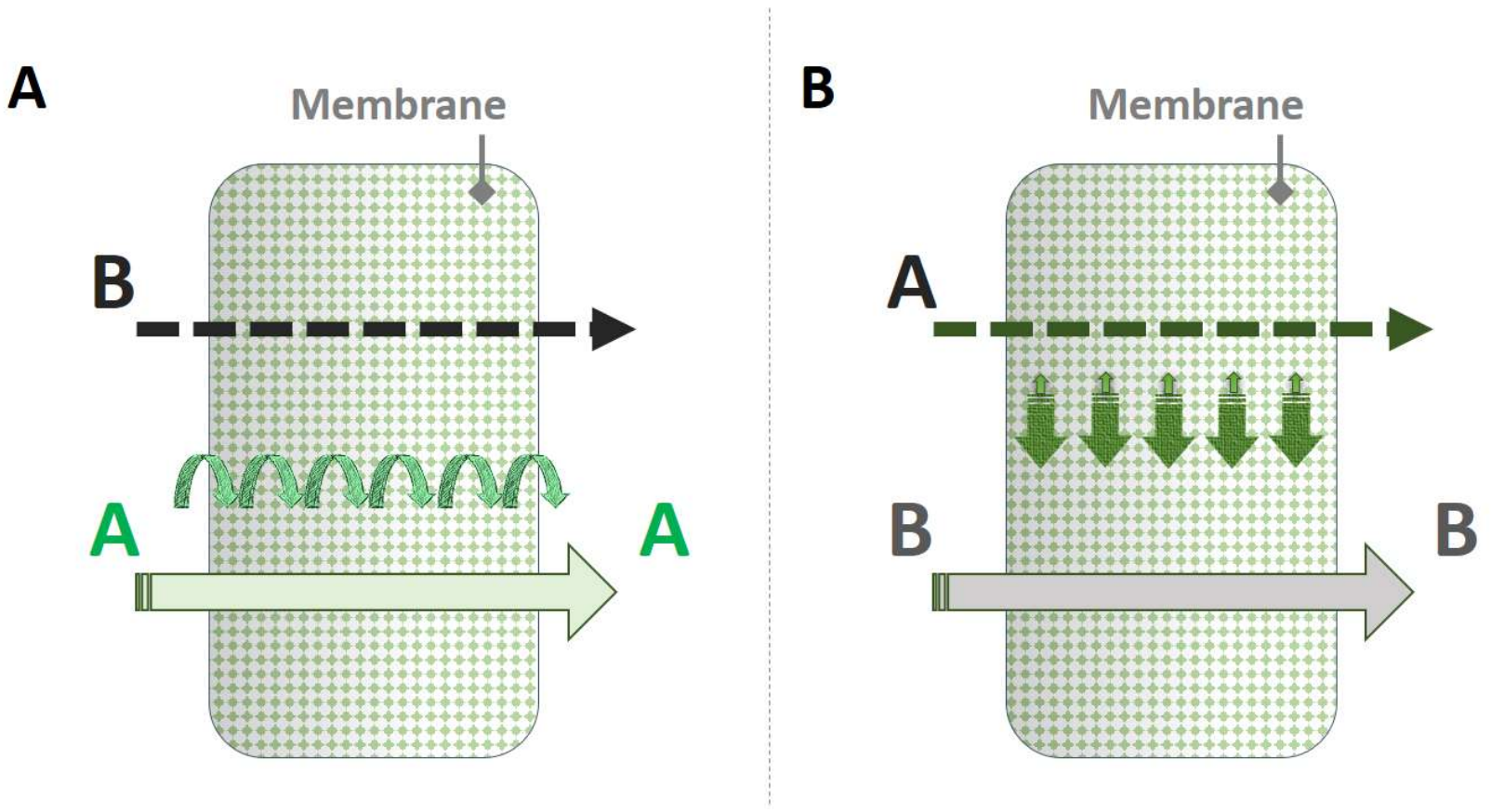

**Fig. 9.** Schematic illustration of (A) facilitated and (B) retarded transport in imprinted membranes.

Fig. 10A shows REE separation performance of an IIM membrane developed by Wu et al. [156] for retarded transport of $Nd^{3+}$ ions. The IIM was found to significantly hinder the transport of $Nd^{3+}$ ions compared with other REEs ($Tb^{3+}$ and $Dy^{3+}$), with selectivity factors $Tb^{3+}/Nd^{3+}$ and $Dy^{3+}/Nd^{3+}$ of 10.99 and 10.17, respectively. On the other hand, nearly similar permeation fluxes were observed for non-imprinted membrane (Fig. 10B) with selectivity factors $Tb^{3+}/Nd^{3+}$ and $Dy^{3+}/Nd^{3+}$ of 1.08 and 1.48, respectively.

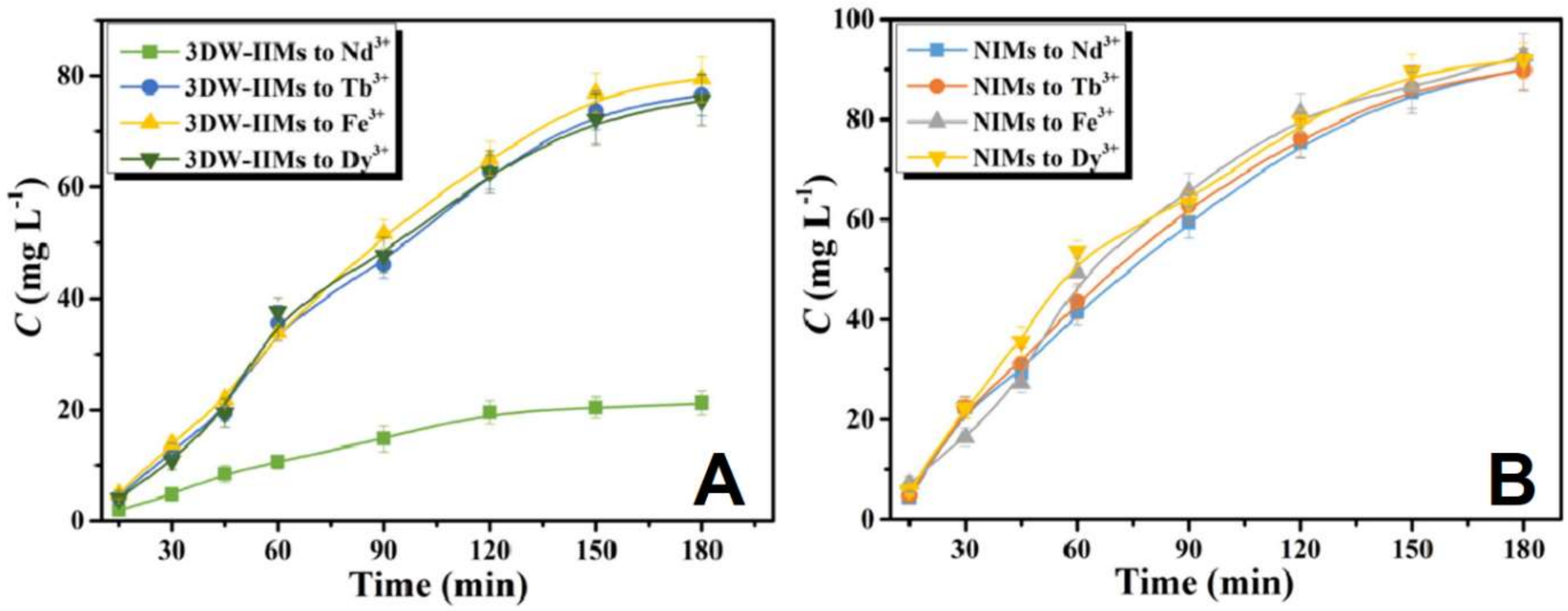

**Fig. 10**. Time-dependent concentrations of $Nd^{3+}$, $Tb^{3+}$, $Fe^{3+}$ and $Dy^{3+}$ crossing (a) a wood-based Nd(III)-imprinted membrane, and (b) a non-imprinted membrane. Reproduced with permission from [156] Copyright © 2021, Elsevier.

### 2.3.Rejection mechanism by membranes without specific carriers or recognition sites

LMs, PIMs and IIMs can suffer from limited stability and/or scaling difficulties, which limits their development on an industrial scale. Polymeric membranes, pure or composite but without the addition of a transporter or specific recognition site, make it possible to overcome these limits. However, the absence of sites or carriers with chemical specificity necessarily limits their selectivity with respect to the various REEs (Fig. 11).

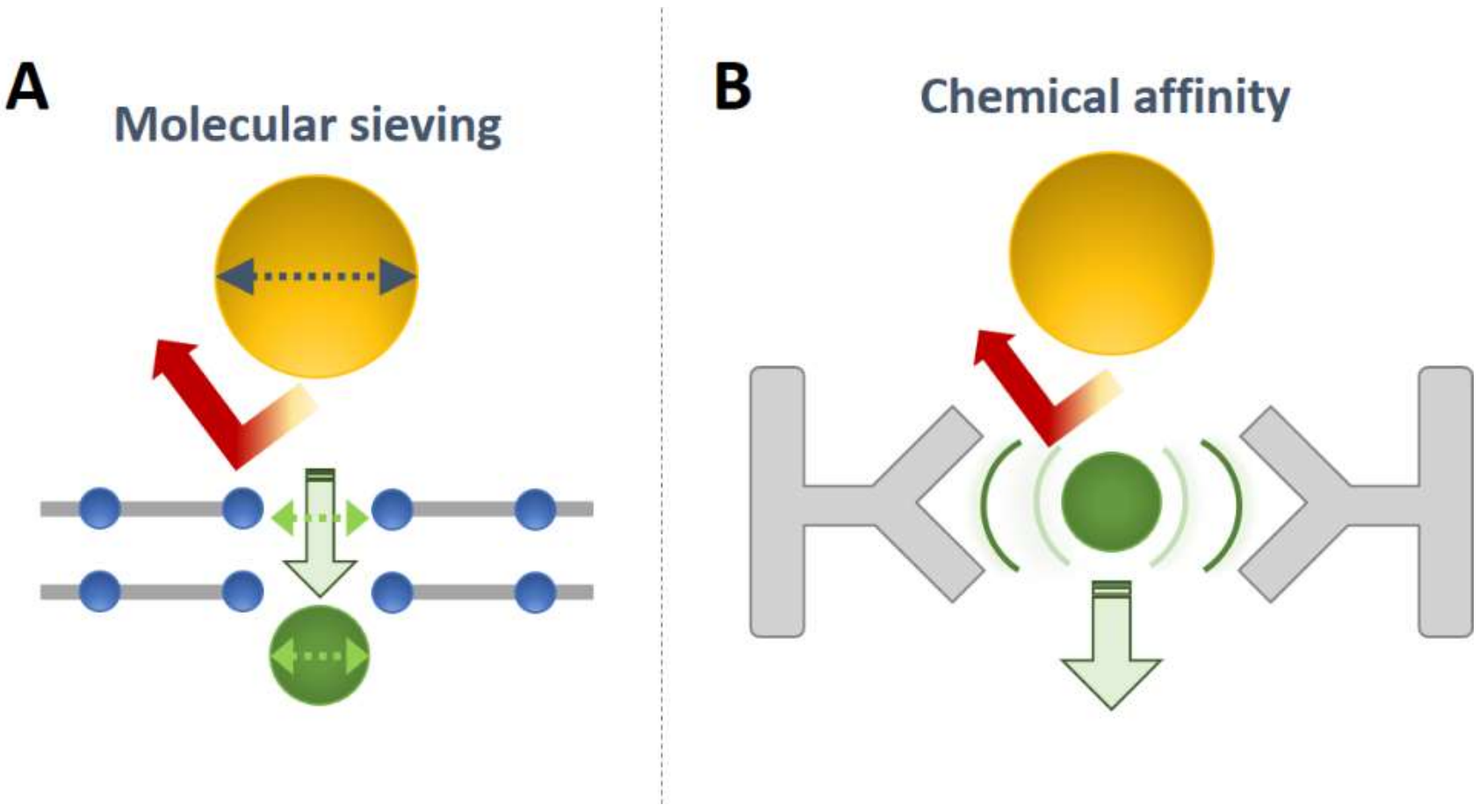

**Fig. 11**. Molecular sieving by a porous membrane (A) and chemical affinity (B).

Porous membranes exhibit selectivity based on the relative size of the compounds to be separated and the membrane pores [147] (molecular sieving, Fig. 11A).

However, the size of the various REEs is extremely similar, with a little difference in radius from La (0.106 nm) to Lu (0.0848 nm). Bearing in mind that polymer membranes do not exhibit a uniform pore size distribution, it is therefore difficult to achieve efficient selectivity between REEs by means of molecular sieving.

Moreover, the typical pore size of microfiltration (MF) and ultrafiltration (UF) membranes is too large (~2-100 nm for UF membranes and >100 nm for MF membranes) to reject REEs dissolved in water by means of these pressure-driven processes. An effective strategy for retaining REEs with a UF membrane is to artificially increase their size in the feed solution using complexing species (Fig. 12).

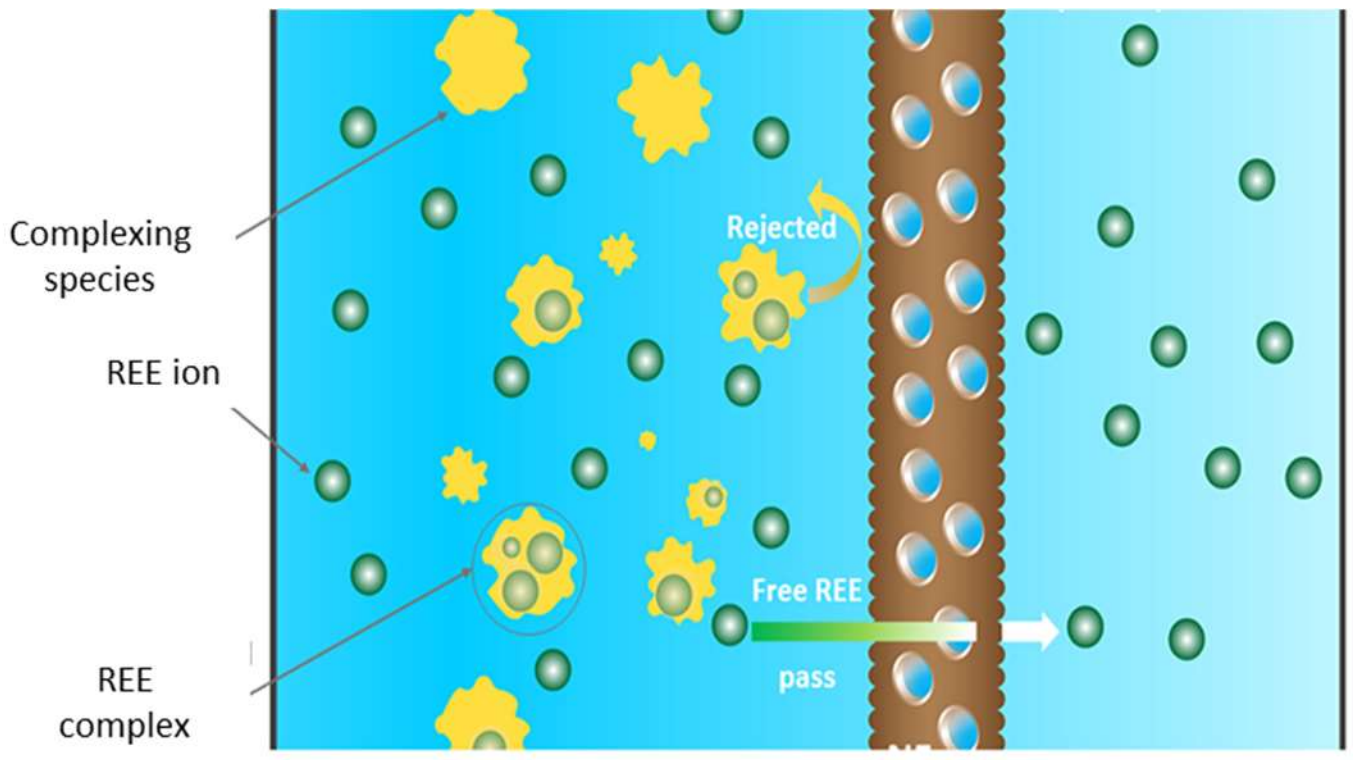

**Fig.12**. Schematic representation of complexation-assisted UF. Reproduced with permission from [160] Copyright © 2005, Elsevier CC license.

Different complexing species have been considered in this complexation-assisted UF strategy, such as polymers (polymer-assisted ultrafiltration (PAUF) [161]), surfactants (micellar-enhanced ultrafiltration (MEUF) [162]), humic matter [163], and organic extractants [164]. For example, Sorin et al. [164] showed that the rejection of gadolinium by a polyamide membrane with a cut-off of 2500 Da (i.e. at the edge between ultrafiltration and nanofiltration) could reach 95% by complexing Gd with diethylenetriaminepentaacetic acid (DTPA), whereas the rejection of uncomplexed Gd (in the form of $Gd(NO_3)_3$ salt) was less than 10% whatever the pH of the feed solution (Fig. 13) [164].

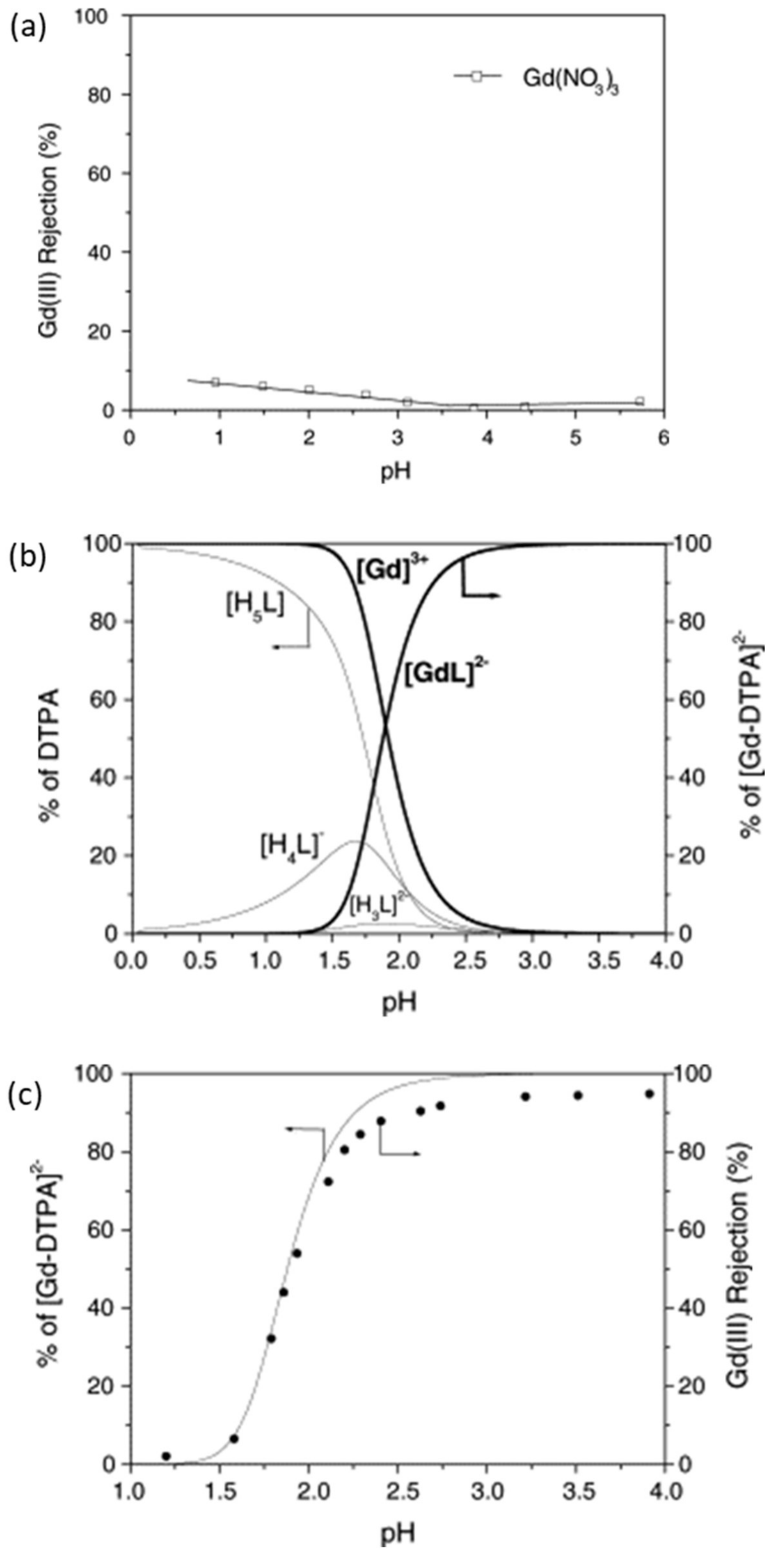

**Fig. 13**. (a) Experimental rejection of Gd (in the form of $Gd(NO_3)_3$ salt; feed concentration: 0.3 mM) at 4 bar by Desal G10 membrane vs. pH; (b) Theoretical speciation diagram for Gd(+III) / DTPA system (L stands for DTPA in the figure); (c) Theoretical speciation of Gd-DTPA complex vs. pH (left axis) and experimental rejection of Gd in the presence of DTPA ([Gd(+III) = [DTPA] = 0.3 mM) at 4 bar by Desal G10 membrane vs. pH (right axis). Reproduced with permission from [164] Copyright © 2005, Elsevier.

Another possible rejection mechanism by membranes is the so-called Donnan exclusion, which refers to electrostatic repulsion between ions bearing an electrical charge of the same sign as the membrane. The mechanism of membrane charge formation can be quite complex, including ionization of surface functional groups (such as carboxylic or sulfonic acids, ammonium) [165-167] as well as adsorption of charged species (ions, charged surfactants, etc.) from the solution onto the membrane surface [168].

The Donnan exclusion mechanism is particularly important for membranes with pore sizes comparable to the Debye length of the solution. It is for instance well-established that mass transport and surface charge effects are strongly interrelated in nanofiltration (NF; a pressured-driven process using membranes pore size in the range 1-2 nm), particularly at low to moderate concentrations of the charged solutes (the membrane charge being progressively screened as ionic strength of solution increases) [169, 170].

Zhao et al. [171] used NF to separate REEs from other cations of different valences and obtained selectivity factors of up to 11.88 between sodium and neodymium and 12.43 between sodium and cerium (Fig. 14). Lopez et al. [172] evaluated the performance of two NF membranes for the recovery of REEs from acid mine water. As in the study by Zhao et al. [171], the authors obtained a fair separation with membrane permeance to REEs much lower than other species with different valence but with low selectivity between REEs (Fig. 15).

Alternatively, Pramanik et al. [173] investigated the potential of the forward osmosis (FO) process for the separation of REEs with the main argument that this osmotic process is less energy intensive than NF as it does not require an external supply of hydrostatic pressure [173] (water diffuses through a dense FO membrane from the feed solution to the draw solution that is characterized by a higher osmotic pressure than the feed solution [174]). Qualitatively similar conclusions to NF were obtained in terms of separation performance, i.e. good REE rejections, from 82 to 96% depending on the solution pH and the membrane orientation, but little selectivity between the different REEs (La, Ce and Dy) [173].

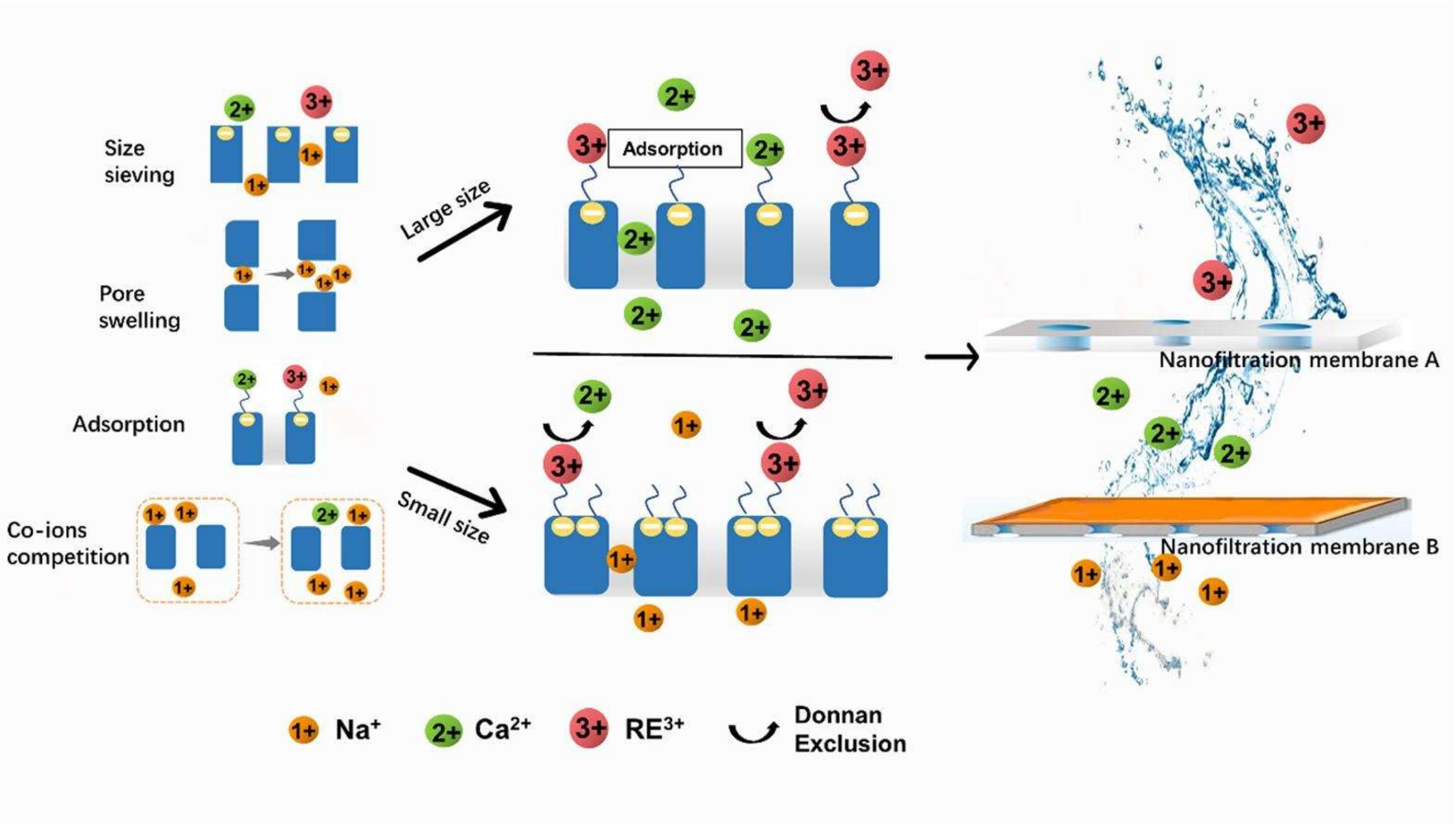

**Fig. 14**. The explanation of ions selectivity of nanofiltration membranes for rare earth wastewater treatment. Reproduced with permission from [171] Copyright © 2022, Elsevier.

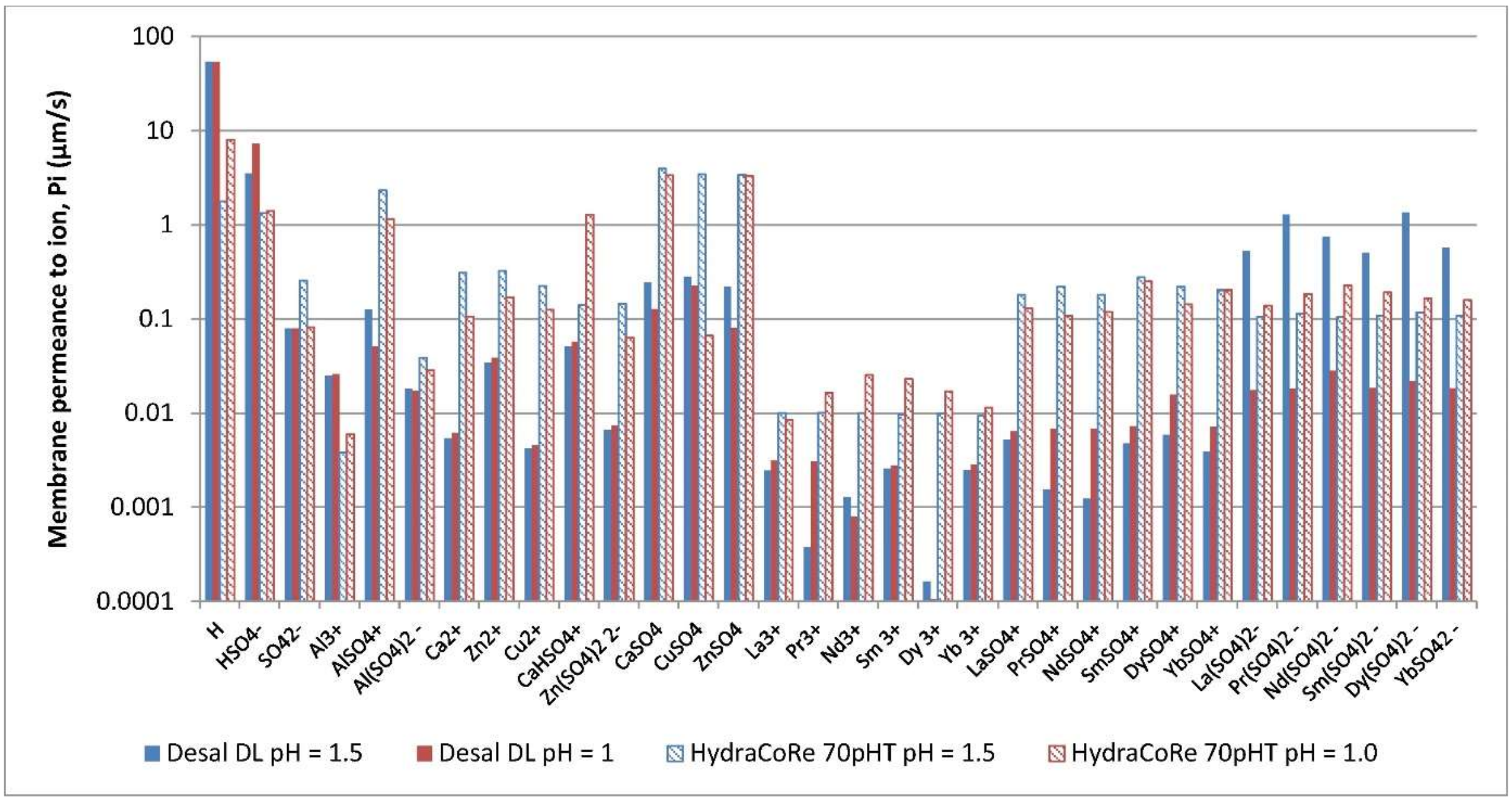

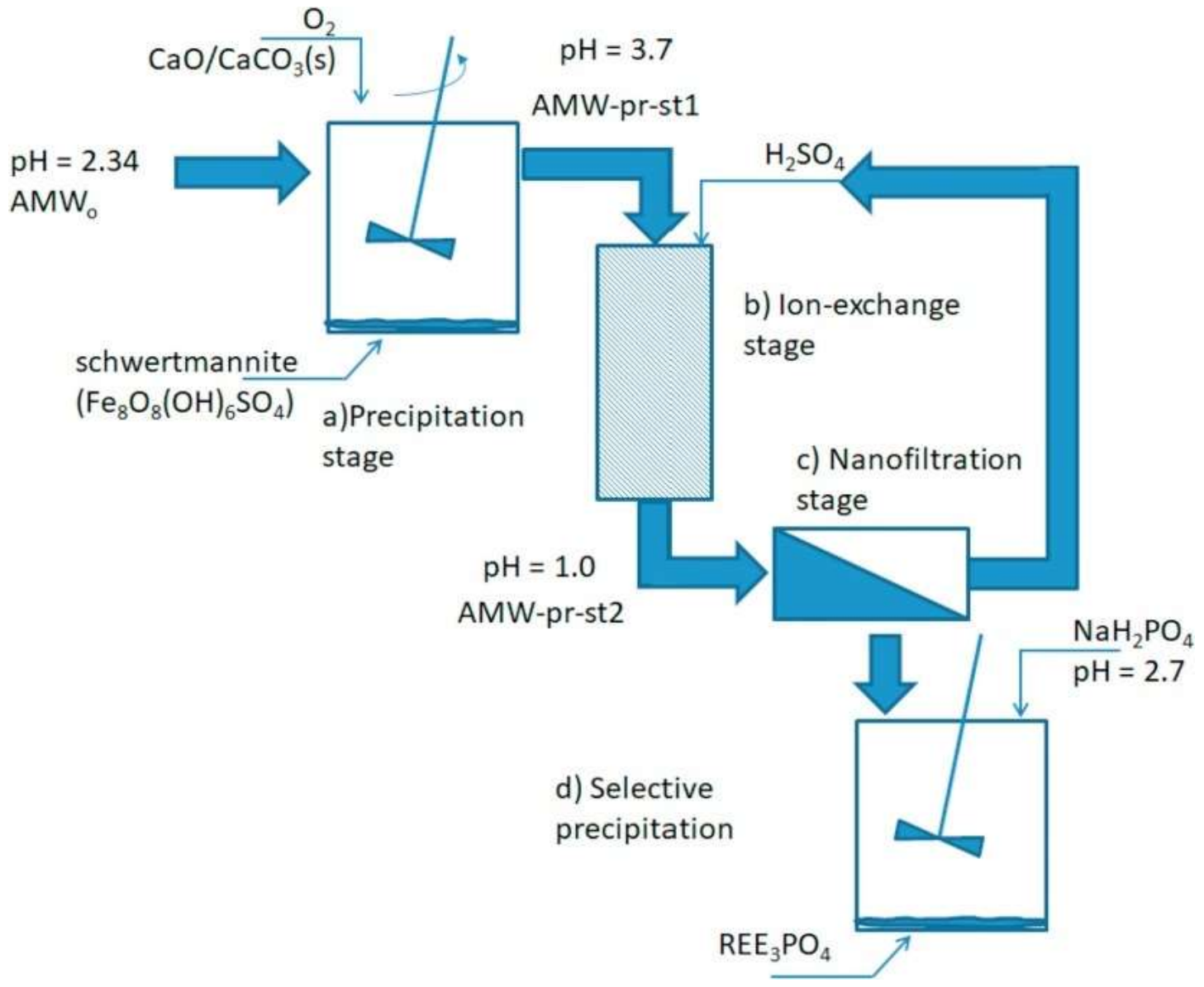

**Fig. 15.** Calculated Desal DL and HydraCoRe 70pHT membrane permeances to ions for NF experiments performed with an acid mine water model solution (upper image). Proposed treatment of an AMW including a) total oxidation of Fe(II) to Fe(III) and precipitation with $CaO/CaCO_3$; b) concentration of valuable metals with IX resins; c) recovery of $H_2SO_4$ and concentration of valuable metals with NF; and d) selective precipitation of REE as phosphates (bottom image). Reproduced with permission from [172] Copyright © 2019, Elsevier.

According to the Donnan exclusion principle, a more efficient rejection would be expected between REE ions and a positively charged membrane due to the strong electrostatic repulsion between trivalent REE ions and the membrane surface of the same sign. Unfortunately, most polymer membranes are negatively charged over a wide range of pH, their isoelectric point being usually reported between pH 2 and 4 [175]. A possible strategy is to lower the pH of the solution below the membrane isoelectric point. The membrane thus becomes positively charged, due to protonation of basic surface groups (e.g. amines) and/or adsorption of protons from the surrounding solution. However, as the ionic strength increases with the medium acidity, the positive surface charge of the membrane is screened, which weakens the Donnan exclusion and results in the increase in REE permeation through the membrane [176]. Hammache et al. [177] developed polymer membranes by mixing cellulose triacetate (CTA) and a cationic polyelectrolyte (polyethylenimine, PEI) with the addition of amino-based or phosphorous-based additives. The isoelectric points of the resulting membranes were found between pH 4.6 and 8.6 due to the presence of positively charged quaternary ammonium groups, making it possible to have positively charged membrane without the need to carry out membrane separation at very low pH (and thus high ionic strength). Membranes were employed to recover neighboring REE elements, Nd and Pr, from electronic waste (magnets of end-of-life computer hard-disk drives) by diffusion dialysis [177]. The strong electrostatic repulsion between the membranes surface and $Nd^{3+}$ and $Pr^{3+}$ ions led to efficient rejection of REEs by the membranes and high selectivity factors between boron (in the form of uncharged boric acid molecules) and REEs, with values up to 3706 for the B / Nd selectivity factor as shown in Fig. 16.

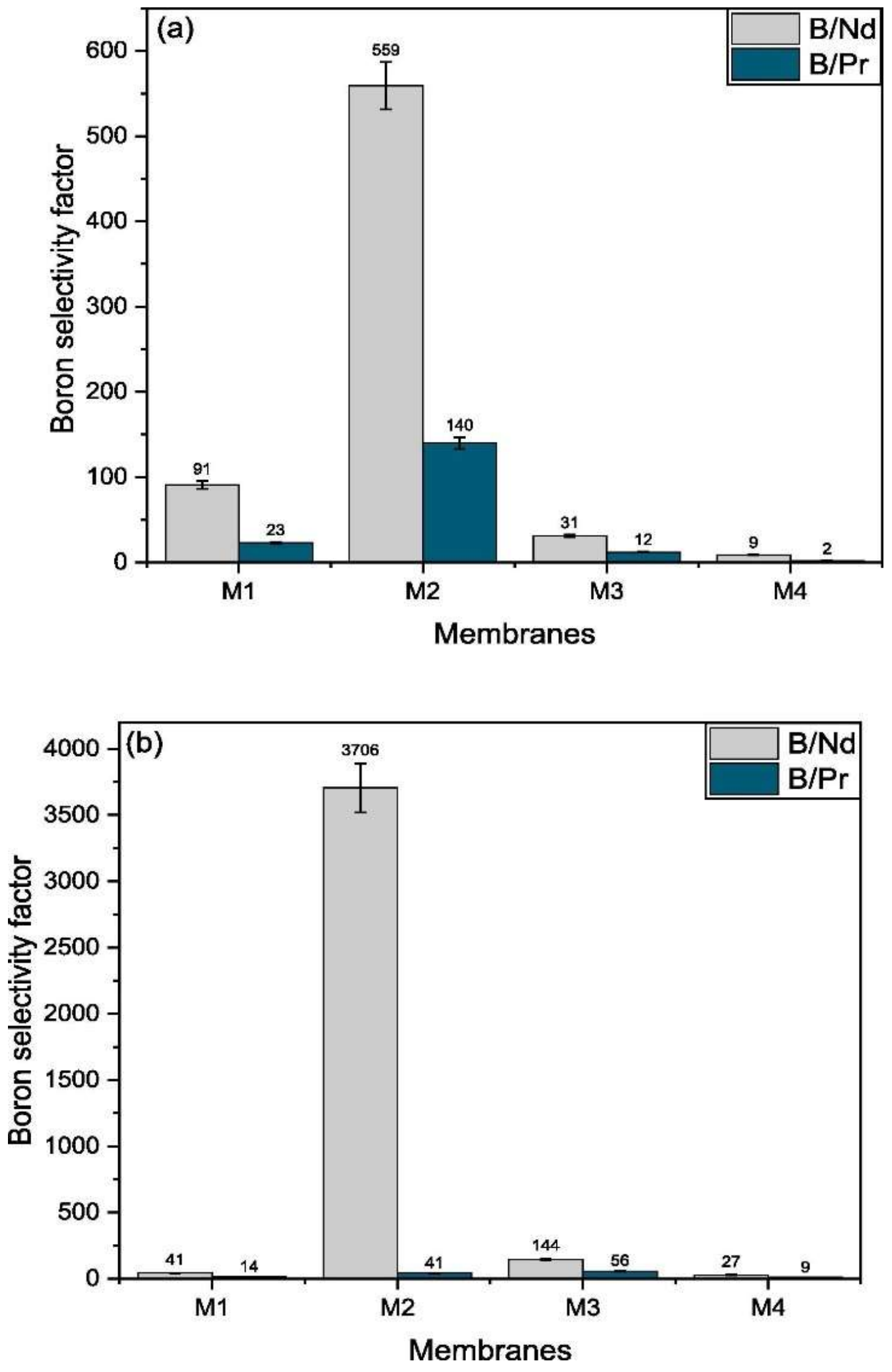

**Fig. 16**. Selectivity factor B/Nd and B/Pr for various CTA-PEI membranes containing amino-based additives (tridodecylamine for M2 and trioctylamine for M3) and phosphorous-based (di-

(2-ethylhexyl) phosphoric acid for M1 and trioctylphosphine oxide for M4) additives. (a) Feed solution composition (mg $L^{-1}$): B: 8.724, Pr: 35.711, Nd: 119.141; (b) Feed solution composition (mg $L^{-1}$): B: 12.630, Pr: 48.142, Nd: 944.721. Reproduced with permission from [177] Copyright © 2021, Elsevier.

Although not reported by the authors, the results in Fig. 16 indicate interesting selectivity between the two neighboring REEs, with Pr/Nd selectivity factors ranging from 2.5 up to 90.4 depending on the membrane and feed solution composition [177].

The dense membranes synthesized by Hammache et al. [177] exhibited higher permeation for $Pr^{3+}$ than $Nd^{3+}$ although both cations are trivalent. As they all had a dense (i.e. non-porous) structure, it is unlikely that solutes could penetrate the membrane phase while keeping fully hydrated [178]. Being charged like $Pr^{3+}$ but slightly smaller, $Nd^{3+}$ has a higher charge density than $Pr^{3+}$ and thus a higher Gibbs energy of hydration (−3280 and −3245 kJ $mol^{-1}$ for $Nd^{3+}$ and $Pr^{3+}$, respectively). The associated excess solvation energy, i.e., the extra-work required to transfer a solute from the bulk solution into the membrane phase, is therefore higher for $Nd^{3+}$ than $Pr^{3+}$. This phenomenon represents another rejection mechanism, usually termed dielectric exclusion in membrane science, that is relevant for dense and non-porous membranes [179-181].

# 3. Extraction and separation methods of REEs

## 3.1. Technological non-membrane-based processes

Cascade [182-184] and fuzzy extraction technologies [49, 185, 186], still overrepresented in the contemporary industry, are also applied as an approach toward scalability of the pre-optimized extraction. This general characteristic stems from the fact that these are multi-stage operations consuming large amounts of energy and organic chemicals, producing large amounts of acidic and/or alkaline wastewater, and therefore leading to severe environmental pollution. Therefore, cascade (Fig. 17) and fuzzy extraction technologies cannot be considered as green solutions; this statement is particularly true since they can be immediately replaced by MSTs. However, a few examples could be recalled in this field though still requiring verification when scalable. For instance, non-aqueous solvent extraction of (heavy) REE hydroxide concentrate from mining waste, redissolved in ethylene glycol + 10 vol.% water, 0.43 M HCl and 0.8 M NaCl, was successfully transformed into 16-stage lab-scale mixer-settlers leading to obtaining thulium and dysprosium group elements of the purity – from originally 34% and 54% to ultimately 99.8% and 98.7%, respectively [49]. Further, REEs were extracted and purified from industrial sludge in a closed-loop system – demonstrated as scalable and economically viable – using two green chemistry washing solutions: (a) [$(NH_4)_2SO_4$, *N*,*N*-bis(carboxymethyl)glutamic acid, tetrabutylammonium bromide, and water], and (b) a porous *β*-cyclodextrin polymer composite (PCDP-M-SHM), with no significant differences in their extracting power. The recycling effectiveness after the purification of leached REEs using PCDP-M-SHM were various for different REEs, in the range of 76% (Gd) to 87% (Pr), and 8% for Ce [187].

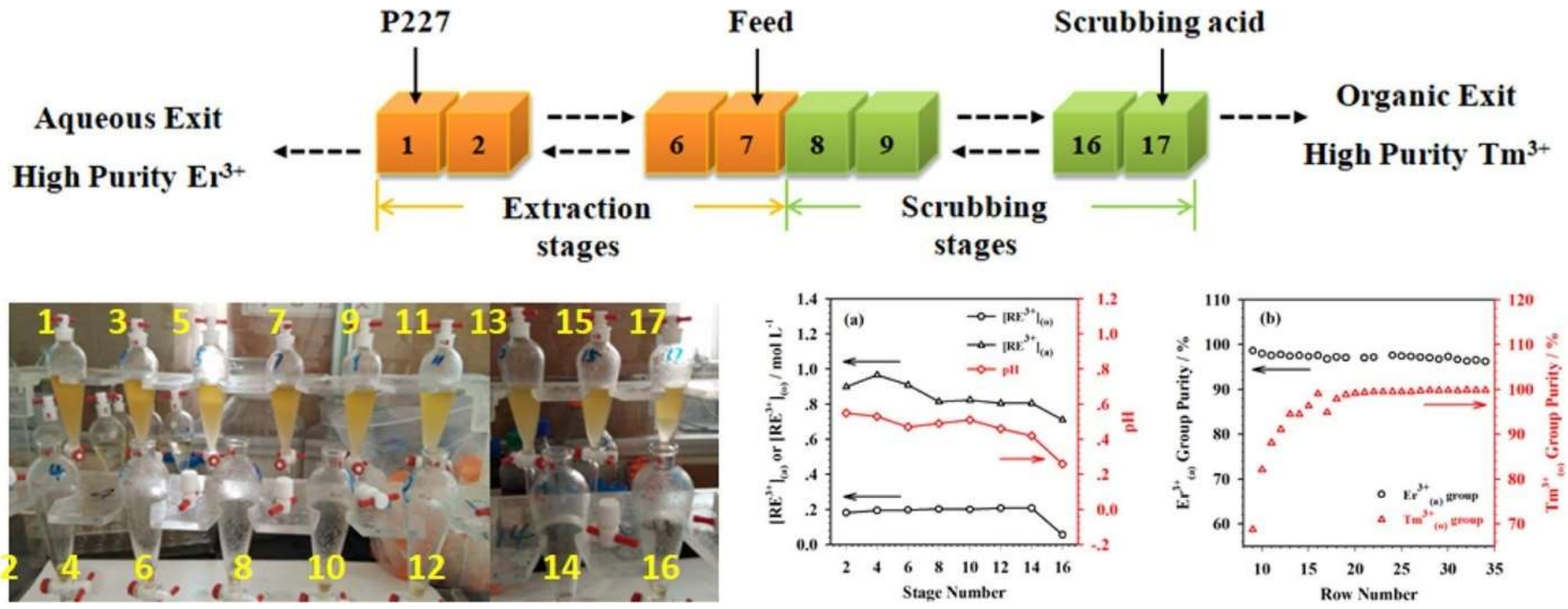

**Fig. 17**. The cascade extraction simulation of $Er^{3+}/Tm^{3+}$ separation. Reproduced with permission from [182] Copyright © 2020, Elsevier.

### 3.2. Solid-liquid separation without membrane implementation

#### 3.2.1. Electrical transformations

An alternative and cleaner than the environmentally hazardous ‘precipitation-roasting’ strategy employing REE chloride-to-oxide transformations, a cationic membrane electrolysis (CME) of 97% current efficiency was studied. This method, characterized by a zero-waste liquid and zero gas emission, yielded pure and fine microparticles of $Nd_2O_3$, $Sm_2O_3$, $Gd_2O_3$, $Yb_2O_3$ and $CeO_2$ after and without calcination, respectively [188]. Furthermore, electro-assisted extraction, i.e., a prolonged, two-compartment electrodialysis allowed to recover REE (from solutes of pH 2) from bituminous (138 ppm) and anthracite (447 ppm, Nd 65 ppm) coal fly ash allowing to extract >70% of REEs [189]. Similarly, but applying orders of magnitude more concentrated secondary source of REEs, i.e., a NdFeB magnet, Nd was solubilized, mobilized, extracted and selectively recovered as oxalate in 95% yield [190].

#### 3.2.2. Solid-phase extraction

Recovery of REE (as oxalates and fluorides) from coal fly ash was achieved by applying choline chloride:*p*-toluenesulfonic acid monohydrate (1:1) deep eutectic solvent (DES) with leachability of >85%, ca. 35% higher than sulfuric(VI) acid [191] (Fig. 18). Yttrium (and other heavy REEs) were extracted from water using liquid *o*-octyloxybenzoic acid (and *o*-ethylhexyloxybenzoic acid) allowing, *via* the optimized fractional extraction, to obtain Y-product with a purity of 99.4% and yield of 96.4% [192]. A previously unexplored as the REE source, silicate-based ore was extracted using sulfuric(VI) acid-baking toward scandium and iron allowing to achieve 13.5% and 65.0% yield, respectively, clearly indicating a room for the future improvement [185]. Concerning the ‘properties-by-design’ approach toward novel extracting agents for REEs, a series of unsymmetrical diglycolamides was proved to act efficiently and release REEs under acidic conditions [193]. Subcritical water extraction (SWE) using citric acid (at 150 °C) as the leaching reagent was proved to rapidly (5 min) and efficiently extract REEs (90.85% of La, 88.84% of Ce, and 90.85% of Nd – as precipitated phosphates) from spent NiMH batteries, locating this approach as technologically valuable [194]. Yttrium and europium, after solid-phase chlorination by $NH_4Cl$ and acidic leaching, have been recovered from a fluorescent lamp waste (as the other secondary resources) by a four-stage cross-flow solvent extraction combining commercially available Cyanex 923 and Cyanex 572 yielding purity at the level ≥94% [195]. Similarly, selective leaching and separation of REEs was performed using $HNO_{3(aq)}$ and Cyanex 923, respectively. The counter-current mixer setter system encompassing three extraction and four stripping stages yielded as the final product yttrium (94.61%)/europium (5.09%) oxides (by calcination of oxalates) [196]. Concerning the other secondary REE resources, i.e. spent NdFeB magnets, an alternative, low-energy and green route to selective recovery of REEs (ca. 100% purity of Nd) from using conc. $ZnCl_{2(aq)}$ (stable up to 4 cycles) was developed and elaborated. The process was characterized by a high Nd/Fe separation factor (>1×10$^5$) upon dissolution of oxides [197]. Also, $(NH_4)_2SO_{4(aq)}$ (300 g L$^{-1}$, 120 °C 180 min, solid/liquid phase ratio 1:3) was applied to recover 99.98% of REEs and ca. 100% of Zn from the spent nickel–metal hydride (Ni–MH) batteries [198].

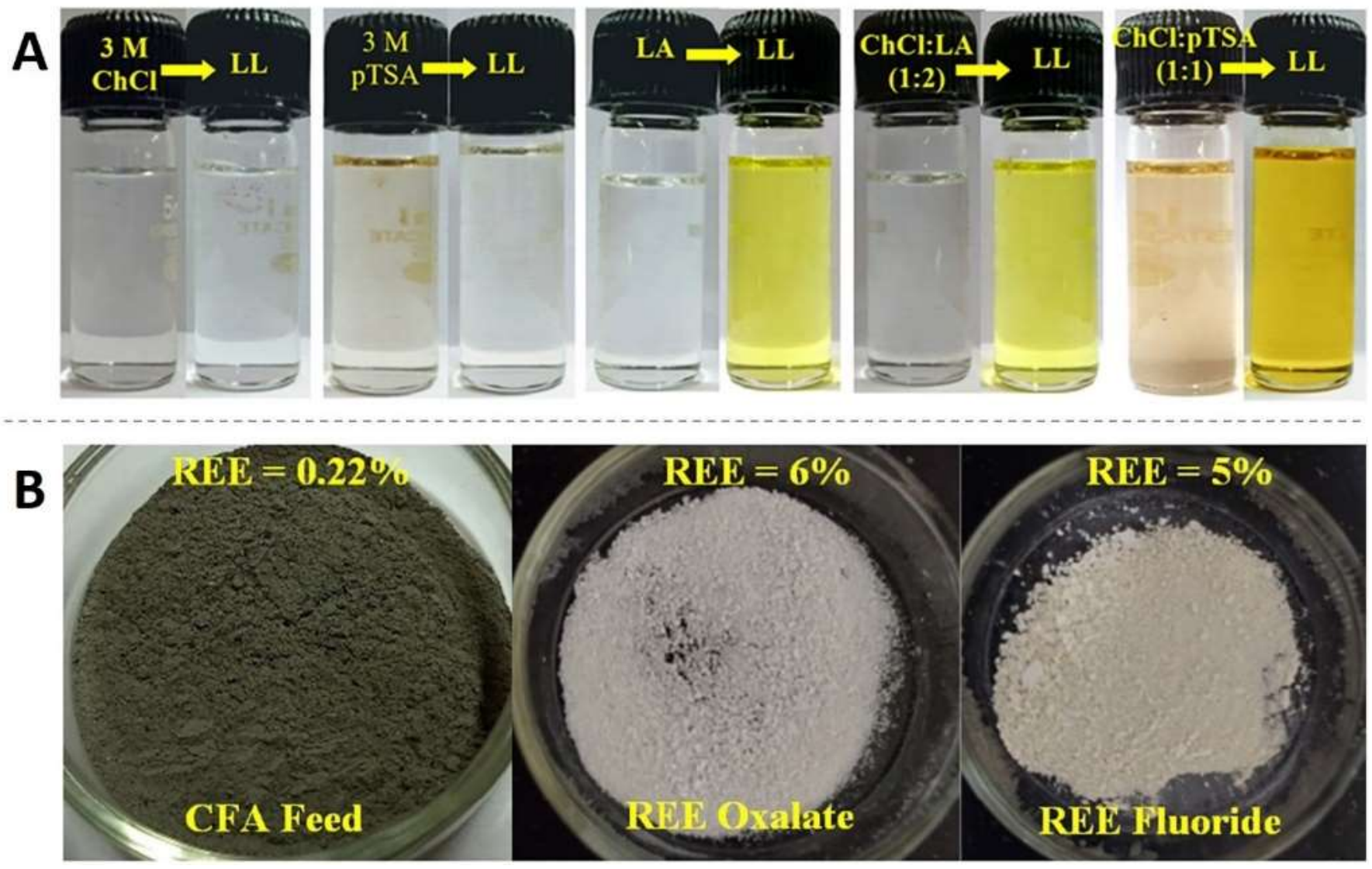

**Fig. 18.** Images of HBD, HBA and DES reagents pre- and post-leaching (LL – Leach Liquor) (A). Precipitated REE-rich products obtained with ChCl:pTSA(1:1) leaching of CFA. (a) CFA, (b) REE-Oxalate and (c) REE-Fluoride. Reproduced with permission from [191] Copyright © 2022, Elsevier.

### 3.2.3. Microorganism- and bio-derived systems

Microorganism systems represent a prospective alternative in the area of biosorption and hence separation of REEs. A method exploring a continuous flow filtration assay, based on the adsorption and pH-dependent desorption of lanthanide to/from the pre-protonated bacterium *Roseobacter AzwK-3b* immobilized on the assay filter, was proven to efficiently concentrate a solution of equal concentrations of each lanthanide to nearly 50% of the three heaviest lanthanides (Tm, Lu, and Yb) in two passes. After protonation of bacteria with 2% $HNO_{3(aq)}$, selectivity was achieved due to a reduced biosorption of the lighter REEs but higher and similar biosorption of the heavier REEs. Similar trends were found for *Shewanella oneidensis*, *Sphingobacterium*, and *Halomonas*.

The above characteristics emerged as surpassing the available industrial processes [199]. Moreover, bioengineered *E. coli* strains were grown and applied to regulate the REE adsorption and recovery by sensing extraneous REEs. Specifically, adsorption capacity of Tb (as a model REE) reached the highest reported value of 41.9 mg $g^{-1}$ per dry cell weight, while its desorption efficiency >90% upon applying three bed volumes of citrate solution [200]. *Ulva lactuca*, an abundant marine macroalgae, allowed to recover Nd and Dy from an artificial seawater – after optimization of biosorbent stock density, ionic strength and contact time – with the removal efficiencies up to 98% and 89% for Nd for Dy, respectively [201]. Staying in the *Kingdom Plantae* and combining ion imprinting techniques with the bio-derived systems, polydopamine-modified basswood 3D-materials were applied to construct selective separation membranes decorated with Nd(III)-imprinted cavities of a high rebinding capacity (120.87 mg $g^{-1}$) and high permselectivity coefficients (>10) (Fig. 19) [156].

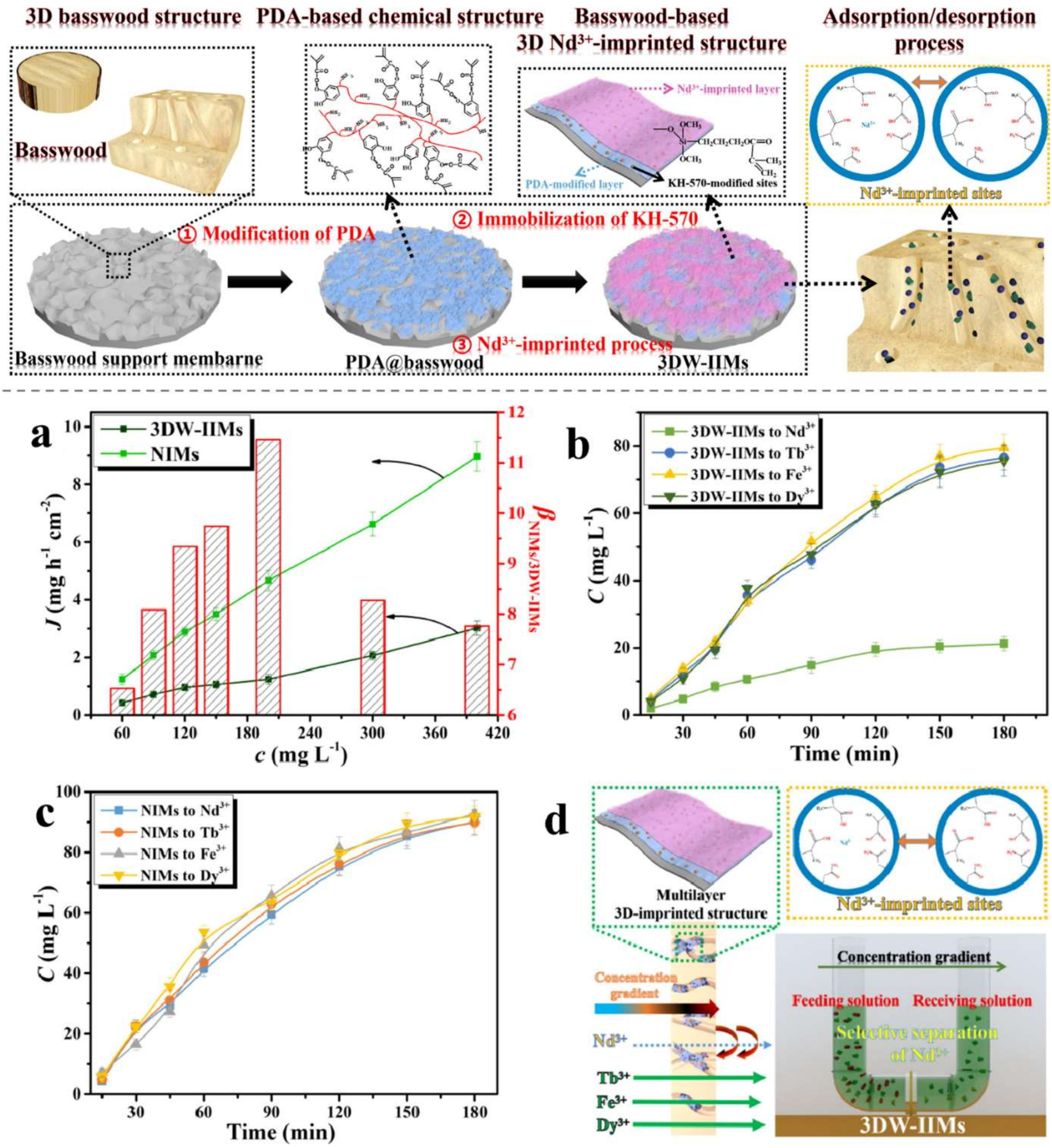

**Fig. 19**. Schematic diagram for the synthesis procedures of 3DW-IIMs (upper part). (a) Isothermal permeation results of 3DW-IIMs and NIMs toward $Nd^{3+}$ and the time-dependent permselectivity curves of $Nd^{3+}$, $Tb^{3+}$, $Fe^{3+}$, $Dy^{3+}$ through (b) 3DW-IIMs and (c) NIMs, (d) schematic representation of the possible permselectivity mechanism of 3DW-IIMs (bottom part). Reproduced with permission from [156] Copyright © 2021, Elsevier.

### 3.2.4. Other hydrometallurgical approaches

A clean and energy-efficient diffusion dialysis was proposed to recover Nd and Pr from NdFeB magnets of end-of-life computer hard disk drives. Precisely, a CTA/PEI/TDDA membrane was found as of the highest B/REE selectivity with values up to 3706 and 140 for Nd and Pr, respectively [177]. Also, $Nd^{3+}$ and $Dy^{3+}$ were recovered from (simulated and real) leaching solution of the NdFeB magnet *via* solid phase extraction (SPE) using $SiO_2$-$NH_2$, EDTA and/or phosphonic groups. For multi-component systems, the affinity of phosphorous/nitrogen adsorbents was found in the order: $Fe^{3+} > Dy^{3+} > Nd^{3+} > Ni^{2+} > Al^{3+}$. Extraction of REEs from supernatant liquid *via* bifunctional mesoporous silica with EDTA and/or phosphonic groups recovered 97.0% of $Nd^{3+}$, with $Ni^{2+}$ and $Al^{3+}$ as impurities, while it was the non-ordered silica functionalized with phosphonic groups which was characterized by the most prospective economy. Also, adsorption capacity by polyacrylonitrile (PAN) nanofibers (both under a batch and continuous mode) impregnated with a commercial organic extractant Cyanex 272, was found as 200 and 400 mg $g^{-1}$ for Y(III) and Eu(III) (from aqueous solution), respectively, revealing a high potential of membranes prepared therefrom [202]. Similarly, electrospun poly(vinyl alcohol) (PVA) nanofibres (d=280 nm) containing hydrothermally synthesized ultrasmall (3 nm) $CeO_2$ NPs (CNPs) (34 wt%) were found as efficient adsorbents of $Eu^{3+}$, $Gd^{3+}$ and $Yb^{3+}$ at from aqueous solutions at pH 5.8 [203]. As an example of other hydrophilic fibrous materials, carboxylated cellulose filters emerged as promising systems dedicated for removal and recovery of La(III) with a high permeation flux (dynamic adsorption at 0.07 kPa) and high adsorption capacity (33.7 mg $g^{-1}$) [204] (Fig. 20).

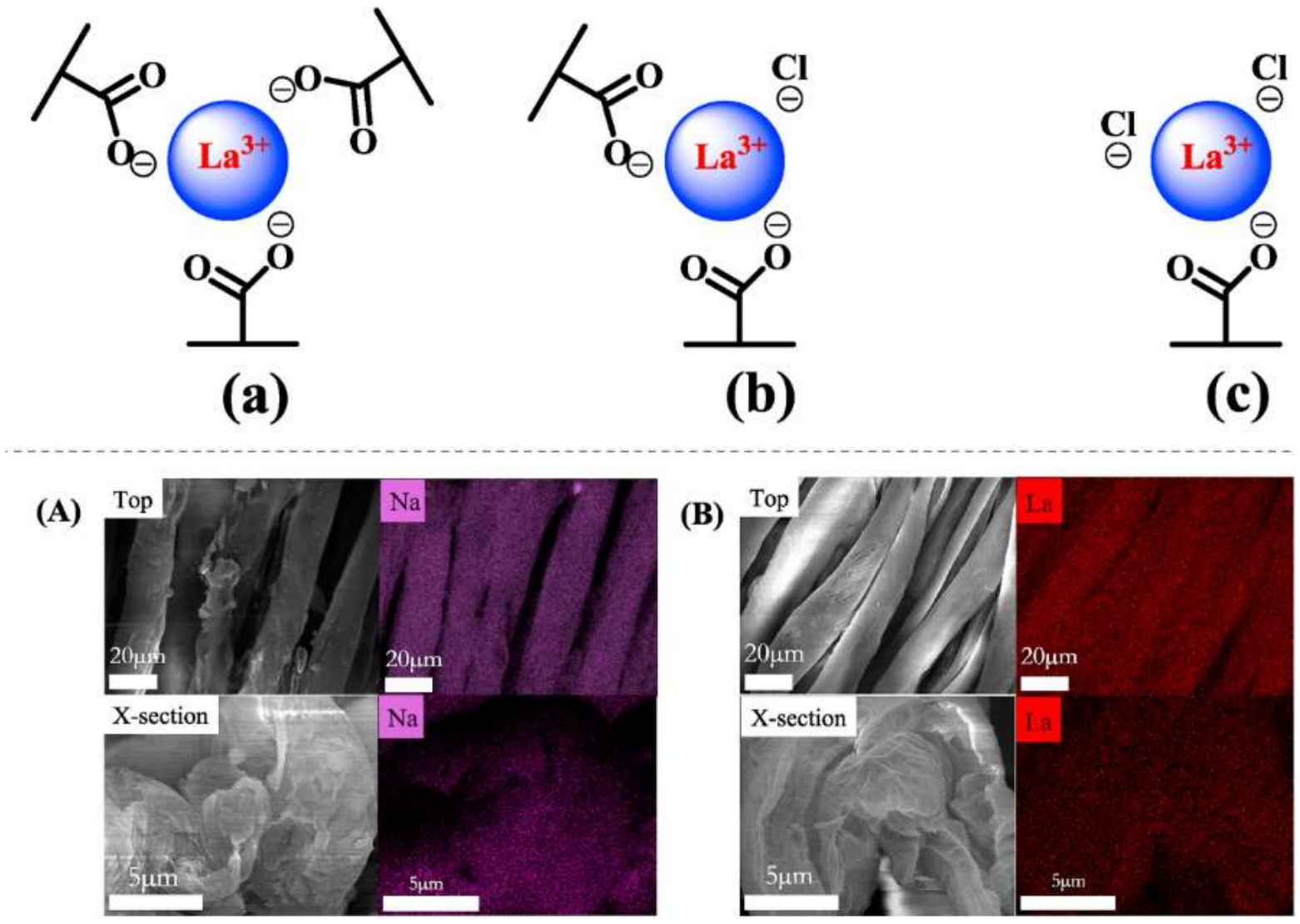

**Fig. 20.** Schematic representation of proposed electrostatic interactions between carboxylate groups located on carboxylated cellulose and La (III) ions (upper image). SEM and EDX mapping images of carboxylated cellulose fibers before (A) and after adsorption of La (III) (B) (bottom images). Reproduced with permission from [204] Copyright © 2020, Elsevier.

## 3.3. Solid-liquid separation with membrane implementation

### 3.3.1. Polymer inclusion membranes (PIM)

Polymer inclusion membranes (PIMs) are a new type of liquid membrane [205]. Usually, PIM consists of a base polymer and a carrier. Polymers offer mechanical properties while the carrier plays key roles in the transport of the targeted chemical species. In some cases, PIMs can also consist of plasticizers which increase the membrane elasticity and modifiers which make the extracted species more soluble in the membrane liquid phase [206]. PIMs are characterized by easy-forming performance, suitable mechanical properties, and a wide range of applications [207].

PIMs simultaneously combine the extraction and stripping processes into a single device. Additionally, PIM could be a promising green alternative for concentrating, separating, and recovering REEs due to its long-term stability, low carrier loss, elimination of significant amounts of diluents, and a lack of phase separation issues [208].

Polyvinylidene fluoride (PVDF) has drawn a lot of interest because of its outstanding thermal stability, chemical resistance, and well-defined film-forming capabilities [209]. Huang et al. [210] used a polymer inclusion membrane functionalized by task-specific ionic liquid di(2-ethylhexyl)phosphinic acid (P227) for the extraction of lutetium(III). PVDF was selected as a polymer-based material, while PP227 was used as a carrier and plasticizer. The first part of this research was related to finding the optimal amount of P227 in the PVDF matrix. Therefore, membranes with the P227 content varying from 20 wt% to 50 wt% were prepared. SEM analysis proved that the membrane containing 40 wt% of P227 demonstrated a certain regularity in the distribution and a hierarchically ordered structure of pores. Contact angle measurements revealed that the contact angle increased with increasing content of P227 in the polymer matrix. Subsequently, membranes were tested in the transport of Lu(III). The highest transport rate of lutetium(III) was observed for the P227@PIM (40 wt%). This PIM demonstrated the transport rate constant equal to 0.4220 $h^{-1}$. Based on the physiochemical analysis and transport rate, it can be concluded that the optimum amount of P227 in PIM was 40 wt% [210]. Huang et al. [211] tested also PIM with 60 wt% of PVDF and 40 wt% of P227 for the separation of Lu(III) from La(III) and Sm(III). It was also noticed that at pH = 1.5 after 5 h, the recovery factors of Lu(III), La(III), and Sm(III) were 85%, 40%, and 4%, respectively. The Sm(III) was successfully isolated from La(III) when the pH was raised to 2.4. This suggested that P227@PIM (40 wt%) can be used as a membrane to separate heavy and light REEs. Moreover, the properties of regenerated PIMs were also investigated. Regenerated PIM was prepared by dissolving the used in a proper amount of DMAc and cast on a glass plate. SEM analysis of the regenerated PIM showed that it was characterized by a similar porous structure to the non-regenerated PIM. Transport experiments proved that regenerated and normal PIMs were characterized by similar Lu(III) transport rate constant [211]. Another task-specific ionic liquid dialkylphosphoric acid [A336][P227] was incorporated into a PVDF matrix and was used for the separation of yttrium [212] (Fig. 21).

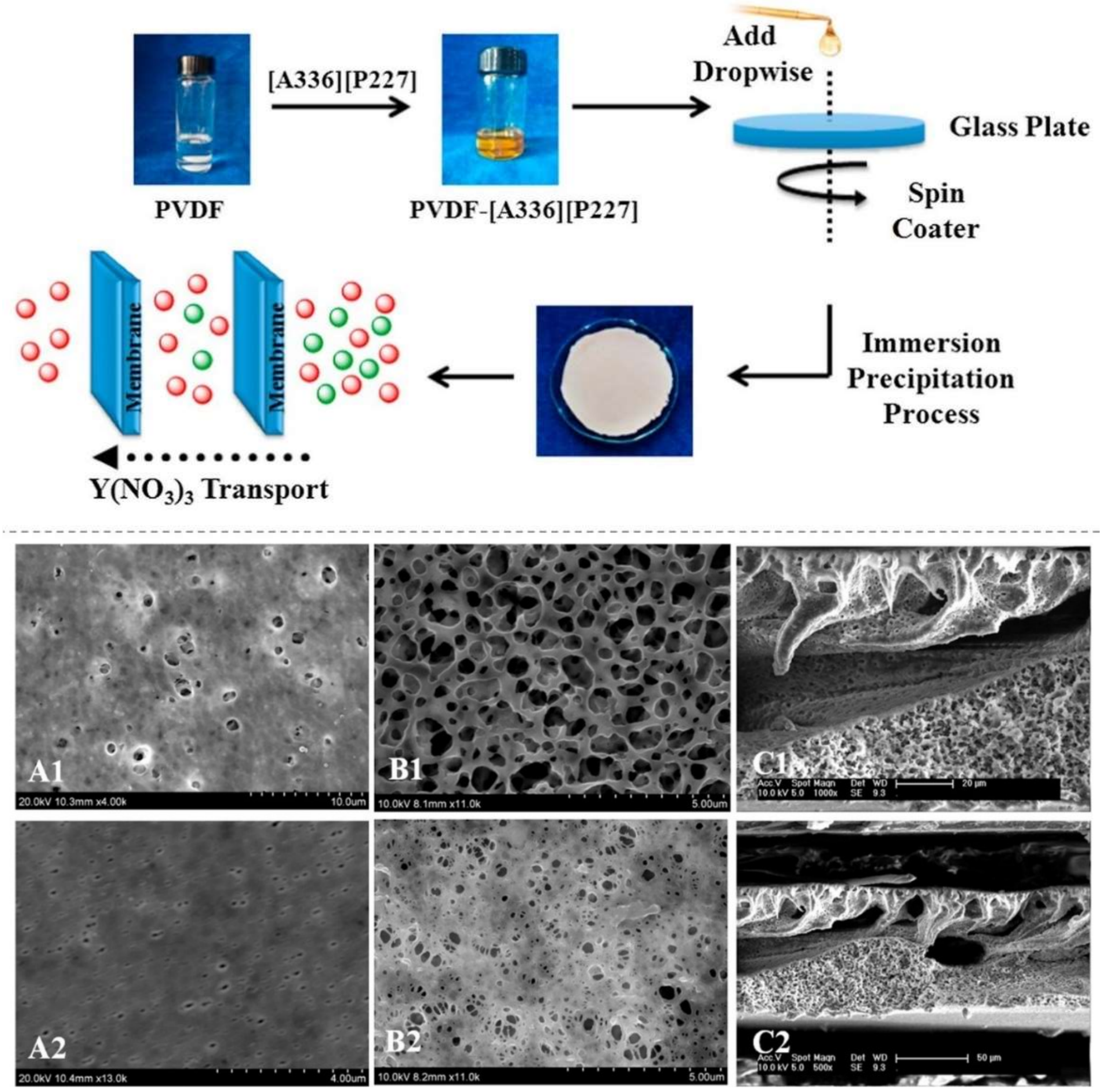

**Fig. 21.** Schematic illustration of the fabrication process of PVDF-[A336][P227] polymer inclusion membrane (upper image). SEM images of PVDF and PVDF-[A336][P227] PIM (A1: large pore side of PVDF, A2: small pore side of PVDF; B1: large pore side of PIM, B2: small pore side of PIM, C1 and C2: cross section of PIM) (bottom images). Reproduced with permission from [212] Copyright © 2018, Elsevier.

Results indicated that prepared PIM (containing 37.5 wt% of [A336][P227] and 62.5 wt% of PVDF) was suitable for the separation of yttrium from holmium and erbium. Additionally, it was also proven that the PIM membrane was stable for 8 cycles. The thickness and flux of the yttrium did not change after the long-term stability test [212]. Chen and Chen [213] incorporated [tricaprylmethylammonium][di(2-ethylhexyl)orthophosphinate] [A336][P507] into PVDF for the separation of Lu(III). SEM analysis demonstrated that a uniform distribution of pores characterized PIM. While the AFM analysis showed that the task-specific ionic liquid was also homogeneously distributed in the PVDF polymer matrix. It was also noticed that various parameters influence the efficiency of the separation of Lu(III). In the case of the content of [A336][P507], results indicated that the permeability coefficient increased with increasing content of carrier in PVDF. The highest permeability coefficient was noticed for the PIM with 60 wt% of [A336][P507] with a stirring speed of 450 rpm. Further increase in stirring speed did not increase the permeability coefficient. Subsequently, PIM was also used in the separation of Yb(III) and Lu(III). Surprisingly, PIM showed better efficiency in the extraction of Yb(III). Long-term stability tests demonstrated that membranes could be used for up to 10 cycles. Moreover, the thickness of the PIM decreased by only 2.6% after 10 cycles [213]. Chen et al. [214] prepared a new type of PIM for the separation of Yb(III) and Lu(III). This new type of PIM consisted of a mixture of PVDF and poly(vinyl-alcohol-co-ethylene) (EVOH) as a base polymer and Cyanex27 as the carrier. SEM analysis of the surface of membranes proved that the incorporation of EVOH into the polymer matrix increased the surface pores and internal channels. Based on the static adsorption dynamics permeation experiments, the PIM with the composition of 25 wt% Cyanex 272 and 10 wt% EVOH was selected for the separation of Yb(III) and Lu(III). Analyzing the results, the membrane demonstrated permeability coefficients of Lu(III) and Yb(III) equal to 41.62 mg $L^{-1}$ and 77.46 mg $L^{-1}$, respectively. While the selectivity ($\beta_{Yb(III)/Lu(III)}$) was equal to 3.78 [214]. Wan et al. [215] tested the PIM membrane consisting of task-specific ionic liquid trihexyl(tetradecyl)phosphonium bis(2,4,4-trimethylpentyl)phosphinate (Cyphos IL 104) (carrier) and PVDF (polymer) in the separation of the Yb(III) and Lu(III). It was found that membranes containing Cyphos IL 104 were characterized by bigger pore sizes compared with the original PVDF membrane. When the content of IL increased from 11.76 wt% to 37.5 wt% the pore size increased from 0.62 µm to 0.97 µm. in view of the separation efficiency of PIMs membranes in the separation of Yb(III) and Lu(III), PIM membranes containing 25 wt% showed the best performances. This membrane demonstrated the permeability of Lu(III) equal to 114.8 μm $s^{-1}$ and Yb(III) equal to 156.0 μm $s^{-1}$. Moreover, the long-term stability of the membranes was also tested. After 8 separation experiments, the permeation flux of the PIM decreased by 24.6% [215].

Makówka and Pośpiech [216] prepared PIM based on cellulose triacetate (CTA) for the separation of lanthanum(III) and cerium(III). Prepared membranes consisted also of ion carriers: di(2-ethylhexyl)phosphoric acid (D2EHPA) or tributyl phosphate (TBP) and plasticizer 2-nitrophenyl octyl ether (NPOE). The prepared membranes contained 30 wt% of CTA, 40 wt% of NPOE, and 30 wt% of D2EHPA or TBP. The obtained results suggested that the prepared membranes could be successfully used for the separation of La(III) and Ce(III). Higher separation efficiency was found for PIM with the D2EHPA ($\beta_{Ce(III)/La(III)}$=2.8) compared with PIM with TBP ($\beta_{Ce(III)/La(III)}$=1.4). Moreover, PIM with the D2EHPA demonstrated twice time higher La(III) flux ($J_{La(III)}$=11.1 μmol $m^{-2}$ $s^{-1}$) and three-time higher Ce(III) flux ($J_{Ce(III)}$=30.7 μmol $m^{-2}$ $s^{-1}$) in comparison with PIM with TBP ($J_{La(III)}$=6.5 μmol $m^{-2}$ $s^{-1}$, $J_{Ce(III)}$=9.2 μmol $m^{-2}$ $s^{-1}$) [216]. Makówka and Pośpiech [217] also separated Ce(III) from the solution containing La(III), Cu(II), Co(II), and Ni(II) using PIM containing CTA, CyphosIL 104, and plasticizer nitrophenyl octyl ether (NPOE). Prepared PIMs differed in the amount of CyphosIL 104 (15 wt%, 25 wt%, 30 wt%) and NPOE (50 wt%, 55 wt%; 65 wt%). It was found that CyphosIL 104 was suitable for the extraction of Ce(III) and La(III) and demonstrated the highest efficiency at pH=4.25. Transport experiments revealed that the highest initial flux was noted for PIM containing 20 wt% of CTA, 55 wt% of NPOE, and 25 wt% of CyphosIL 104 (PIM2). At the high content of ion carrier, PIM was characterized by too high viscosity to reach a sufficient value of transport of ion-carrier complex through the membrane. The PIM2 demonstrated a higher efficiency during the extraction of Ce(III) ($P$=3.1 μm $s^{-1}$, $RF$=68.1%) in comparison with the efficiency of removal of La(III) ($P$=0.5 μm $s^{-1}$, $RF$=16.2%). Subsequently, PIM2 was tested in the separation of Ce(III) from the solution containing La(III), Cu(II), Co(II), and Ni(II). Results indicated that prepared PIM2 was able to selectively extract Ce(III) from the mixture of various ions. The highest removal factor was found for Ce(III) (RF=67%). The removal factor of other ions was in the range of 3.6% (Ni(II)) to 15.7% (La(III)) [217]. Sharaf et al. [218] also selected cellulose triacetate as a base polymer for the preparation of PIM. Dioctylphalate (DOP) and 2-nitrophenyloctyl ether (2NOPE) were selected as plasticizers, while the PC-88A and Versatic 10 were used as carriers. In the preliminary study, the optimization of the PIM composition was performed. It was noticed that the quantitative extraction of the Sc(III) increased with increasing the content of PC-88A or Versatic 10 in the PIM. PC-88A demonstrated better ability in the extraction of Scandium in comparison with Versatic 10. Versatic 10 showed a greater efficiency in the back-extraction. Therefore, the influence of the mixture of ion carriers on the extraction was investigated. Based on the obtained results it was concluded that the highest efficiency of extraction and back-extraction was noticed for PIM containing 4 wt% of PC-88A and 36 wt% of Versatic 10. In the case of plasticizer, the best efficiency was found for the PIM with 40 wt% of plasticizer. Subsequently, the optimized PIM membrane was used in transport experiments. The optimized PIM was characterized by permeability and flux of Sc(III) equal to $6.77 \cdot 10^{-3}$ m $h^{-1}$ and $1.88 \cdot 10^{-7}$ mol $m^{-2}$ $s^{-1}$, respectively [218]. Yoshida et al. [219] prepared a series of PIM based on CTA with various ion carriers ((*N*-[*N,N*-di(2-ethylhexyl)aminocarbonylmethyl]glycine (D2EHAG), *N*-[*N,N*-di(2-ethylhexyl)aminocarbonylmethyl]phenylalanine (D2EHAF), *N*-[*N,N*-di(2-ethylhexyl)aminocarbonylmethyl]sarcosine (D2EHAS), D2EHPA, Versatic 10, and TOPO)). During the preliminary studies, the efficiency of various carriers in the extraction and stripping of Sc(III) at acidic pH was investigated. The results indicated that D2EHAG and D2EHAF were able to extract Sc(III) while D2EHPA showed sufficient extraction of Sc(III) at pH below 0.5. Such low pH may be the problem with the stripping of Sc(III). In the case of Versatic 10 and TOPO carriers, these two carriers demonstrated poor efficiency in the extraction of Sc(III).

Based on the preliminary studies, D2EHAG and D2EHAF were selected for the extraction of Sc(III) from the solution containing various common metal ions such as Fe(III), Ni(II), Al(III), Co(II), Mn(II), Cr(III), Ca(II), and Mg(II). It was noticed that PIM with a D2EHAF carrier was able to separate Sc(III) from Ni(II), Al(III), Co(II), Mn(II), Cr(III), Ca(II), Mg(II) and partially from Fe(III). PIM with D2RHAF was characterized by Sc(III) flux equal to $1.9 \cdot 10^{-7}$ mol $m^{-2}$ $s^{-1}$ and a much lower flux of other ions [219]. Ansari et al. [220] used PIM based on CTA and *N,N,N′,N′*-tetraoctyl-3-oxapentanediamide(TODGA) for the facilitated transport of La, Eu, and Lu. Additionally, plasticizer NPOE was added to obtain a more flexible membrane. Performed transport experiments revealed that the highest flux was noticed for La(III) ($7.3 \cdot 10^{-8}$ mol $m^{-2}$ $s^{-1}$) while the smallest was for Lu(III) ($2.1 \cdot 10^{-8}$ mol $m^{-2}$ $s^{-1}$). It can be concluded that the transport of lanthanides followed the order of their ionic potential (Lu(III)>Eu(III)>La(III)). Moreover, it was also noticed that after 5 h, PIM quantitatively transported Lu(III) while only 60% of La(III) was transported [220].

Croft et al. [149] tested the possibility of using the PIM based on the poly(vinyl chloride) (PVC) and D2EHPA for the selective extraction of the La(III), Gd(III), and Yb(III). Ion carrier D2EHPA was selected due to its enhanced capability for extraction and separation of lanthanides. PIM with 45 wt% of D2EHPA was selected for the extraction of lanthanides. The efficiency of the extraction of Yb(III), Gd(III), and La(III) was performed at pH equal to 0.45, 1.35, and 2.50, respectively. The highest flux was noticed during the extraction of La(III) ($7 \cdot 10^{7}$ mol $m^{-2}$ $s^{-1}$) while the smallest one was for the Yb(III) ($4 \cdot 10^{7}$ mol $m^{-2}$ $s^{-1}$). Moreover, according to calculations, the thermodynamic extraction constants for Yb(III), Gd(III), and La(III) were 92,7, 85.5, and 0.9, respectively [149]. Zaheri and Ghassabzedeh [221] prepared PIM based on PVC and a mixture of D2EHPA and Cyanex272 for the selective extraction of Eu. The optimization study revealed that the best performance (extraction and back-extraction) was observed for PIM containing 35.16 wt% of carriers (60 mg of D2EHPA and 53.9 mg of Cyanex272). The incorporation of plasticizers is a very common way to improve the flexibility of PIM. Therefore, the influence of the incorporation of poly(oxyethylene alkyl ether) (POE) on the efficiency of D2EHPA-Cyanex272-PVC@PIM in the extraction of Eu was studied. The highest value of flux of Eu was noticed for PIM containing 18.52 wt% of POE. Optimized PIM (35.16 wt% of D2EHPA+Cynanex272, 18.52 wt% of POE, and 46.31 wt% of PVC) is characterized by the flux of Europium equal to $2.7 \cdot 10^{-6}$ mol $m^{-2}$ $s^{-1}$. Moreover, the long-term experiments showed that up to 5 cycles PIM demonstrated constant efficiency in the extraction of Eu. In further experiments, the effectiveness of the membrane gradually decreased [221].

**Table 2**. Comparison of the efficiency of various PIM in the extraction of REE.

| Membrane matrix | Carrier | REE | Results | Ref. |
|---|---|---|---|---|
| PVDF | P227 | Lu(III) | P227@PIM (40 wt%) demonstrated the transport rate constant equal to 0.422 $h^{-1}$. | [210] |
| PVDF | P227 | Lu(III), Sm(III), La(III) | The recovery factors of Lu(III), La(III), and Sm(III) were 85%, 40%, and 4%, respectively. The Sm(III) was successfully isolated from La(III) when the pH was raised to 2.4. | [211] |
| PVDF | [A336][P227] | Yb(III) | PIM was suitable for the separation of yttrium from holmium and erbium. | [212] |
| PVDF | [A336][P507] | Lu(III) | The best permeability coefficient was noticed for the PIM with 60 wt% of [A336][P507] with a stirring speed of 450 rpm. Long-term stability tests demonstrated that membranes could be used for up to 10 cycles. | [213] |
| PVDF-EVOH | Cyanex 272 | Lu(III), Yb(III) | PIM demonstrated permeability coefficients of Lu(III) and Yb(III) equal to 41.62 mg $L^{-1}$ and 77.46 mg $L^{-1}$, respectively. The selectivity ($\beta_{Yb(III)/Lu(III)}$) was equal to 3.78. | [214] |
| PVDF | Cyphos IL 104 | Lu(III), Yb(III) | The membrane demonstrated the permeability of Lu(III) equal to 114.8 µm $s^{-1}$ and Yb(III) equal to 156.0 µm $s^{-1}$. Moreover, the long-term stability of the membranes was also tested. After 8 separation experiments, the permeation flux of the PIM decreased by 24.6%. | [215] |
| CTA | D2EHPA and TBP | La(III), Ce(III) | PIMs could be successfully used for the separation of La(III) and Ce(III). Higher separation efficiency was found for PIM with the D2EHPA ($\beta_{Ce(III)/La(III)}$=2.8) compared with PIM with TBP ($\beta_{Ce(III)/La(III)}$=1.4). | [216] |
| CTA | Cyphos IL 104 | La(III), Ce(III) | PIM demonstrated higher efficiency during the extraction of Ce(III) ($P$=3.1 µm $s^{-1}$, $RF$=68.1%) in comparison with the efficiency of La(III) extraction ($P$=0.5 µm $s^{-1}$, $RF$=16.2%). | [217] |
| CTA | PC-88A and Verstic 10 | Sc(III) | The best efficiency of extraction and back-extraction was noticed for PIM containing 4 wt% of PC-88A and 36 wt% of Versatic 10. The membrane was characterized by permeability and flux of Sc(III) equal to $6.77 \cdot 10^{-3}$ m $h^{-1}$ and $1.88 \cdot 10^{-7}$ mol $m^{-2}$ $s^{-1}$, respectively. | [218] |
| CTA | D2EHAG, D2EHAF, D2EHAS, D2EHPA, Versatic 10, and TOPO | Sc(III) | Results indicated that D2EHAG and D2EHAF were able to extract Sc(III). It was noticed PIM with a D2EHAF carrier was able to separate Sc(III) from Ni(II), Al(III), Co(II), Mn(II), Cr(III), Ca(II), Mg(II) and partially from Fe(III). | [219] |
| CTA | TODGA | La(III), Eu(III), Lu(III) | The highest flux was noticed for La(III) ($7.3\cdot10^{-8}$ mol $m^{-2}$ $s^{-1}$) while the smallest was for Lu(III) ($2.1\cdot10^{-8}$ mol $m^{-2}$ $s^{-1}$). The transport of lanthanides followed the order of their ionic potential (Lu(III)>Eu(III)>La(III)). | [220] |
| PVC | D2EHPA | La(III), Gd(III), Yb(III) | The efficiency of the extraction of Yb(III), Gd(III), and La(III) was performed at 0.45, 1.35, and 2.50, respectively. The highest flux was noticed during the extraction of La(III) ($7\cdot10^{7}$ mol $m^{-2}$ $s^{-1}$) while the smallest one was for the Yb(III) ($4\cdot10^{7}$ mol $m^{-2}$ $s^{-1}$). | [221] |
| PVC | D2EHPA and Cyanex272 | Eu(III) | The optimized PIM (35.16 wt% of D2EHPA+Cynanex272, 18.52 wt% of POE, and 46.31 wt% of PVC) is characterized by the flux of Eu(III) equal to $2.7\cdot10^{-6}$ mol $m^{-2}$ $s^{-1}$. The long-term experiments showed that up to 5 cycles PIM showed constant efficiency in the extraction of Eu(III). | [221] |

### 3.3.2. Molecular and ion-imprinted membranes (IIM)

The ion-imprinting technique (IIT) is a type of molecular imprinting technology inspired by the interaction of natural receptors and ligands [222]. The preassembly is initially used by ligands, template ions, and functional monomers to generate the ternary complexes during a standard ion-imprinting polymerization [223]. The ternary complexes are then crosslinked and form polymers in the presence of a cross-linker. After the removal of template ions by protonation, precise recognition sites capable of adsorbing target ions may be created on the so-called ion-imprinted polymers (IIPs) [224].

Chen et al. [139] prepared a yttrium(Y) ion-imprinted membrane (IIM). IIM was prepared by immersing PIM (containing PVDF, Cynanex272, and EVOH) in the solution of $YCl_3$ and itaconic acid. The template of Y(III) was removed by immersing the membrane in EDTA solution for 24 h. SEM analysis proved that IIM was characterized by a lower pore size (0.5 µm) compared with the PIM (1-1.5 µm). Some aggregates of polymer on the surface of IIM were detected. Moreover, forming an imprinted layer reduced the contact angle from 79.4° to 20.3°. Subsequently, prepared IIM was tested in the separation of Y(III), Ho(III), and Er(III). Furthermore, the initial Y membrane fluxes of PIMs and Y-IIMs were 3.97 µmol $m^{-2}$ s and 3.09 µmol $m^{-2}$ s, respectively. It can be concluded that the initial flux was not reduced significantly. The highest flux was observed during the extraction of Y(III) (3.09 µmol $m^{-2}$ $s^{-1}$) while the lowest one was for the Ho(III) (1.84 µmol $m^{-2}$ $s^{-1}$). The $\beta_{(Y/Ho)}$ and $\beta_{(Y/Er)}$ relative separation factors were determined to be 1.32 and 1.45, respectively. A long-term experiment, revealed that the IIM showed stable flux up to 10 cycles [139]. Chen et al. [157] also used IIM based on electrospinning PVDF for the separation of the neighboring heavy rare earth ion Y(III) from Ho(III), and Er(III). SEM analysis indicated that also in this case agglomeration on the surface of IIM. These agglomerates were formed by the strong interaction between Y ions and Cyanex272. IIM based on the electrospinning PVDF is characterized by static adsorption greater than 7. The formation of the ion-immobilized layer on the surface of the electrospinning PVDF significantly increased the $\beta$ from 1.24 to 2.01 for the extraction of Y(III). Based on the obtained results, it can be concluded that the prepared IIM a sufficient number of imprinting sites [157] (Fig. 22).

Zhang et al. [225] used an ion-imprinted chitosan composite membrane for the adsorption of Nd(III). FTIR analyses proved successful synthesis of IIM based on chitosan while the $N_2$ adsorption-desorption showed that the IIM was characterized by the specific surface are ($S_{BET}$) equal to 38.1 $m^2$ $g^{-1}$ and the pore size equal to 5.89 nm. The optimization experiments proved that the highest adsorption capacity of IIM was found for a pH equal to 5. Moreover, it was also proved that the IIM based on chitosan adsorbed more Nd(III) compared to PIM. It can be concluded that ion-imprinting technology can significantly improve the adsorption efficiency of the membrane. IIM showed a single metal adsorption capacity equal to 43.6 mg $g^{-1}$. Both membranes reached the maximum adsorption of Nd(III) after 120 min. The composite membrane was also reusable. After 5 cycles, IIM is characterized by an efficiency of 86% [225]. The three-dimensional macroporous wood-based ion-imprinted membrane was used for the selective separation of Nd(III) [156]. The membrane was prepared in a four-stages procedure. The first step was related to the self-polymerization of dopamine on the basswood chips. Subsequently, the membrane with poly(dopamine) layer was treated for immobilization of KH-570 (3-(trimethoxysilyl)propyl methacrylate). In the last step, PDA@basswood (KH-570/PDA@basswood) membrane was immersed into the solution of Nd(III) nitrite, methylacrylic acid, acrylamide, and ethylene glycol dimethacrylate. SEM images showed that the roughness increased after the creation of poly(dopamine) layer which means that the layer of poly(dopamine) was successfully created. Subsequently, the membrane was tested in the separation of Nd(III). Adsorption studies showed that the 3D-IIM membrane demonstrated higher adsorption capacity (120.9 mg $g^{-1}$) compared with the PIM (33.6 mg $g^{-1}$).

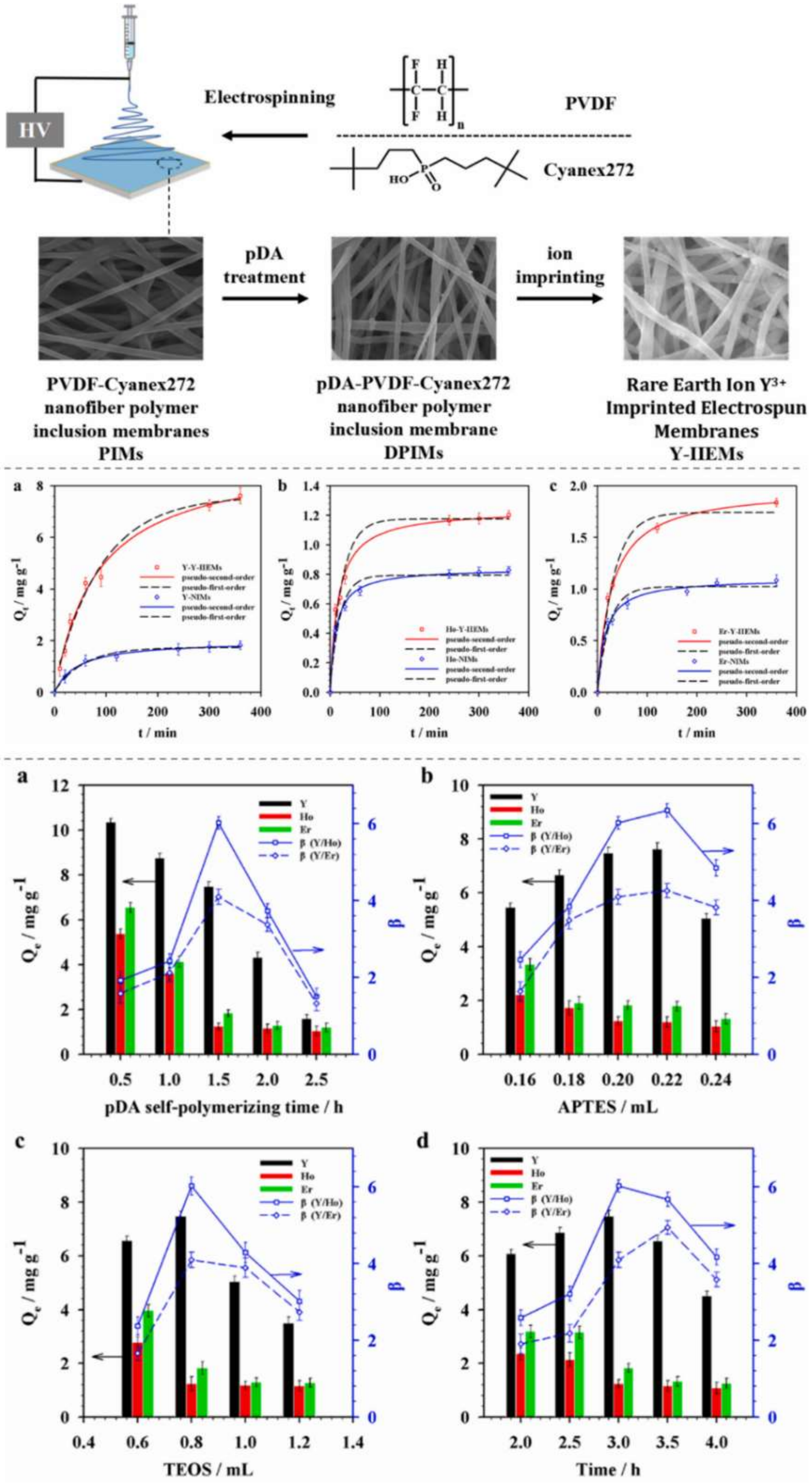

**Fig. 22**. Illustration on the synthetic process of Y-IIEMs (upper image). (a) pDA processing time; (b) Dosage of APTES; (c) Dosage of TEOS; (d) Influence of reaction time on Y(III),

Ho(III), and Er(III) adsorption properties of Y-IIEMs (left vertical axis is corresponding to the Qe value represented by the bar data, and right vertical axis is corresponding to the separation factor represented by the point plots) (middle image). Effect of time on the adsorption properties of (a) Y(III), (b) Ho(III) and (c) Er(III) by Y-IIEMs and NIMs (liquid phase: 20 mg $L^{-1}$REC13, 298.15K) (bottom image). Reproduced with permission from [157] Copyright © 2022, Elsevier.

Equilibrium adsorption was reached at 120 mg $L^{-1}$ of Nd(III). Additionally, the properties of the membrane were tested in the mixture of Nd(III), Tb(III), Fe(III), and Dy(III). During these experiments, the mixed solution containing 120 mg $L^{-1}$ of Nd(III), Tb(III), Fe(III), and Dy(III) was used. It was found that 3D-IIM demonstrated higher adsorption capacity to Nd(III) compared with non-template ions (Tb(III), Fe(III), and Dy(III)). The imprinting factor (IF) of 3D-IIM/PIM toward Nd(III) was close to 4.0 while the selective adsorption coefficient of $\alpha_{Nd(III)/Tb(III)}$, $\alpha_{Nd(III)/Tb(III)}$, and $\alpha_{Nd(III)/Tb(III)}$ were 5.1, 4.8, and 5.2, respectively [156].

Cui et al. [226] prepared Gd(III)-imprinted membranes based on carbon nanotubes and graphene oxide (GO) modified by poly(dopamine). TEM and SEM analysis showed after the polymerization of dopamine roughness of GO had become lower and less transparent. This observation proved the successful polymerization of dopamine on the GO. Selective adsorption experiments revealed that the membrane demonstrated the highest adsorption capacity for the Gd(III) and much lower for Eu(III) and La(III). The adsorption selectivity coefficients of Gd(III)/Eu(III) and Gd(III)/La(III) onto GIMs reach 1.83 and 3.39, respectively, Additionally, it was also found that the Langmuir model better described the adsorption of Gd(III) compared with the Freundlich model. It can be concluded that the adsorption sites were spread evenly on the surface of the prepared membrane. Subsequently, the membrane was tested in the separation of Gd(III) from La(III) and Eu(III). Analyzing the obtained results it was noticed that the membrane demonstrated a permeation selectivity coefficient of La(III)/Gd(III) and Eu(III)/Gd(III) was equal to 2.91 and 2.49, respectively [226]. Zheng et al. [227] used mesoporous carboxymethyl chitosan membrane for the selective extraction of Gd(III). The effect of various parameters such as pH and temperature on the efficiency of the extraction was also investigated. Considering the effect of pH on extraction efficiency, it was noticed that the amount of adsorbed Gd(III) increased with increasing pH. Based on the results, pH=7 was chosen as the optimal pH for the further experiments. In the case of the temperature effect, it was detected that temperature had a positive influence on the adsorption process. The amount of adsorbate per gram increased when the temperature of the experiment also increased. Under the optimal adsorption conditions, the so-prepared IIM demonstrated an adsorption capacity equal to 25.37 mg $g^{-1}$ and the distribution coefficient ($K_d$) was ca. 640 mL $g^{-1}$. The long stability test displayed that membrane lost 21% of its initial performance after 5 consecutive runs [227].

Lu et al. [228] applied an Eu(III) imprinted membrane based on the GO and modified silicon dioxide ($SiO_2$) for the separation of Eu(III) from La(III), Gd(III), and Sm(III). The membrane was also modified by the attachment of Ag particles to increase the anti-fouling properties. Analyzing the SEM micrographs, it was found that after the incorporation of Ag particles, no changes in the surface structure of the membrane were noticed. In the case of IIM membranes, the tumor-like polymers around nanospheres on the surface were detected which confirmed the preparation of the Eu(III) imprinted membrane. Antifouling studies showed that after being buried in natural soil for 20 days, the membrane without Ag particles was completely damaged while the membrane with Ag particles was almost undamaged. Subsequently, the influence of pH on the rebinding capacities and regeneration properties was investigated. It was observed that the best rebinding and regeneration properties membrane possessed at pH equal to 7. In the last part of this research, the transport and selective properties were investigated. The significantly lower flux and permeability of Eu(III) were noticed compared with the fluxes and permeabilities of other ions (Gd(III), La(III), and Sm(III)). These results indicated that Gd(III), La(III), and Sm(III) were transported through the membrane while the Eu(III) was captured. Membrane showed permselectivity coefficients of La(III)/Eu(III), Gd(III)/Eu(III), and Sm(III)/Eu(III) equal to 3.82, 3.47, and 3.34 respectively [228]. PVDF/1-butyl-3-methylimidazolium tetrafluoroborate (RTIL) nanofiber was used for the dynamic recovery of Eu(III) [229]. Physiochemical analysis showed that the incorporation of RTIL increased surface roughness from 84.5 nm to 114.9 nm. The imprinted cavity sites provide hierarchical roughness and height, resulting in the formation of homogeneous pits for improved Eu(III). Adsorption experiments showed that Freundlich ($R^2$=0.99) isotherm is the most representative model to describe the adsorption process. IIM PVDF/RTIL nanofiber exhibited excellent absorption efficiency, up 90% of Eu(III) was recovered after 3 h of experiment. Additionally, IIM was characterized by a remarkable reusability with no appreciable decrease in efficiency over 5 consecutive runs [229].

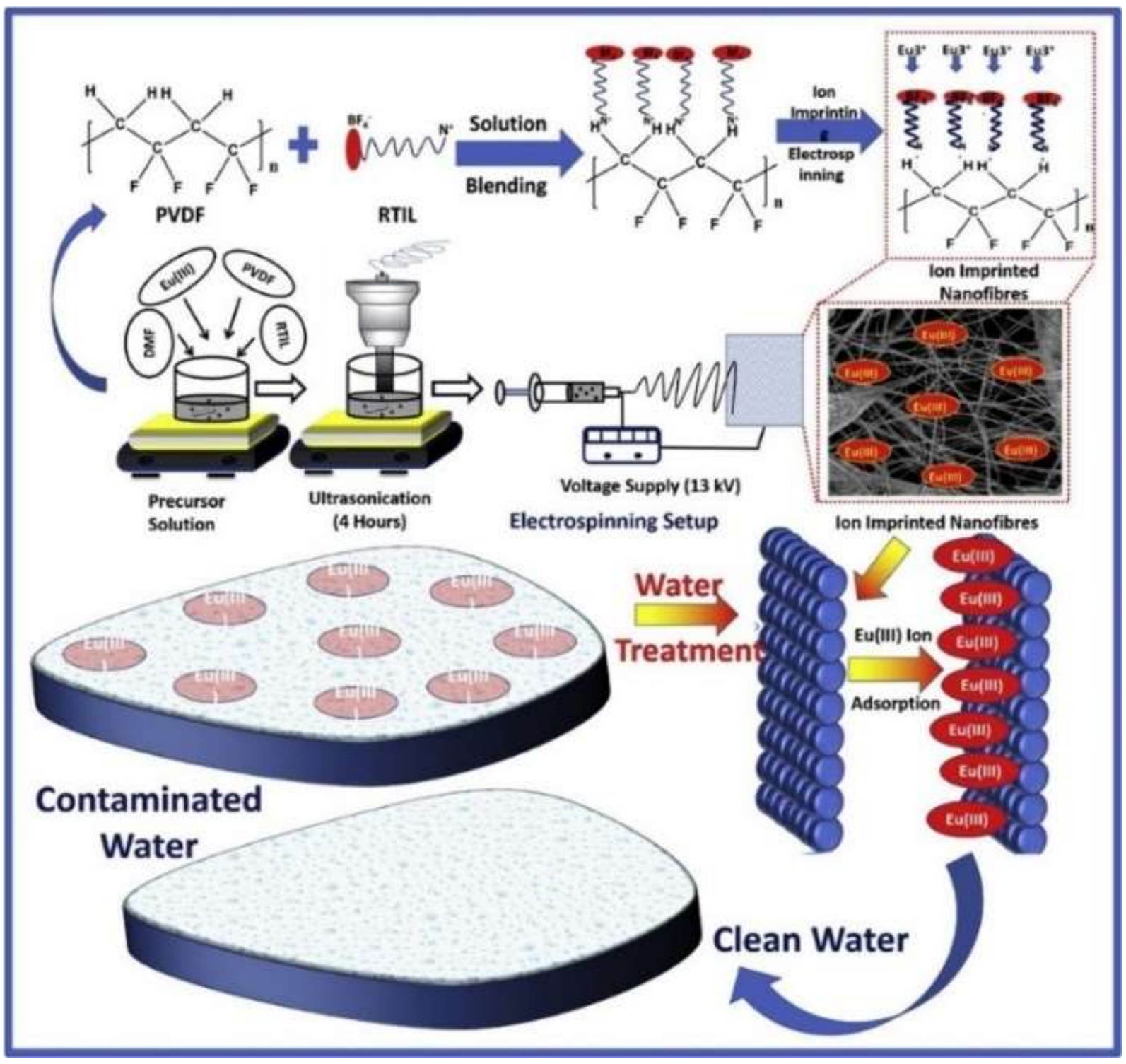

**Fig. 23**. Ion-imprinted electrospun PVDF nanofibers functionalized with ionic liquid, i.e. RTIL, for effective removal of europium (III) ions. Reproduced with permission from [229] Copyright © 2019, Elsevier.

IIM based on the methacrylic acid (MAA) monomer was prepared and used in selective solid-phase extraction of lanthanides from the tap and river waters [230]. Optimization of the solid-phase extraction protocol showed that the optimum pH for the experiments is 6.5 and 1M $HCl_{(aq)}$ was selected as an elution medium. Subsequently, the membrane was tested in the extraction recovery of eight Ln(III) (La(III), Ce(III), Nd(III), Sm(III), Gd(III), Dy(III), Er(III), and Lu(III)) from the tap and river waters. The efficiency of recovery of IIM was also compared with the efficiency of the non-ion-imprinted membrane (NIM). Experiments were performed for the tap and river waters containing 30 ng of each Ln(III). In the case of tap water, the highest extraction recoveries were noticed for Dy(III) (94%) while the lowest for Sm(III) and Lu(III) (both 82%). The prepared IIM demonstrated slightly lower efficiency of recovery of Ln(III) from river water. The highest and lowest efficiency of recovery was observed for Nd(III) (80%) and Lu(III) (47%), respectively. A comparison of extraction recoveries of IIM and NIM revealed that in both cases (tap and river waters) IIM displayed improved performances in the recovery of Ln(III) [230].

Liu et al. applied an ion-imprinted macroporous chitosan membrane for the absorption of Dy(III) [231]. The membrane was prepared in three steps process. The first step was related to the preparation of silica particles. Subsequently, the chitosan was modified by *o*-nitrobenzaldehyde. In the last step, the membrane was cast from the solution containing modified chitosan, silica particles, and $DyCl_3 \cdot 6H_2O$. After that, the membrane was washed in 1M $HCl_{(aq)}$ to remove Dy(III) and obtain imprinted cavities. Under the optimal conditions (pH=7 and 25°C), IIM showed adsorption capacity toward Dy(III) equal to 23.3 mg $g^{-1}$ and the adsorption equilibrium was reached in 150 min. Analyzing the results of selectivity studies, it was noticed that IIM demonstrated a higher absorption ability towards Dy(III) in comparison with competitive ions (Nd(III), Pr(III), and Tb(III)). IMM demonstrated a distribution coefficient for Dy(III) equal to 500 mL $g^{-1}$ while for the competitive ions was equal to 150 mL $g^{-1}$. Long-term stability revealed that the adsorption capacity of IMM was reduced by only 8.4% after 5 runs. This indicated that MAC was a very effective adsorbent for the recovery of Dy(III) [231].

Zheng et al. [232] applied dual-layer ionic imprinted bilayer mesoporous membrane based cellulose nano-crystallites for the recovery of Nd(III) and Dy(III). The membrane was prepared using the dual template docking oriented ionic imprinting. The obtained membrane is characterized by a high specific area and ability for the selective adsorption of Nd(III) and Dy(III). IIM possessed Janus structure therefore membrane could adsorb Nd(III) and Dy(III) at the same time. It was observed that IIM demonstrated the adsorption capacity of Nd(III) and Dy(III) equal to 12.15 mg $g^{-1}$ and 17.50 mg $g^{-1}$, respectively. Subsequently, the selective adsorption properties of IIM towards Nd(III) and Dy(III) in the presence of competitive ions (Tb(III), Pr(III)). The properties of the membrane were evaluated by using the distribution coefficient ($K_d$). The results revealed that IIM demonstrated higher $K_d$ for Nd(III) (ca. 225 mL $g^{-1}$) and Dy(III) (ca. 310 mL $g^{-1}$) compared with other ions (ca. 75 mL $g^{-1}$). Moreover, it should be also noted that IIM could be very easily washed and reused again. It was noticed that the membrane reduced its adsorption capacity to Nd(III) and Dy(III) by 17.05% and 19.66%, respectively. TGA and XRD analysis also confirmed the structure of the membrane and functional group did not change after 5 cycles [232].

**Table 3**. Comparison of the efficiency of various IIM in the extraction of REE.

| Target ion | Membrane substrate | Competitive ions | Results | Ref. |
|---|---|---|---|---|
| Y(III) | PVDF, Cynanex272, and EVOH | Ho(III), Er(III) | The highest flux was observed during the extraction of Y(III) (3.09 µmol $m^{-2}$ $s^{-1}$) while the lowest one was for the Ho(III) (1.84 µmol $m^{-2}$ $s^{-1}$). The $\beta_{(Y/Ho)}$ and $\beta_{(Y/Er)}$ relative separation factors were determined to be 1.32 and 1.45, respectively. | [139] |
| Y(III) | PVDF, Cyanex272 | Ho(III), Er(III) | The creation of the ion-immobilized layer on the surface of the electrospinning PVDF significantly increased the β from 1.24 to 2.01 for the extraction of Y(III). | [157] |
| Nd(III) | chitosan | Nd(III), Dy(III), Pr(III), La(III) | IIM showed a single metal adsorption capacity equal to 43.6 mg $g^{-1}$. Membranes reached the maximum adsorption of Nd(III) after 120 min. The composite membrane was also reusable. After 5 cycles, IIM is characterized by an efficiency of 86%. | [225] |
| Nd(III) | three-dimensional macroporous wood | Tb(III), Dy(III) | 3D-IIM membrane demonstrated an adsorption capacity equal to 120.9 mg $g^{-1}$ and higher adsorption capacity towards Nd(III) compared with non-template ions (Tb(III) and Dy(III)). | [156] |
| Gd(III) | carbon nanotubes and graphene oxide (GO) modified by poly(dopamine) | Eu(III), La(III) | IIM demonstrated the highest adsorption capacity for the Gd(III) and much lower for Eu(III) and La(III). The adsorption selectivity coefficients of Gd(III)/Eu(III) and Gd(III)/La(III) reach 1.83 and 3.39, respectively, | [226] |
| Gd(III) | mesoporous carboxymethyl chitosan | Pr(III), Dy(III), Tb(III) | Under optimal adsorption conditions, prepared IIM demonstrated an adsorption capacity equal to 25.37 mg $g^{-1}$ and the distribution coefficient ($K_d$) was ca. 640 mL $g^{-1}$. | [227] |
| Eu(III) | GO and modified silicon dioxide ($kSiO_2$) | La(III), Gd(III), Sm(III) | Membrane showed permselectivity coefficients of La(III)/Eu(III), Gd(III)/Eu(III), and Sm(III)/Eu(III) equal to 3.82, 3.47, and 3.34 respectively. | [228] |
| Eu(III) | PVDF/RTIL | - | IIM PVDF/RTIL nanofiber exhibited excellent absorption efficiency up 90% of Eu(III) was recovered after 3 h. | [229] |
| 8 Ln(III) | macroporous chitosan | - | Better efficiency in the extraction of lanthanides from tap water compared with river water. In the case of tap water, the highest extraction recoveries were noticed for Dy(III) (94%) while the lowest for Sm(III) and Lu(III) (both 82%). | [230] |
| Dy(III) | macroporous chitosan | Nd(III), Pr(III), Tb(III) | IIM showed adsorption capacity toward Dy(III) equal to 23.3 mg $g^{-1}$ and distribution coefficient for Dy(III) equal to 500 mL $g^{-1}$ while for the competitive ions was equal to 150 mL $g^{-1}$. | [231] |
| Nd(III), Dy(III) | cellulose nano-crystallines | Tb(III), Pr(III) | IIM demonstrated the adsorption capacity of Nd(III) and Dy(III) equal to 12.15 mg $g^{-1}$ and 17.50 mg $g^{-1}$, respectively. Membraned showed higher $K_d$ for Nd(III) (ca. 225 mL $g^{-1}$) and Dy(III) (ca. 310 mL $g^{-1}$) compared with other ions (ca. 75 mL $g^{-1}$). | [232] |

### 3.3.3. Metal organic frameworks (MOFs) and covalent organic frameworks (COFs) materials for REE ions separation

#### 3.3.3.1. MOF adsorbents and MOF membranes for REE separation

Metal organic frameworks (MOFs) are porous crystalline materials containing metal clusters coordinated with organic ligands [140]. MOFs have shown tremendous potential for the separation of REEs due to their intrinsic properties e.g., large surface area, tunable pore size, tunable surface chemistry, and unsaturated metal sites [233, 234]. Paz et al. [235] synthesized Cu-BTC MOF *via* the solvothermal reaction for the adsorption of Samarium (Sm), Lanthanum (La), and Erbium (Er) from aqueous solution. The effects of pH, the initial concentration of lanthanide ions, the adsorbent mass, and the adsorption time on the adsorption performance of Cu-BTC MOF were investigated. It was found that the prepared MOF showed the highest adsorption capacity for lanthanide ions under the optimal experimental conditions (adsorbent amount = 20 mg, $t_{eq}$ = 120 min, pH = 6, $T$ = 298 K, $C_0$ = 500 mg $L^{-1}$). The maximum adsorption capacities for Sm(III), La(III), and Er(III) are 248.4, 235.4, and 131.4 mg $g^{-1}$, respectively. Moreover, the prepared MOF showed higher adsorption capacity for Sm(III) with the presence of other metal ions such as Al(III), Fe(III), Cr(III), Mg(II), Co(II), and Na(I). The high adsorption performance for lanthanide ions was resulted from the involvement of free carboxylic groups and the ion-exchange mechanism in the adsorption process. Ammari Allahyari et al. [236] synthesized MOF $[Zn(bim)_2(bdc)]_n$ for La(III) separation from aqueous solution. In the batch experiments, the prepared MOF exhibited the highest La(III) adsorption capacity equal to 130 mg $g^{-1}$ under the optimized conditions (adsorbent dosage = 2500 mg $L^{-1}$, $t_{eq}$ = 150 min, pH = 7, $T$ = 298 K). From the kinetic adsorption study, it was found that the pseudo-first-order kinetics well described the adsorption kinetics and more than 97% of La(III) ions were adsorbed onto the prepared MOF. In the fixed bed column mode, the dynamic adsorption capacity of La(III) ions onto the synthesized MOF was 20% higher than that from the batch experiments. It was found that the synthesized MOF showed higher adsorption capacity and higher adsorption efficiency when the flow rate in column mode was lower. Khalil et al. [237] synthesized cobalt MOFs containing diallylamine as the ligand by using the sol-gel method. The synthesized Co-MOF demonstrated high adsorption efficiency for Ce(III) and Eu(III) separation from aqueous solution. It was found that he adsorption efficiency for Ce(III) and Eu(III) were 93.3% and 27.4%, respectively when the pH of solution was 5.1. The adsorption behavior of the synthesized Co-MOF was well described by the pseudo-second-order kinetic and the Langmuir adsorption model. Under the optimized conditions (adsorbent dosage = 5000 mg $L^{-1}$, $t_{eq}$ = 30 min and 180 min, pH = 5.1, $T$ = 298 K, $C_0$ = 800 mg $L^{-1}$), the adsorption capacity of Ce(III) and Eu(III) were 102.24 mg $g^{-1}$ and 52.93 mg $g^{-1}$. The synthesized Co-MOF showed higher adsorption capacity for Ce(III) because the Co-MOF possessed a cavity size which is similar to Ce(III) ionic radii, which allows the Ce(III) ions to enter the Co-MOF and interact with the amine groups on the Co-MOF wall. Zhang et al. [238] combined ZIF-8 and UiO-66-$NH_2$ to synthesize a hybrid material U6N@ZIF-8-20 possessing a 3D-agaric like core-shell structure for the adsorption of REEs. It was found that the maximum adsorption capacities of U6N@ZIF-8-20 for Nd(III), Eu(III), Gd(III), and Er(III), were 249.90 mg $g^{-1}$, 295.28 mg $g^{-1}$, 316.22 mg $g^{-1}$, and 340.95 mg $g^{-1}$, respectively. The high adsorption capacity of U6N@ZIF-8-20 for REEs are attributed to the abundant amino, hydroxyl and carboxyl groups, which have high affinity for REEs. What is more, the high surface area, high porosity, the agaric core-shell structure, and the abundant adsorption sites resulted in the high adsorption capacity of U6N@ZIF-8-20 for REEs as well. Comparing with the sing ZIF-8 and UiO-66-$NH_2$, the synthesized U6N@ZIF-8-20 showed enhanced thermal stability, water stability and reusability.

In order to improve the adsorption capacity and the selectivity of MOFs for the separation of the target REEs, further modifications are generally conducted. The modification of MOFs with functional groups is a favorable way to prepare more efficient and stable porous adsorption materials. Pei et al. [239] prepared carboxyl functional poly(ionic liquid)s@MOF composite (PIL@MIL-101) *via* the *in situ* polymerization of ILs monomers in MOF pores. The prepared PIL@MIL-101 was used to separate La(III), Sm(III), and Nd(III) from aqueous solution. It was found that PIL@MIL-101 exhibited high optimized adsorption efficiency of 99.8% for La(III), Sm(III), and Nd(III) ions. Furthermore, the adsorption performance of PIL@MIL-101 was again well described by the Langmuir model and the pseudo-second-order model. The high adsorption performance of PIL@MIL-101 was attributed to the electrostatic interaction and the coordination of metal ions by the carboxyl group. Ahmed et al. [240] prepared [$C_4$mim]@UiO-66 by entrapping ionic liquid 1-butyl-3-methylimidazolium bromide ([$C_4$mim]$^+$[Br]$^-$) into the cavities of UiO-66 *via* a ship-in-a-bottle technique. The prepared [$C_4$mim]@UiO-66 was used to capture Gd(III) from aqueous solution. Under the optimized experimental conditions ($t_{eq}$ = 180 min, pH = 6, $T$ = 298 K, $C_0$ = 75 mg $L^{-1}$), [$C_4$mim]@UiO-66 exhibited the maximum adsorption capacity equal to 85 mg $g^{-1}$, which was significantly higher than the pristine UiO-66 (17 mg $g^{-1}$). This is because the strong interaction between Gd(III) ions and the imidazole rings of $C_4$mim in $C_4$mim@UiO-66 *via* coordination and interaction with the aromatic π-electron cloud of imidazole. What is more, $C_4$mim@UiO-66 showed a high selectivity towards Gd(III) ions in the presence of other transitional metal ions, and alkali ions. Li et al. [241] synthesized lanthanum-based MOF (LaBDC) by using the hydrothermal method and modified LaBDC with polyethyleneimine (PEI) by using the impregnation method. It was found that the modified LaBDC@50%PEI showed the highest adsorption capacity of 181.77 mg $g^{-1}$ for Gd(III), which is 5 times higher than that of pristine LaBDC. The adsorption isotherm and the adsorption kinetics of LaBDC@50%PEI were well-defined by the Langmuir and pseudo-second-order kinetic model, respectively. The modified LaBDC@50%PEI showed high Gd(III) adsorption capacity because the PEI modification enriched the LaBDC surface with the abundant amino groups, carboxyl groups, and hydroxyl groups, which overall showed a high affinity to Gd(III) ions.

In order to improve the recyclability of MOF-based materials, MOF crystals are formed on substrates, such as fibers [242], membranes [243], and nanofibrous mats [244]. To increase the stability of polyacrylonitrile fibers (PANF) and overcome the difficulty of separating of MOFs from solution, Hua et al. [243] synthesized the MOF nanofibrous membranes (NFMs) for the separation of Tb(III) and Eu(III) from aqueous solution by embedding the UiO-66-$(COOH)_2$ nanoparticles into polyacrylonitrile (PAN) nanofibers using colloid-electrospinning technique (Fig. 24).

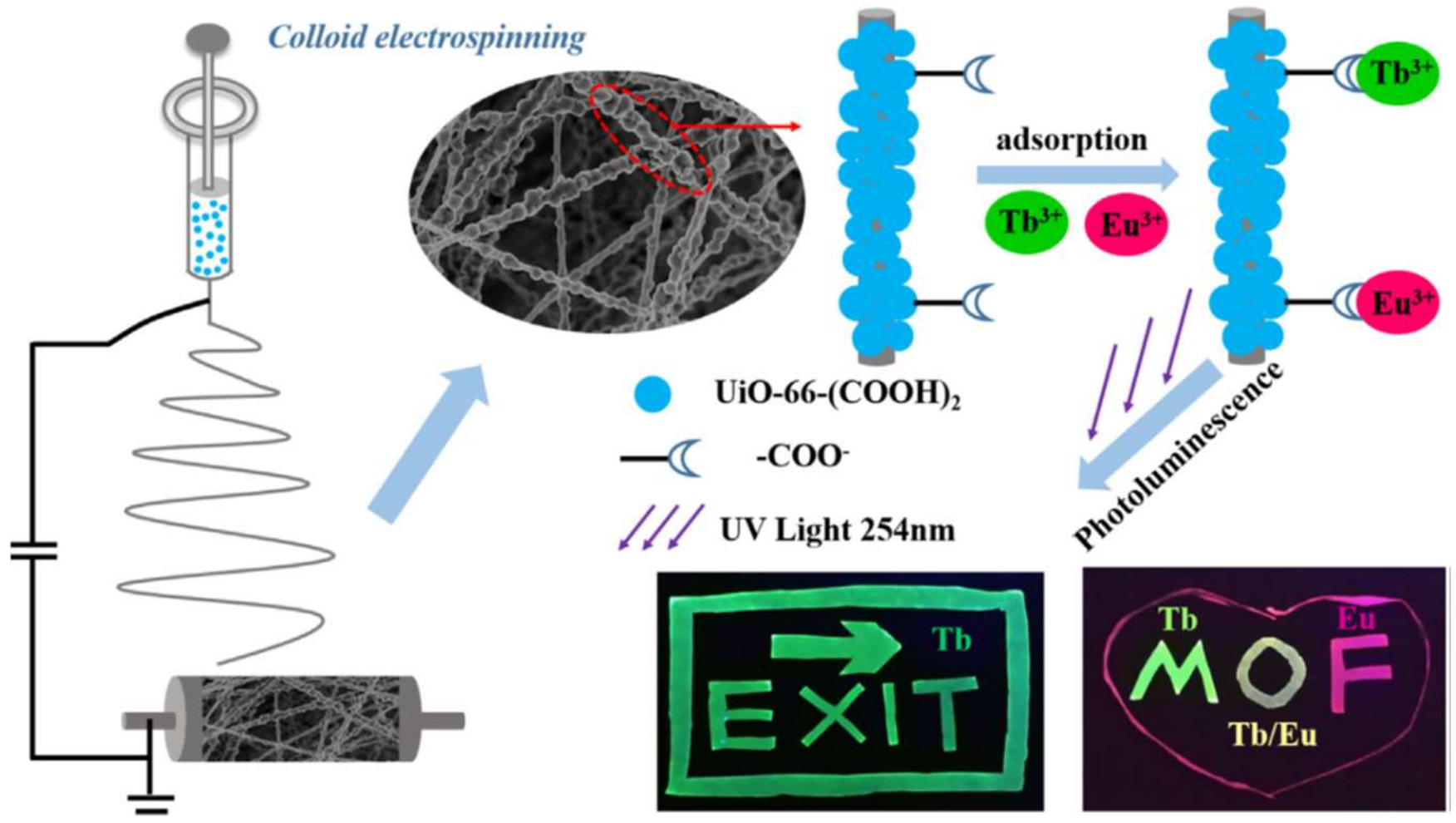

**Fig. 24**. Schematic illustration of the fabrication procedure of PAN/UiO-66-$(COOH)_2$ nanofibrous membranes for the adsorption of $Tb^{3+}$ and $Eu^{3+}$ ions. Reprinted with permission from [243] Copyright ©2019 Elsevier.

The UiO-66-$(COOH)_2$ possessed a high number of free carboxyl functional groups which showed high affinity and strong chelating effect to Tb(III) and Eu(III), while PAN nanofibers provided high specific surface area, good flexibility, and easy separation. The prepared PAN/UiO-66-$(COOH)_2$ containing 60 wt% MOFs showed high adsorption capacity equal to 203.4 mg $g^{-1}$ and 181.3 mg $g^{-1}$ for Tb(III) and Eu(III), respectively under the optimal conditions ($t_{eq}$ = 240 min, pH = 6, $T$ = 298 K, $C_0$ = 200 mg $L^{-1}$). The adsorption process was well described by the Langmuir isotherm and pseudo-second-order kinetic model. Mahmoud et al. [245] prepared a microporous MOFs/polymer hybrid material Zn(Glu)-SiNH/PPC consisting of silica-amine modified MOFs and poly(piperazine-cresol) (PPC) for La(III) adsorption separation from aqueous solution. It was found that the synthesized Zn(Glu)-SiNH/PPC showed the maximum adsorption capacity of 289.3 mg $g^{-1}$ for La(III) under the optimized conditions (adsorbent amount = 20 mg, $t_{eq}$ = 20 min, pH = 5, $T$ = 298 K, $C_0$ = 0.01 – 0.1 mol $L^{-1}$). The adsorption isotherm and the adsorption kinetics of Zn(Glu)-SiNH/PPC were well described by the Langmuir model and pseudo-second-order kinetic model, respectively.

Apart from the preparation of MOF-based materials as adsorbents for the separation of REEs, the fabrication of MOF-based membranes is also important. Even the utilization of MOF-based membranes for the separaiton of REEs is not common, the study on their application for the separaiton of REEs is required. Yao et al. [246] fabricated H-UiO-66-PF2 membranes for the simultaneous removal of Gd(III) and oil from wastewater.

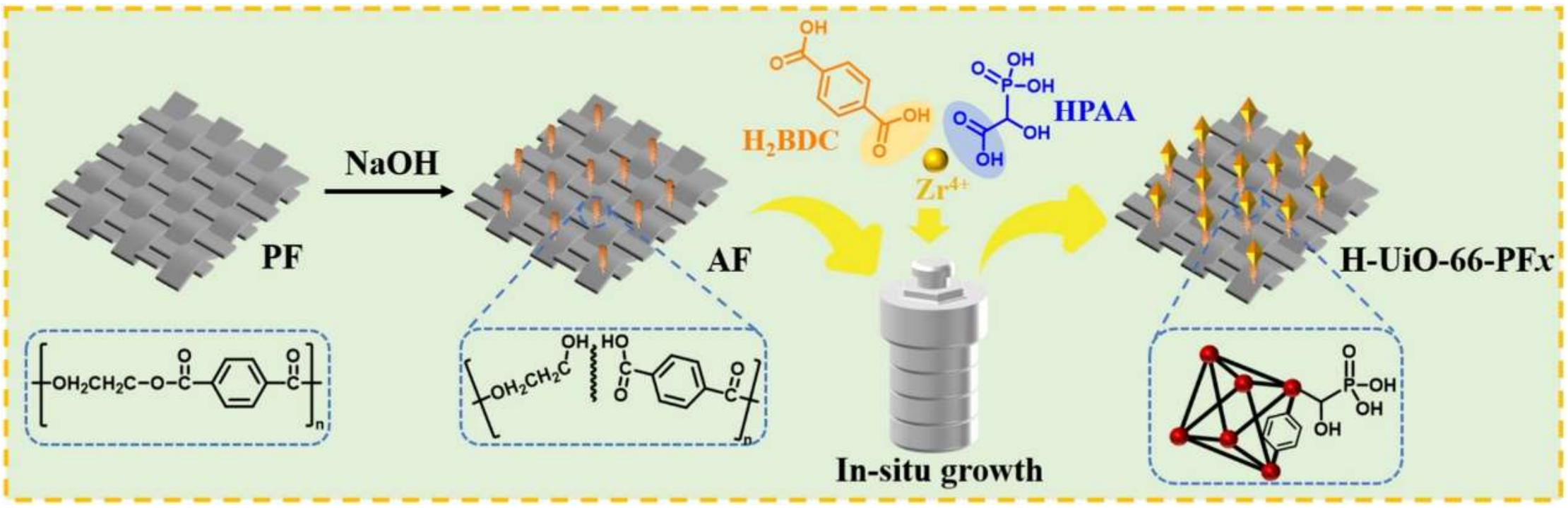

**Fig. 25**. Illustration of the fabrication process for H-UiO-66-PFx. Reprinted with permission from [246] Copyright 2022 Elsevier.

The H-UiO-66-PF2 membranes were fabricated by incorporating the 2-hydroxyphosphono- acetic acid modified UiO-66 crystals on polyester fabric (PF) using the in-situ growth method (Fig. 25). Comparing with the pristine PF membranes, the modified one showed enhanced hydrophilicity, high oil/water separation performance with a water flux up to 126300 L $m^{-2}$ $h^{-1}$, and a high maximum Gd(III) adsorption capactiy equal to 156.6 mg $g^{-1}$ under the optimized experimental conditions (adsorbent amount = 10 mg, $t_{eq}$ = 60 min, pH =

6, $T$ = 298 K, $C_0$ = 250 mg $L^{-1}$), which is attributed to the chelation effect between the abundant phosphate and carboxy groups and Gd(III) ions. The prepared H-UiO-66-PF2 membranes possessed high stability and recyclability. Most importantly, the Gd(III) ions were completely captured from wastewater *via* the filtration process.

Liang et al. [247] manufactures 2D vertical heterostructure membranes by inserting a 2D MOF similar to the [002] crystal plane of ZIF-8 between the GO layers for the lanthanide separation. The 2D MOF was *in-situ* synthesized in the interlayers of the GO membranes. It was found that the prepared membranes were highly selective to La(III) owing to the stronger attractive between La(III) and the prepared membranes, resulting from the selective electrostatic interaction between the small pores on the surface of the MOF and the lanthanide ions. Therefore, the prepared membranes exhibited high selectivities equal to 55.97 and 6.02 for La(III)/Yb(III) and La(III)/Ce(III), respectively. Moreover, the prepared membrane showed a high water permeability of 14.20 L $m^{-2}$ $h^{-1}$ $bar^{-1}$, which is 7-fold higher than that of the GO membrane. It was also found that the generated membrane was characterized by a high stability in strong acids since the 2D MOF was stabilized by the entrapment of graphene and the interlayer spacing of GO was expanded and fixed by 2D MOF.

**Table 4.** The separation of REEs by using MOF-based materials.

| Material | Target REEs | Adsorption capacity (mg $g^{-1}$) | Optimal experimental conditions | Selectivity | Ref. |
|---|---|---|---|---|---|
| [$C_4$mim]@UiO-66 | Gd(III) | 85.0 | $t_{eq}$ = 180 min, pH = 6, $T$ = 298 K, $C_0$ = 75 mg $L^{-1}$ | - | [240] |
| PAN/UiO-66-$(COOH)_2$ nanofibers | Tb(III)<br>Eu(III) | 203.4<br>181.3 | $t_{eq}$ = 240 min, pH = 6, $T$ = 298 K, $C_0$ = 200 mg $L^{-1}$ | - | [243] |
| H-UiO-66-PF2 membrane | Gd(III) | 156.5 | Adsorbent amount = 10 mg, $t_{eq}$ = 60 min, pH = 6, $T$ = 298 K, $C_0$ = 250 mg $L^{-1}$ | - | [246] |
| UiO-66-$NH_2$@ZIF-8 | Nd(III)<br>Eu(III)<br>Gd(III)<br>Er(III) | 249.5<br>295.3<br>316.2<br>341.0 | Adsorbent amount = 5 mg, $t_{eq}$ = 10 min, pH = 5, $T$ = 303 K, $C_0$ = 20 – 500 mg $L^{-1}$ | - | [238] |
| Cu-BTC MOF | Sm(III)<br>La(III)<br>Er(III) | 248.4<br>235.4<br>131.4 | Adsorbent amount = 20 mg, $t_{eq}$ = 120 min, pH = 6, $T$ = 298 K, $C_0$ = 500 mg $L^{-1}$ | - | [235] |
| Zn(Glu)-SiNH/PPC nanocomposite | La(III) | 289.3 | Adsorbent amount = 20 mg, $t_{eq}$ = 20 min, pH = 5, $T$ = | - | [245] |
| | | | 298 K, $C_0$ = 0.01 – 0.1 mol $L^{-1}$ | | |
| Ce-BTC MOF | La(III) | 99.0 | Adsorbent dosage = 6 mg/L, $t_{eq}$ = 120 min, pH = 6, $T$ = 335 K, $C_0$ = 100 – 500 mg $L^{-1}$ | - | [248] |
| LaBDC@50%PEI | Gd(III) | 181.2 | Adsorbent dosage = 200 mg/L, $t_{eq}$ = 180 min, pH = 5.5, $T$ = 298 K, $C_0$ = 70 mg $L^{-1}$ | - | [241] |
| $[CdL(bipy)]_n$ | La(III)<br>Gd(III)<br>Nd(III)<br>Sm(III) | 143.0<br>211.0<br>153.0<br>179.0 | Adsorbent dosage = 100 mg/L, $t_{eq}$ = 360 min, pH = 5, T = 298 K, $C_0$ = 1 – 40 mg $L^{-1}$ | - | [249] |
| $[Zn(bim)_2(bdc)]_n$ | La(III) | 156.7 | Adsorbent dosage = 2500 mg/L, $t_{eq}$ = 150 min, pH = 7, $T$ = 298 K | - | [236] |
| Co-MOF | Ce(III) | 102.2 | Adsorbent dosage = 5000 mg/L, $t_{eq}$ = 30 min, pH = 5.1, $T$ = 298 K, $C_0$ = 500 mg $L^{-1}$ | - | [237] |
| Co-MOF | Eu(III) | 52.9 | Adsorbent dosage = 5000 mg/L, $t_{eq}$ = 180 min, pH = 5.1, $T$ = 298 K, $C_0$ = 800 mg $L^{-1}$ | - | [237] |
| Organophosphorus modified MIL-101(Cr) | Er(III) | 57.5 | Adsorbent dosage = 1000 mg/L, $t_{eq}$ = 180 min, pH = 5.5, $T$ = 298 K, $C_0$ =200 mg $L^{-1}$ | Er/Nd = 10<br>Er/Gd = 5 | [250] |
| EDTA–CS@ZIF-8 | La(III)<br>Eu(III)<br>Yb(III) | 256.4<br>270.3<br>294.1 | Adsorbent dosage = 1.5 mg/L, $t_{eq}$ = 24 h, pH = 6, $T$ = 298 K, $C_0$ =5 – 45 mg $L^{-1}$ | - | [251] |

$t_{eq}$ – equilibrium time, $C_0$ – initial concentration, $T$ – temperature,

#### 3.3.3.2. COF adsorbents and COF membranes for REEs separation

Covalent organic frameworks (COFs) are reticular crystalline framework materials connected by covalent bonds [234]. Due to their regular pore structure, high specific surface area and high stability, the utilization of COFs in the separation of REEs from aqueous solutions has drawn great attention [252]. Xiao et al. [253] synthesized TpPa COFs from 1,3,5-triformylphloroglucinol, and *p*-phenylenediamine using the deep eutectic solvent (DES) as the reaction medium at room temperature. The prepared TpPa COFs had a high crystallinity, a uniform mesoporous structure, and an excellent chemical stability in acidic and alkaline solutions. They exhibited the maximum La(III) adsorption capacity of 84.7 mg $g^{-1}$ under the optimized experimental conditions (adsorbent dosage = 1000 mg $L^{-1}$, $t_{eq}$ = 40 min, pH = 3.5, $T$

= 298 K, $C_0$ = 125 mg $L^{-1}$). It was found that the good adsorption performance of TpPa COFs was attributed to the coordination between the oxygen anions and REEs within the uniform 1D channel. Zhang et al. [254] synthesized N-rich COFs (COF-PA-CC and COF-ML-CC) by using the one-step solvothermal method for La(III) recovery from aqueous solution. The COF-PA-CC was synthesized from p-phenylenediamine (PA) and cyanuric chloride (CC); and the COF-ML-CC was synthesized from melamine (ML) and cyanuric chloride (CC). It was found that COF-PA-CC owned a nanowire morphology while the COF-ML-CC were spherical. The prepared COF-PA-CC and COF-ML-CC showed maximum adsorption capacities of 150.88 mg/g and 168.19 mg/g, respectively for La(III) under the optimized conditions (adsorbent dosage = 500 mg $L^{-1}$, $t_{eq}$ = 120 min, pH = 5.5, $T$ = 338 K, $C_0$ = 900 mg $L^{-1}$). The adsorption isotherm and the adsorption kinetics of COF-PA-CC and COF-ML-CC were well defined by the Langmuir model and pseudo-second-order kinetic model, respectively. The high La(III) adsorption capacities for the synthesized COFs resulted from the complexation interaction between La(III) and the unoccupied *N*-sites of *N*-rich COFs.

The COF-based membranes were fabricated for the separation of REEs from aqueous solution. Lai et al. [255] manufactured PEI-BDSC/TpPa/PES nanofiltration membranes with an excellent acid resistance and a high trivalent rare-earth ions ($RE^{3+}$) separation performance by using the *in-situ* interfacial polymerization (IP) technique. As shown in Fig. 26, the TpPa COF layer was firstly formed on the surface of polyethersulfone (PES) ultrafiltration membrane from 2,4,6-triformylphloro-glucinol (Tp) and phenylenediamine. Subsequently, the polysulfonamide (PSA) layer was formed on the COF layer from polyethyleneimine (PEI) and 1,3-benzenedisulfonyl dichloride (BDSC). The permeance and salt rejection of the membranes were evaluated by a laboratory-scale cross-flowing nanofiltration rig with a 21 $cm^2$ effective area at both neutral (pH = 6.8) and acidic (pH = 1) conditions. The prepared PEI-BDSC/TpPa/PES composite membranes displayed a high rejection of 92.9%, 92.8%, 92.8%, 92.3%, and 92.2% for La(III), Nd(III), Gd(III), Yb(III), and Y(III), respectively, along with the water permeance of >43.3 L $h^{-1}$ $m^{-2}$ $bar^{-1}$ at pH=6.8 and pH=1.0. The high trivalent rare-earth ions ($RE^{3+}$) separation performance of the prepared composite membrane was attributed to the interlaced stacking between the COF and PSA layers while the COF regulated IP process. Xiong et al. [256] synthesized [$NH_4^+$] [COF-$SO_3^-$] and COF/PES mixed matrix membranes for the removal of Th(IV) and REEs from aqueous solution. It was found that the synthesized COF showed much higher adsorption capacity for Th(IV) than other rare earth elements REEs due to the cation exchange between $[NH_4]^+$ and Th(IV) ions and the strong binding between $SO_3^-$ and Th(IV) ions *via* coordination interaction. The prepared COF showed the maximum Th(IV), Eu(III), and Ce(III)adsorption capacity of 385 mg $g^{-1}$, 43 mg $g^{-1}$, and 39 mg $g^{-1}$, respectively under the optimized conditions (adsorbent dosage = 500 mg $L^{-1}$, $t_{eq}$ = 60 min, pH = 2.8, $C_0$ = 125 mg $L^{-1}$). In the filtration process, the prepared COF/PES membranes showed 72.4% removal efficiency for Th(IV), 16.5% for Ce(III), and 19.2% for Eu(III).

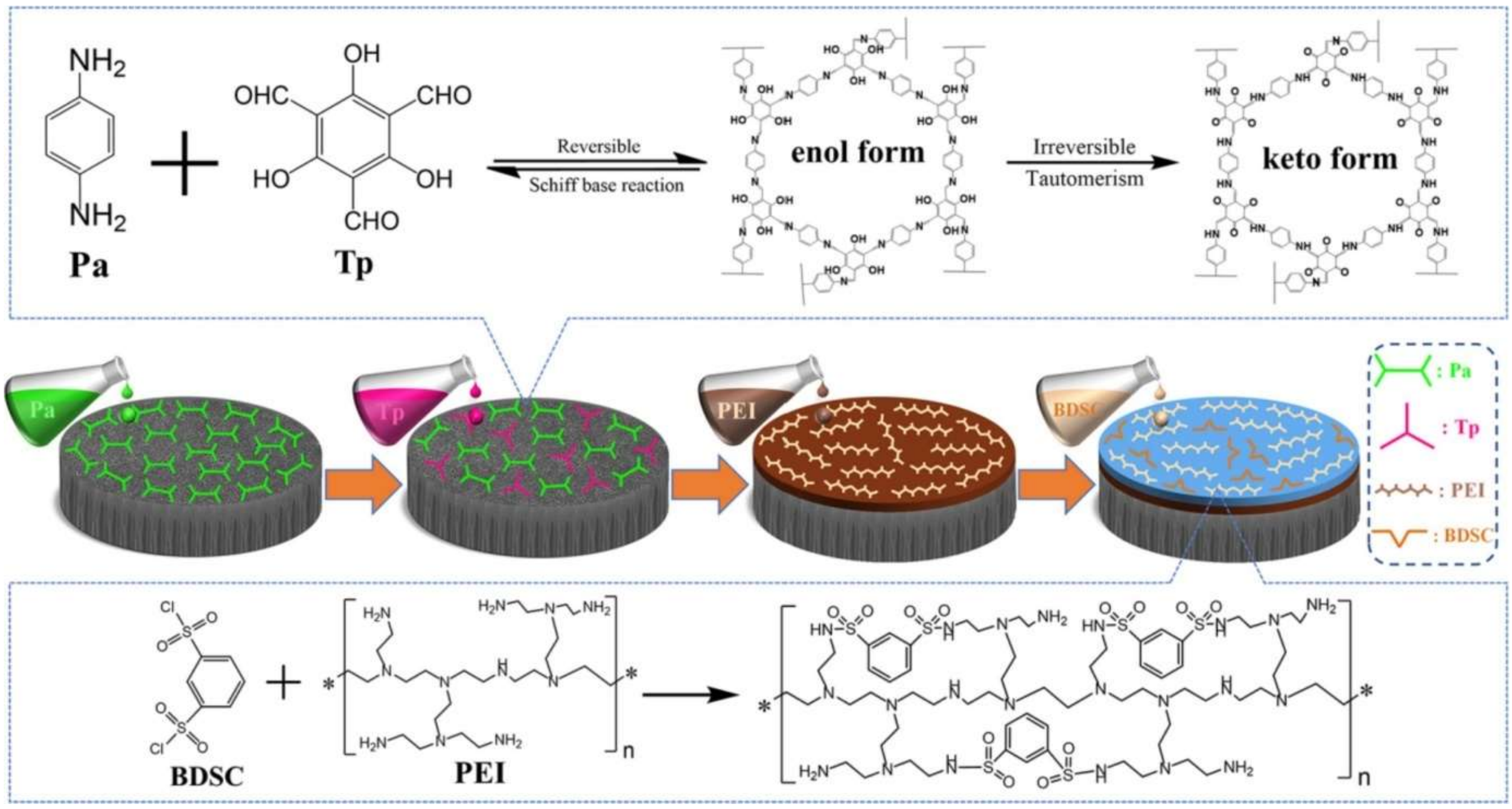

**Fig. 26**. Fabrication of the PEI-BDSC/TpPa/PES composite membrane. Reprinted with permission from [255] Copyright © 2022 Elsevier.

**Table 5**. The separation of rare earth elements REEs by using COF-based materials.

| Material | Target REEs | Adsorption capacity (mg $g^{-1}$) | Optimal experimental conditions | Selectivity | Ref. |
|---|---|---|---|---|---|
| COF-TZ-TP | La(III) | 165.6 | Adsorbent dosage = 500 mg $L^{-1}$, $t_{eq}$ = 180 min, pH = 6, $T$ = 308 K, $C_0$ = 1000 mg $L^{-1}$ | La/Sm = 2<br>La/Er = 2<br>La/Lu = 6 | [257] |
| COF-TA-TP | | 89.8 | | La/Sm = 5<br>La/Er = 2<br>La/Lu = 3 | |
| TpPa COFs | La(III) | 84.7 | Adsorbent dosage = 1000 mg $L^{-1}$, $t_{eq}$ = 40 min, pH = 3.5, $T$ = 298 K, $C_0$ = 125 mg $L^{-1}$ | Eu/Yb = 15<br>Eu/Tm = 15<br>Eu/La = 11 | [253] |
| $[NH_4^+]$ $[COF\text{-}SO_3^-]$ | Th(IV)<br>Eu(III)<br>Ce(III) | 395.0<br>43.0<br>39.0 | Adsorbent dosage = 500 mg $L^{-1}$, $t_{eq}$ = 60 min, pH = 2.8, $C_0$ = 125 mg $L^{-1}$ | - | [256] |
| P-COP-1 | Nd(III) | 321.0 | Adsorbent dosage = 1000 mg $L^{-1}$, $t_{eq}$ = 30 min, pH = 5, $T$ = 298 K, $C_0$ = 300 mg $L^{-1}$ | Nd/Ce = 20<br>Nd/La = 15 | [258] |
| P-COP-2 | | 175.6 | | | |
| COF-PA-CC | La(III) | 150.90 | Adsorbent dosage = 500 mg $L^{-1}$, $t_{eq}$ = 120 min, pH = 5.5, $T$ = 338 K, $C_0$ = 900 mg $L^{-1}$ | - | [254] |
| COF-ML-CC | La(III) | 168.20 | | - | [254] |

$t_{eq}$ – equilibrium time, $C_0$ – initial concentration, $T$ – temperature,

Table 4 and Table 5 summarize the MOF-based materials and COF-based materials which were utilized for the separation of REEs from aqueous solutions. It can be found that the MOF-based materials and COF-based materials were mostly used as adsorbents to capture REEs in the batch adsorption experiments. The experimental conditions, such as pH of solutions, the adsorbent dosage, the contact time, the temperature, and the initial concentration of REEs play a crucial role in the adsorption process. More works related to the application of MOFs for the adsorption of REEs were found in the literature that that of COFs. Therefore, more attention should be paid on the synthesis and characterization of COFs for the adsorption separation of REEs. In addition to the utilization of MOF-based adsorbents and COF-based materials in the adsorption processes, the application of MOF-based membranes and COF-based membranes in the membrane separation processes are also important since membrane technology has many advantages e.g., simple operation, environmental friendliness, low cost, and high separation efficiency [131]. Although, the exploration of MOF-based membranes and COF-based membranes with a high separation performance for REEs separation is still challenging. MOF-based membranes, such as MOF/GO membranes [247], and MOF/polymer membranes [246, 259] and COF-based membranes, such as the COF/polymer membranes [255, 256, 260] are believed to be the new generation of separation materials and represent the future of the separation of REEs.

### 3.3.4. Nanocomposite membranes

#### 3.3.4.1. Separation of REEs by physical and chemical adsorption

Separation of REEs is challenging due to their similar physical and chemical properties [261]. Fortunately, adsorption-based methods can be very helpful. It is usually accepted that adsorption (i.e., accumulation of a component at the interphase) can be chemical (i.e., occurring *via* the formation of new chemical bonds) and/or physical – here, van der Waals forces dominate [262]. Sometimes it is not easy to distinguish both these processes because physical adsorption can occur on a chemically adsorbed monolayer. Adsorption from solution is a competitive phenomenon defined by the so-called Gibbs excess [263]. In the majority of cases, the adsorption of REEs ions from solution is an endothermic process, however, the sign of enthalpy refers to the whole process, and it is not easy to determine its values for substages. Considering the major factors determining adsorption from solution, one should mention temperature, concentration, ionic power, and pH [264]. The latter factor determines the form of a solute as well as the surface charge. Usually, Langmuir (L) and/or Freundlich (F) models are applied for the theoretical description of adsorption from solution data. Owing to chemical inertness and the ability for modification, active and activated carbons are the most popular adsorbents widely applied in the industry [265], thus the attention will focus in this chapter on this adsorbent. Recently Gismondi et al. [266] endeavored to address some similarities in the REEs adsorption. Oligo-grafted mesoporous carbons were applied as adsorbents, and $Lu^{3+}$, $Dy^{3+}$, $La^{3+}$ adsorption was studied. It was concluded that adsorption increased with the ionic radii, and coordination by oxygen ions occurred. Based on the presented results, one can conclude that the adsorption of REEs ions is usually an ion-exchange process with the participation of chemical bonds and chelating (Fig. 27).

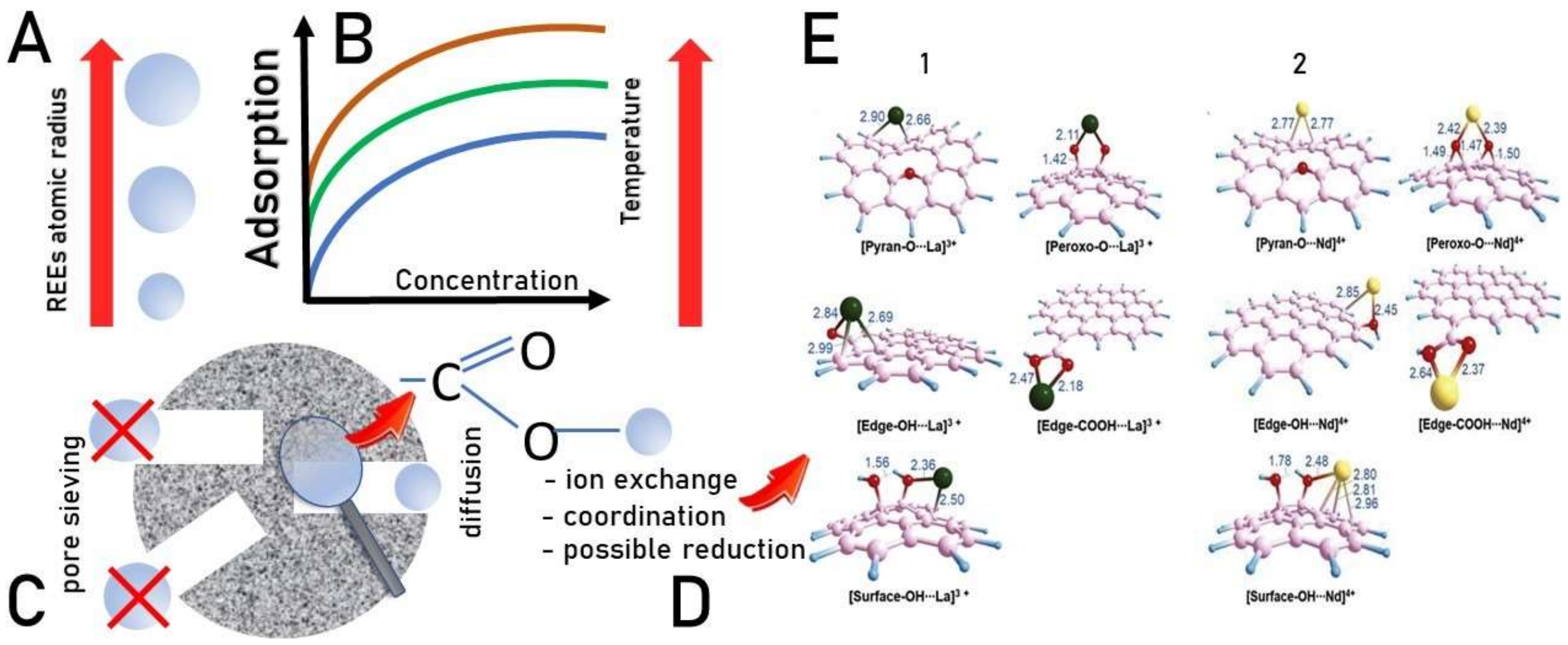

**Fig. 27.** The rise in REEs adsorption with atomic radius (A), and with temperature (B) for three isotherms (temperature rises from blue to brown). Schematic representation of carbon-containing porous adsorbent, with pores enabling/disabling REEs diffusion (C), schematic representation of carboxylic surface group with possible adsorption - increasing interactions (D) and the DFT – calculated configurations of graphene oxide functional groups – $La^{3+}$ (E 1) and $Nd^{2+}$ (E 2). Reprinted with permission from [266] Copyright © 2022 Elsevier.

Below a short discussion of the results of optimal REEs carbon-containing adsorbent searching is presented. Fundamental properties of modern carbon nanomaterials were recently described by Santana–Mayor et al. [267].

Xiong et al. [268] presented the hydrothermal method of P-doped activated bagasse-based carbon for $La^{3+}$ separation from the mixtures with $Na^{+}$ and $Ca^{2+}$. Here, capacitive deionization technology was used. The process of P-doping introduced mesopores to the system as well as increased surface hydrophilicity. L- and F-models were applied for the theoretical description of adsorption isotherms, and the comparison with other adsorbents (activated carbons (ACs) and carbon nanotubes (CNTs)) led to the conclusion about very high maximum adsorption capacity (all provided in this section adsorption capacities were calculated using L-model; 140 mg $g^{-1}$). Electro-sorption [269] was used for $La^{3+}$,$Nd^{3+}$, and $Ce^{3+}$ uptake on modified AC. Electro-sorption with the application of a laser-induced graphene (G) film [270] allowed to obtain extremely high sorption capacities for $Nd^{3+}$ (2349.25 mg $g^{-1}$), $Ce^{3+}$ (2150.75 mg $g^{-1}$), and $La^{3+}$ (2510.5 mg $g^{-1}$). The obtained values are significantly larger than those recorded for other materials, for example nanohydroxyapatite, CNTs or calcium alginate beads.

Overdose of $La^{3+}$ present in food can cause serious problems with human health. Thus a $Fe_3O_4/C_3N_5$ magnetic framework material was demonstrated as effective/selective for detection of $La^{3+}$ in food samples [271] (Fig. 28). Adsorption capacity (39.2 mg $g^{-1}$) was higher than observed for CNTs and $TiO_2$ nanotubes. Also $Fe_3O_4$ and graphene oxide (GO) – containing, La-imprinted polymer [272] was shown to be effective $La^{3+}$ adsorbent (with a capacity equal to 111 mg $g^{-1}$). Similar adsorbent ($Fe_3O_4/MnO_2/GO$) was shown as effective for the separation of the $La^{3+}/Ce^{3+}$ mixture from aqueous solution while the capacities were equal to 1016 and 981 mg $g^{-1}$ for $La^{3+}$ and $Ce^{3+}$, respectively. Electrostatic attraction between negatively charged surface and metal cations was proposed as the basic phenomenon determining the adsorption mechanism.

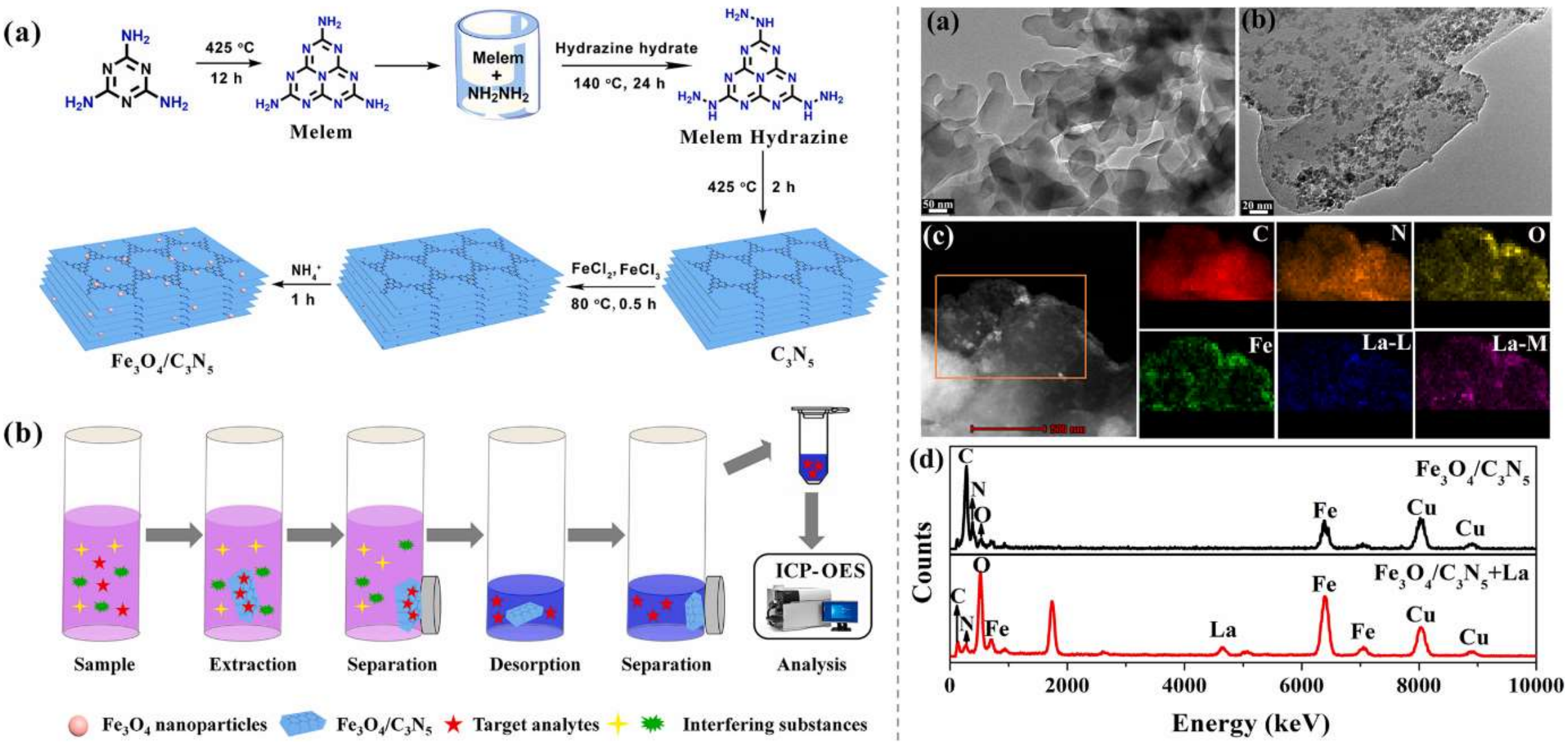

**Fig. 28**. Synthetic route of $Fe_3O_4/C_3N_5$ and its application for MSPE of La(III) (left part). TEM images of (a) $C_3N_5$ and (b) $Fe_3O_4/C_3N_5$ material; (c) TEM image and elemental mapping of $Fe_3O_4/C_3N_5$ after La(III) adsorption; (d) EDS spectra of $Fe_3O_4/C_3N_5$ before and after La(III) adsorption. Reprinted with permission from [271] Copyright © 2021 Elsevier.

Yang et al. [273] proposed the application of GO/poly-(*N*-isopropyl acrylamide-maleic acid) cryogel for $La^{3+}$ adsorption. Adsorption capacity (33.1. mg $g^{-1}$) and high selectivity of this cryogel can be adapted for $La^{3+}$ recovery from wastewaters *via* ion-exchange and chelation mechanism. Covalently functionalized (by 1,3-bis(tris(hydroxymethyl)methylamino)propane) GO was successfully applied for the separation of $La^{3+}$ and $Er^{3+}$ from heavy metals and organic contaminants, since the affinity of this adsorbent to REEs was negligibly small [274] (Fig. 29).

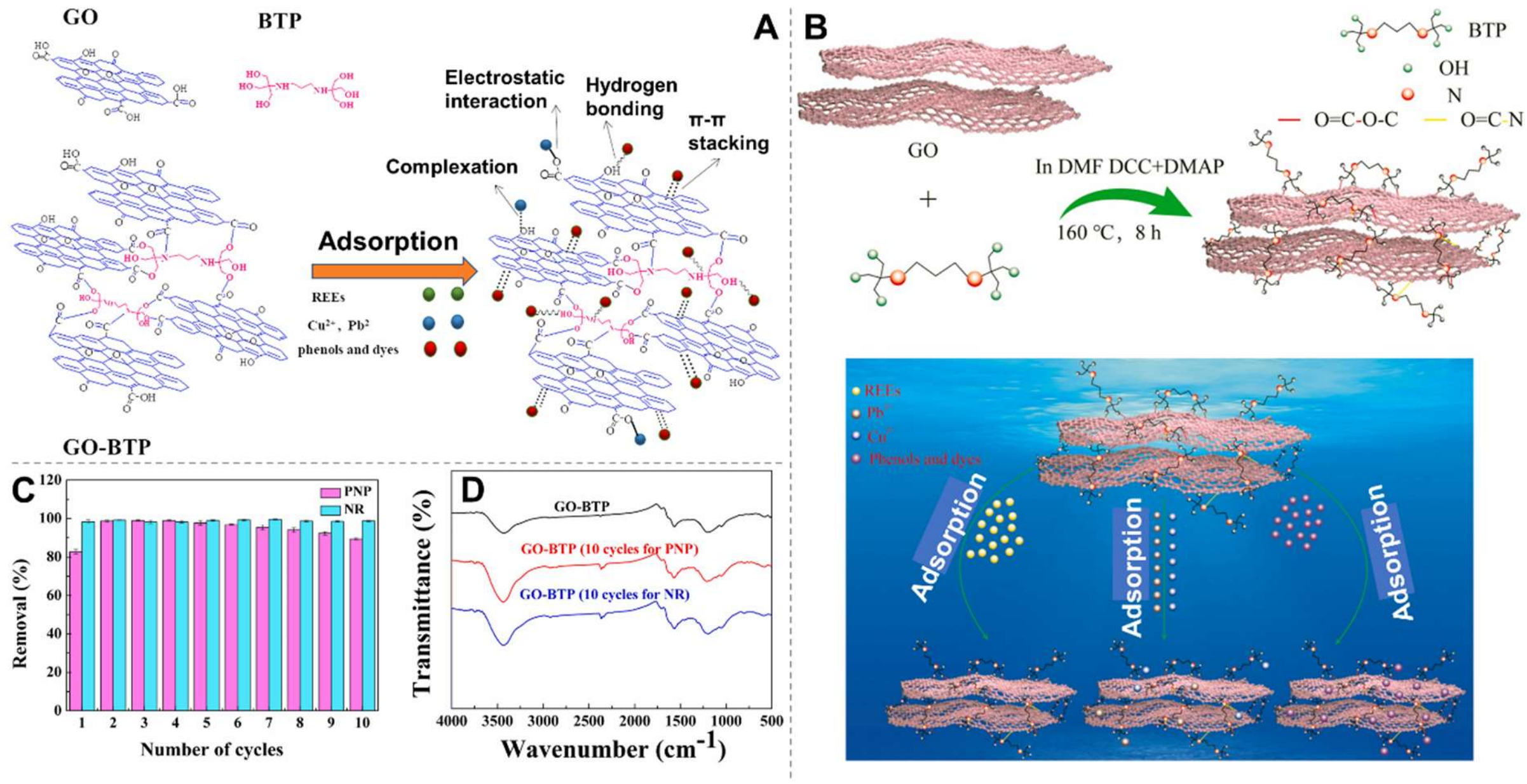

**Fig. 29**. Construction of a novel nitrogen- and oxygen-containing GO-based composite with specific adsorption selectivity (A). The preparation of GO-BTP composite for removal of various contaminants (B). Reusability of GO-BTP composite (C), FT-IR spectra of unused GO-BTP composite and GO-BTP composite after 10 adsorption-desorption cycles toward PNP or NR (D). Reprinted with permission from [274] Copyright 2021 Elsevier.

$MnO_2$/GO composite [275] was utilized for adsorption of $Ce^{3+}$ and $Eu^{3+}$ (with adsorption capacity 102 and 103 mg $g^{-1}$, respectively). Hua et al. [276] proposed the application of carbon nanospheres combined with PAN *via* electrospinning to obtain $La^{3+}$ adsorbing membrane. The determined adsorption capacity (174.5 mmol $g^{-1}$) was higher than observed for functionalized diatomite, polydopamine membranes, or silica nanocomposite. The mechanism was based on electrostatic interactions between $La^{3+}$ and surface carboxylic groups. Li et al.

[277] studied mechanistic aspects of $La^{3+}$ and $Nd^{2+}$ adsorption on GO using DFT calculations. Two types of surface oxygen groups were studied at two different locations. It was concluded that the adsorption mechanism was based on charge transfer and reduction for $La^{3+}$ and reduction/covalent interactions for $Nd^{2+}$. It was also proved that GO-based adsorbents can be efficiently applied for the adsorption of REEs from nuclear power plants waste [278]. Hao et al. [279] reported the results on $Ce^{3+}$ adsorption on GO/cellulose composite with an adsorption capacity of 225.8 mg $g^{-1}$. Ion-exchange mechanism determined the adsorption process. $Dy^{3+}$, $Nd^{3+}$, and $Pr^{3+}$ were adsorbed on cellulose-based gel modified using GO and polyethyleneimine [280]. $Dy^{3+}$ selectivity was increased by imprinting. The process of chemisorption was observed with the capacity of 36.5 mg $g^{-1}$. Similarly, oxidized cellulose nanocrystals-based composite adsorbent, reinforced by GO and oxidized CNTs [281], was shown to be a selective $Dy^{3+}$ adsorbent (average capacity ca. 40 mg $g^{-1}$). Also, a GO-based adsorbent was applied for adsorption of $Nd^{3+}$ and $Ce^{3+}$. In this case GO was modified with sodium carboxy-methyl cellulose applying tetraethyl orthosilicate as a linker [282]. High adsorption capacities i.e., 654 and 431 mg $g^{-1}$ for $Nd^{3+}$ and $Ce^{3+}$ were recorded, respectively, being larger than for the observed different clays, GO materials and composites. An experimental study on $Eu^{3+}$ adsorption on oxidized carbon nanomaterials, in the pH range of 2-7 [283], led to the conclusion about chemisorption and complexation between metal cations and oxygen surface functionalities. $Nd^{3+}$ adsorption from water [284] was studied on activated micro-mesoporous carbons obtained by a pyrolysis of ZIF-8 and subsequent $HNO_3$ oxidation. In this case it was postulated that high adsorption capacity (175 mg $g^{-1}$) was caused by creation of coordination bonds between $Nd^{3+}$ and surface carboxylic groups. The obtained carbon possessed higher capacity than mesoporous silica, selected covalent organic frameworks and layered hydroxide. A zinc-trimesic acid MOF/graphene nanocomposite [285] was shown applicable for the separation of $Ce^{3+}/Lu^{3+}$ and $Nd^{3+}/Pr^{3+}$ mixtures. The mechanism of this process was based on

a similarity of the composite pore diameters and the diameters of REEs ions, enabling penetration of MOF channels and coordination with the oxygen moieties forming pore walls. Combined MOF/GO nanocomposite adsorbent was proposed by Chen et al. [261] for REEs adsorption with capacity 340 mg $g^{-1}$ and high $Sc^{3+}/Tm^{3+}$ and $Sc^{3+}/Er^{3+}$ mixtures separation. It should be mentioned that the critical review on the $Sc^{3+}$ recovery was published recently [286] and some high-capacity adsorbents were discussed. Light REEs ($Nd^{3+}$, $La^{3+}$ and $Ce^{3+}$) were successfully adsorbed on the magnetite (20%)/carbon black (80%) composite. Maximum adsorption (around 385 - 400 mg $g^{-1}$) was recoded at pH=7.0. On the other hand, the data reported by Abdollahi et al. [287] are particularly interesting because the authors provided the results of modelling leading to conclusion about the leading role of electronegativity and molecular mass of LREEs during adsorption. $La^{3+}$ and $Ce^{3+}$ adsorption (capacity 86 and 200 mg $g^{-1}$) and separation was studied on graphite and graphite/alginate composite [288]. The adsorption properties were discussed considering electronegativities, hydration energies and hydrated radii [288].

A separate group of adsorption methods is based on the application of superparamagnetic nanomaterials in so-called magnetic nanohydrometallurgy for REEs separation. This method, together with the mechanisms of REEs separation, has been recently reviewed by Molina-Calderon et al. [289] (Fig. 30).

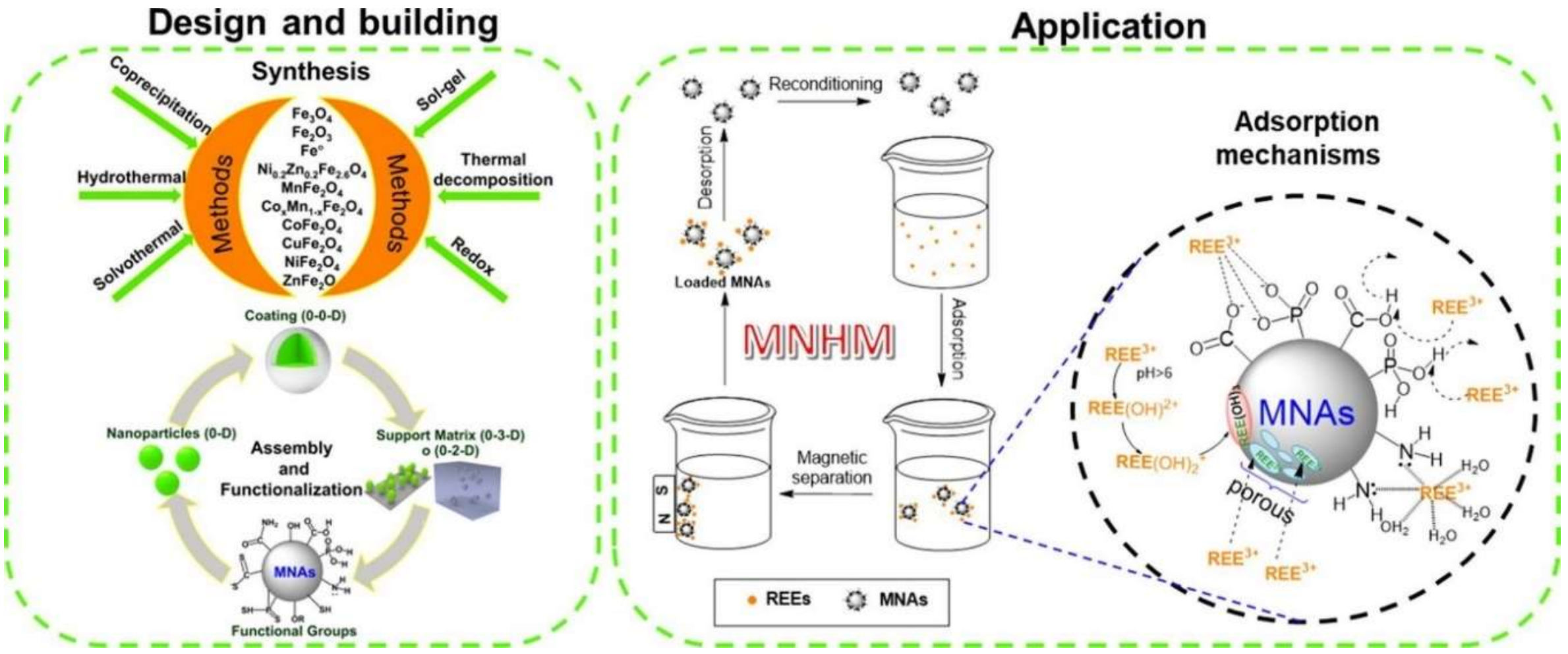

**Fig. 30**. Advances of magnetic nanohydrometallurgy using superparamagnetic nanomaterials as rare earth ions adsorbents. Reprinted with permission from [289] Copyright 2022 Elsevier.

### 3.3.4.2. Carbon containing membrane-based separation

Tan et al. [290] proposed the application of N-doped G[30] with tunable porosity for REEs separation. The pores with controlled diameters (from few to several tens of nanometers) and pyrrolic N determined membrane properties, especially high selectivity during $Sc^{3+}$ separation from other REEs (Fig. 31).

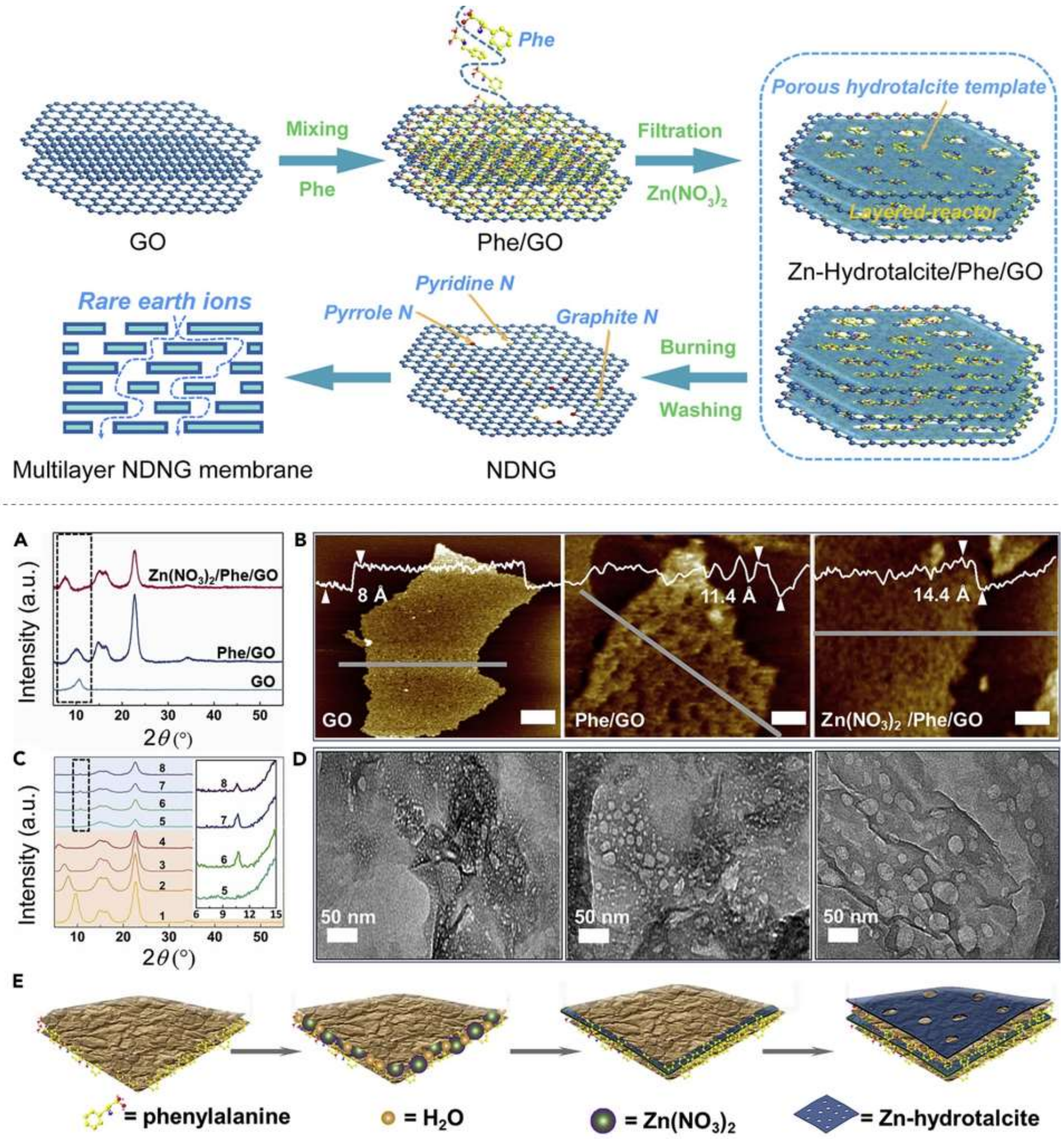

**Fig. 31**. Synthesis of NDNG through multiple confinement strategy and then NDNG membrane was prepared for rare earth elements separation (upper part). Fabricating mechanism of Zn-hydrotalcite/Phe/GO composites with sandwich structure (bottom part). Reprinted with permission from Ref [290] Copyright 2022 Cell. CC Licence.

A combined membrane, based on the 2D analog of ZIF-8 MOF synthesized between (stabilizing the MOF) GO layers [247], was proposed to separate $La^{3+}$/ $Ce^{3+}$ and $La^{3+}$/ $Yb^{3+}$ mixtures. The chemical composition of the MOF and the diameter of pores in this membrane allowed $La^{3+}$ diffusion, hindering the diffusion of other ions. GO-based membrane was applied for the adsorption of $Dy^{3+}$ [291]. The highest performance (26.27 mg $g^{-1}$) was observed ad pH = 5.0. Wang et al. [292] described the method of sol-gel preparation of thin cellulose-tetraethyl orthosilicate film. This film was carbonized and modified using 3-aminopropyl triethoxysilane and 1-(2-pyridylazo)-2-naphthol. The so-formed mesoporous membrane was applied for the selective $Er^{3+}$ ions separation from a mixture of metal ions usually contained in rare-earth wastewater.

The hybrid materials possessing in their structure supramolecules, e.g. MOFs and carbon-based materials are in the spotlight of the researcher for REEs separation [293]. Tursi and co-workers [294] generated a novel bioMOF-based single-walled carbon nanotube bucky paper (SWCNTBP). The formed composite material (BioMOF@SWCNT-BP) was applied for recovery of the endangered REEs from aqueous systems. After incorporating such MOFs with hexagonal functional channels decorated with threonine amino acid residues pointing toward the accessible void spaces, the capture properties of the final membrane were successfully improved, providing an adaptable functional environment to interact with lanthanides. The researchers studied the ability of SWCNT-BPs and BioMOF@SWCNT-BPs in recovering lanthanides from water solutions in static and dynamic conditions as a function of pH values and initial lanthanide concentrations. It was revealed that the adsorption is not impacted by pH of lanthanide solutions. Moreover, BioMOFs in SWCNT-BPs play a beneficial role in the increase of $Ce^{3+}$ recovery at higher concentrations owing to the alcohol functionalities of threonine, decorating the MOF pores. It was found that the relative recovery percentage after a 7 days recirculation in higher concentration (50 ppm of cerium) solutions—recirculating through the same membrane—with a 263.30 mg of cerium adsorbed per gram of BioMOF@SWCNT-BP. The spectrum of these materials has been extended also to the selective lead decontamination [295]. Cheng and co-workers [296] generated 2D-MOF/graphene oxide membranes as highly efficient adsorbents for the removal of $Cs^+$ from aqueous solutions. The researchers pointed out that the dominant interaction mechanism was interface or surface complexation and electrostatic interaction, the maximum adsorption amount of $Cs^+$ was 88.4 %.

#### 3.3.4.3.Other separation-related applications

Composite MOF-based nanomembrane, containing La and GO for adsorption of P from water (utilizing high affinity between La and P) was described by Wei et al. [297] La was introduced *via* impregnation into a MOF structure and next GO was added to increase the water purification ability of a membrane (Fig. 32). Also Gd-doped (for improving photocatalytic activity), G-containing $BiSO_4$ composite material for application in photocatalytic removal of methylene blue was reported [298]. Since radioactive Th coexists with REEs, it was proposed to apply a polyethylenimine-scaffolded and functionalized G aerogel [299] for adsorption of $Th^{4+}$, separation from REEs, and/or electro-sorption on carbon-based electrodes [300]. Usually, strongly oxidized microporous AC is sufficient for the purification of REEs *via* $Th^{4+}$ adsorption [301]. The process takes place in the pH range of 3-5, because in this range, REEs elements occur as cations while $Th^{4+}$ forms hydrolysates ($Th(OH)^{3+}$ and $Th(OH)_2^{2+}$) having larger ionic diameters than $Th^{4+}$. A mesoporous graphite carbon nitride can be used as an adsorbent for $Th^{4+}$ separation from monazite (containing REEs phosphates) [302], while calcium/alginate-GO nanocomposite, for separation of $Th^{4+}$, $U^{6+}$ and $Fe^{3+}$ [303]. Also peptide-carbon hybrid membranes [304] were synthesized to remove not only $Th^{4+}$ but also $U^{6+}$ that are present in concentrates of REEs minerals as defects in crystal lattices. It turn, the method of $LiFePO_4$ recovering from used batteries was presented by Hu et al. [305]. It was based on high-gradient and induced roll magnetic separators with a high recovery (ca. 99%). Adsorption – based separation on cationic resins was used for recovery of REEs from processed carbonaceous shale [306]. Yuan et al. [307] reported the lack of $La^{3+}$, $Eu^{3+}$ and $Tm^{3+}$ adsorption on graphdiyne and this can lead to the preparation of a membrane separating REEs from other cations. The lack of adsorption was caused by the positive adsorption energy.

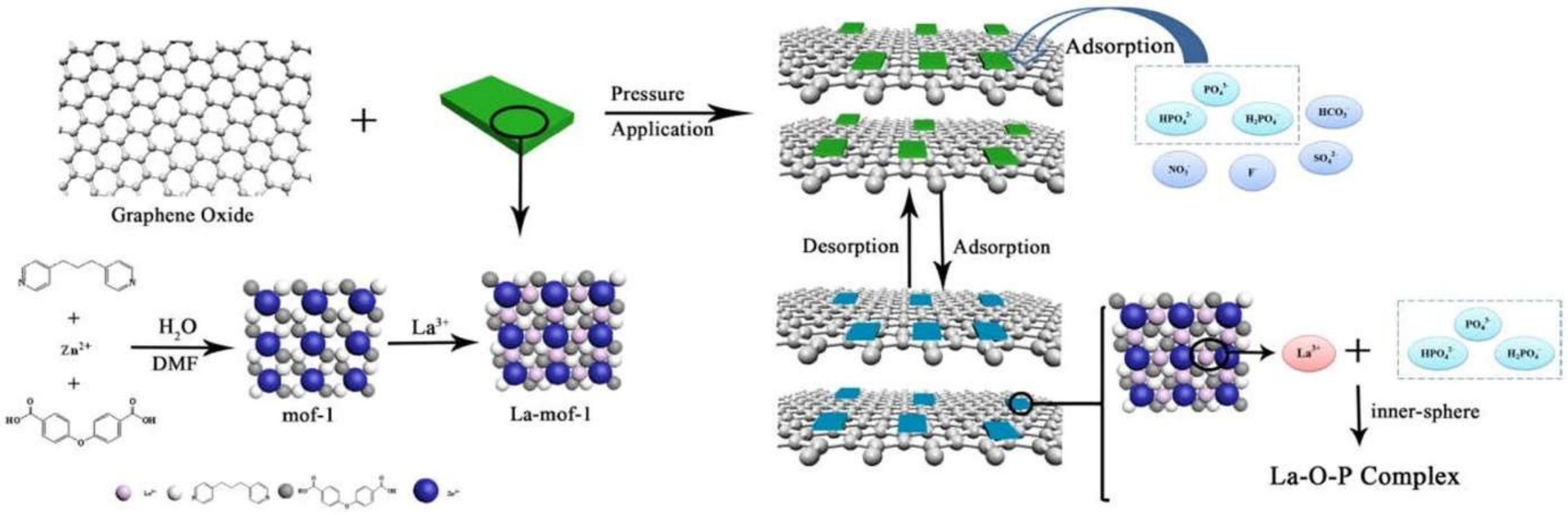

**Fig. 32.** Construction of lanthanum modified MOFs graphene oxide composite membrane for water purification**.** Reproduced from [297] Copyright © 2020, Elsevier, CC License.

## 4. Health issue

Although REEs are broadly applied in the biomedical sectors, e.g., bioimaging, sensing, photothermal and photodynamic therapy [308, 309], drug delivery [310, 311], and optogenetics [312, 313], their extensive utilization and high demand have raised concerns about their safety. The topic of the hazard of REEs to human health is emerging, and further more focused comprehensive research must be done. In the scientific literature, the main concerns on REEs health problems are related to REEs acquaintance over the food chain, gadolinium contrast agents instigating health problems, cytotoxicity of REEs, environmental exposure, and endangering REEs due to lifestyle. Recently, much more attention has been also paid to the toxic effect of REEs on the aquatic organisms [314]. Disposing of old REE-containing equipment, using REE-containing phosphate fertilizers, mining, and dispersion from native rocks may all enhance the possibility of REEs pollution in water fauna and flora. As a result, pollution may lead to their release into the nearby ecosystems. Owing to the lack of safety regulations, legislation, and widespread utilization in the numerous sectors, REEs' diverse effects on aquatic life forms have been noticed. To solve the abovementioned problem, a better understanding of accumulation, bioavailability, cytogenetic effects, organ-specific toxicity, growth inhibition, and embryotoxicity criteria of REEs should be done. Furthermore, the reuse and recycling technologies of REEs need to be implemented to reduce the environmental and human impacts.

Based on the available data, it was pointed out that REEs can enter the human body *via* several routes, e.g., ingestion, inhalation, skin contact, and food chain [315, 316] (Table 6 and 7). The latter is related to the REEs utilization in producing pesticides, fertilizers, herbicides, and feed additives for animals [317, 318]. Moreover, REEs accumulate in soil [319], roots, and tops of plants [320].

The most severe concern regarding REE toxicity in humans is that these elements can damage DNA, originate the production of reactive oxygen species (ROS) (nanoparticles of Ce, $CeO_2$-Gd, and $Gd_2O_3$) [321], and cause cell death. For instance, Gd and La can be mutagenic to human primary peripheral lymphocytes [322, 323]. On the other hand, the chloride of these REE, i.e. $LaCl_3$ and $CeCl_3$ inhibited the proliferation of the leukemic cell lines HL-60 and NB4 [324].

**Table 6**. Examples of the effects of REE exposure.

| Species/Model | Endpoints /effects | Ref. |
|---|---|---|
| **Mice** | $CeCl_3$ oral administration → liver and lung toxicity | [325] |
| **Rats** | $CeO_2$ acute pulmonary and thrombotic effects → lung fibrosis | [326] |
| **HepG2 and HT−29 cell lines** | $CeCl_3$ and $LaCl_3$ affect gene regulation detected by RT-PCR based arrays | [327] |
| **Human spermatozoa** | Nanoparticles of $CeO_2$ | [328] |
| **Sea urchin embryos and sperm** | Proportional toxicities of different REEs to early development, cytogenetic damage and oxidative stress fertilization success, offspring damage. | [329] |

| Exposed population | Main endpoints | |
|---|---|---|
| **Housewives exposed to indoor air pollution** | Excess REE levels in scalp hair, associated with indoor air pollution and related to hypertension risk | [330, 331] |
| **Industrial city in central China, Zhuzhou** | Excess REE levels in street dust and linked to health risks | [332] |
| **Residents in Baiyun Obo mining area (China)** | Excess scalp hair levels of REEs, heavy metals and U | [332] |
| **Exposure to REEs in the workplace in humans** | **Main endpoints** | |
| **Manufacturing Ce and La oxide** | Excess Ce and La levels in urine | [333] |
| **REE miners** | Excess REE level in hair and dysregulation of the protein expression | [334] |
| **e-waste processing** | Reduced level of hemoglobin in REE-exposed | [335] |

**Table 7**. Associated health disorders with selected REE.

| REE | Associated health disorders | Ref. |
|---|---|---|
| **Ce** | Increasing risk of anemia | [335] |
| **Ce** | Endomyocardial fibrosis, concentration in primary teeth | [336] |
| **Ce** | Risk of acute myocardial infarction | [337] |
| **Ce** | Risk of stroke | [338] |
| **La** | Affecting pregnancy, concentration in hair | [339] |
| **Gd** | Acute renal failure | [340] |
| **La, Nd** | Orofacial clefts to infants | [341] |
| **Ca, Yb** | Increasing TSH levels in infants | [342] |
| **La, Ce, Gd, Lu** | Accumulation in brain-tumor tissues | [343] |
| **REE** | Subclinical organ damage | [344] |
| **REE** | Leaving close to high REE area – observed disorders: anorexia, indigestion, abdominal distension, diarrhea, fatigue, and weakness. Effect for human body: lower levels of total protein, albumin, globulin, and serum-glutamic pyruvic transaminase and higher levels of IgM in their blood serum. | [345] |

## 5. Impurity elimination in the course of REEs separation

Owing to the complex matrix from which the REEs are separated, removing the impurities and generating highly pure REEs is crucial for the advanced applications. The presence of additional elements and molecules substantially affects the quality of the final product and the processing effectiveness. The most common impurities during the acquisition process of REEs are Ca, Mg, Fe, Bi, Pb, Cu, Co, Mo, and Mn. Nevertheless, the most problematic are Al, Fe, Th, and U [346]. The most common techniques used for the separation non-REEs from REE are ion exchange, adsorption, solvent extraction, selective precipitation, protein-based process [347] as well as membrane technologies [346, 348].

Kim and co-workers [349] presented the successful implementation of membrane-assisted solvent extraction (MSX) with hollow fiber membranes for the selective extraction of REEs from commercial NdFeB magnets and industrial scrap magnets. The benefit of the MSX technique is the possibility to exclude the drawback of common equilibrium-based solvent extraction methods, e.g., third-phase formation, loading, and flooding [208, 350-352]. The separation during the membrane-assisted solvent extraction process is enhanced owing to the non-equilibrium conditions. The available conditions can be kept thanks to the high driving forces during the long-term process compared to the equilibrium-restricted conventional solvent extraction. The great advantage of MSX is the fact that both circulating the feed and strip solutions steps are continues without dispersion of phases. However, during the classical process, solvent extraction carries out extraction and stripping separately [353]. Moreover, applying hollow fiber membrane modules in the MSX system gives a large contact surface area per unit volume, resulting in a high REE extraction rate. The presented solution is also economically friendly, owing to the minimal loses of solvent and was highly effective for selective recovering pure REEs (Nd, Dy, Pr) from the industrial scrap magnets containing also Fe and B.

Another, the most important and challenging is the separation of REEs from thorium and uranium. Previously, the membrane techniques were implemented to separate individually Th or U from other metal ions in the liquid solutions [354, 355] using $SiO_2$ or GO membranes [356]. Formation of a complex between the element of interest and the membrane caused such a recovery. Membranes were effective and selective under acidic pH (4-5.5) for U and $< 4$ for Th. Nevertheless, an additional modification of the membrane was required to enhance the efficiency of Th and U from the matrix containing REEs. Such a functionalization boosted selectivity towards the target elements, e.g., Th and U versus REEs (Eu, Sm, and Nd) at ambient conditions and pH <2 [357].

Xiong et al. [256] presented for the first time a highly effective separation of Th(IV) from U(VI) and REEs ($Ce^{3+}$ and $Eu^{3+}$) with the implementation of the breakthrough experiment and membrane-based separation technique. The membranes were prepared with COF of a high affinity to the target elements. Moreover, it was possible to separate Th(VI) solution with high purity > 93% (Fig. 33).

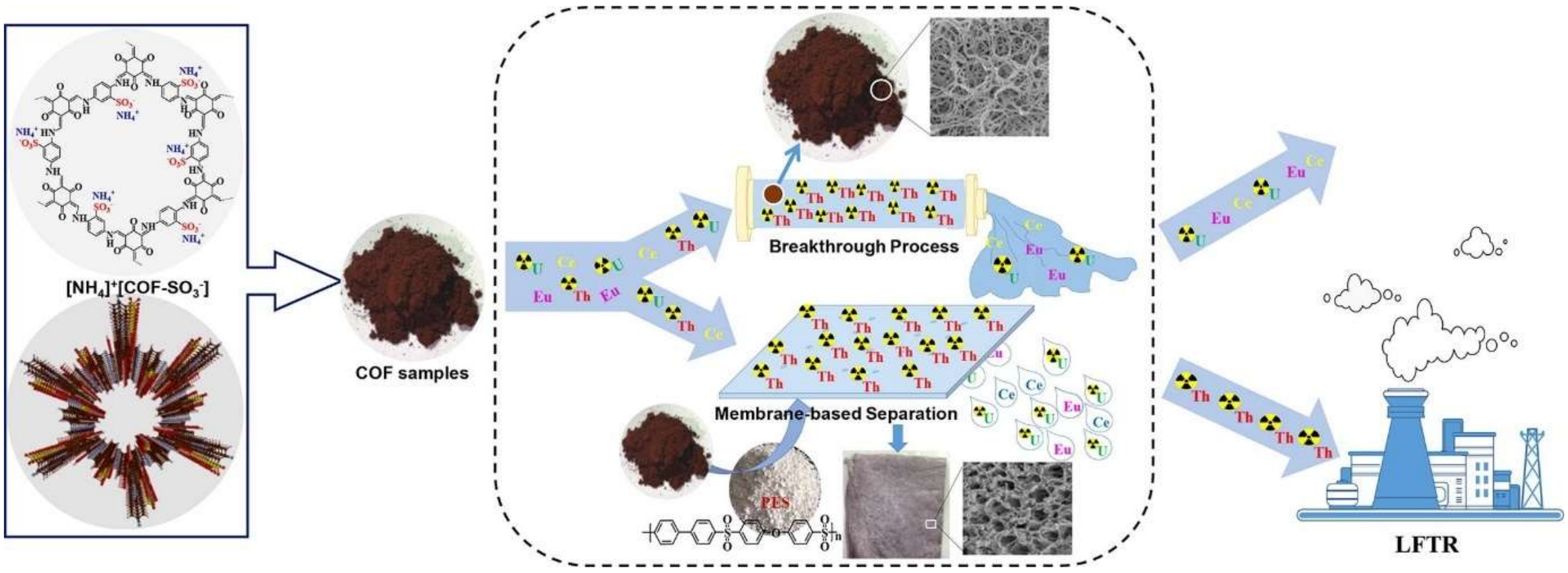

**Fig. 33.** Selective extraction of Th from U and REEs using sulfonated COF and its membrane derivate. Reproduced from [256] Copyright © 2020, Elsevier.

## 6. Conclusions and critical perspectives

Since the demand for REEs is growing rapidly worldwide, mainly owing to their end-use applications (e.g., in wind turbines, electric vehicles, advanced electronic products, and displays), new technologies for extracting and recovering these precious metals appear to be critical.

Among the generally implemented separation methods, the classic stepwise procedure for obtaining a single REE necessitates numerous dissolution-crystallization steps. Considering the production cost and cycle, there is no room to employ the method for separation a single REE. To overcome this challenge, the MOF selective crystallization technology has slowly become more popular owing to the reduction of steps during the process and its higher selectivity to a single REE. However, the current study on this issue is still in its infancy, and whether it can be employed in the industrial manufacture remains an open question. Considering the ion exchange method, producing a single REE with a very high purity is possible. However, the continuous processing capacity is limited. Specifically, the production cycles are long and the ion exchange resin has a mediocre stability, hence often requiring the replacement. These operations generate costs for the overall process. For that purpose, the replacement with membrane-based separation technologies is giving an amazing opportunity to shift to more economically and environmentally friendly processes.

The solvent extraction method is broadly utilized commercially for the REE separation. The high effectiveness of the method, continuous production, and a large processing capacity caused their implementation in industrial production. Although the solvent extraction process possesses many benefits, the method consumes massive amounts of organic solvents for the extraction, and the regeneration of the organic phase after stripping is very challenging. At this point, membrane technology is becoming very practical, particularly non-liquid membranes owing to their high stability compared to the liquid counterparts.

In the case of working with low-concentration REE solutions, it is not recommended to apply the solvent extraction. Here, the adsorption method could be a suitable technique due to enrichment effect on the rare earth ions in the solution. Nevertheless, this method cannot achieve high selectivity and high adsorption capacity at the same time. In turn, the advantages of electrochemical separation of metal components are low energy consumption, continuous and easy operation, and environmental protection. This method can be applied for large-scale industrial extraction of REEs from compounds or minerals. However, to be able to apply the method for the separation of a single REE, the accomplishment of comprehensive research is still required. Indeed, there is a room for further enhancement of REE separation approaches and boosting the separation effectiveness.

Based on the presented data, it can be highlighted that membranes and membrane-based separation technologies can integrate green and sustainable solutions for effective REEs separation and purification.

To sum up, LM methods have improved over the previous few decades, resulting in outstanding permeation and separation capabilities, however, there has been no significant progress in the scale-up application. For that reason, the non-liquid membranes turn out to be very promising and worthy of further investigation. PIMs demonstrated higher transport flux, faster permeation, lower carrier usage, and more advanced adaptability and stability than SLMs. Its scalability should be pursued, and more sensitive designs for precisely recognizing target species should be encouraged. The imprinting approach is an excellent approach for separating adjacent REEs, but the durability of the template cavity should be considered for the long-term operation. The future research should focus on discovering novel, tailored and hence unique functional monomers with task-specific recognition abilities and efficient synthesis routes in the aqueous phase. Despite significant advancements in the membrane techniques over the last few decades, their applicability in REEs separation is still limited. Numerous sophisticated membrane approaches, such as nanocomposite membranes with inorganic nanoparticles and MOF membranes, have received significant consideration in gas and liquid separation procedures. Moreover, they can be unequivocally indicated as the most interesting new non-liquid membrane approaches for the REEs separation. Such a highly efficient and integrated solutions are urgently needed for the greener REE separation.

**Acknowledgements**

A.P.T gratefully acknowledges the financial support from NCN Opus 22 project: UMO-2021/43/B/ST5/00421.

S.B. greatly acknowledges the supporting action from EU's Horizon 2020 ERA-Chair project ExCEED, grant agreement No. 952008.

**List of abbreviations:**

2NOPE - 2-nitrophenyloctyl ether; BDC - bis-(9-octylamino(2-dimethylaminoethyl)acridine-4-carboxamide; BDCS - 1,3-benzenedisulfonyl dichloride; BLM - bulk liquid membrane; BTC - benzene-1,3,5-tricarboxylic acid; CC - cyanuric chloride; CME - cationic membrane electrolysis; CNT - carbon nanotubes; COF – covalent organic framework; CTA - cellulose triacetate; D2EHPA - di(2-ethylhexyl)phosphoric acid; DES - deep eutectic solvent; DFT - density functional theory; DOP - dioctylphthalate; DTPA - diethylentriaminepentaacetic acid; ELM - emulsion liquid membrane; EVOH - poly(vinyl-alcohol-co-ethylene); f-CNTs - functionalized carbon nanotubes; G – graphene; GO - graphene oxide; IIMs - ion imprinted membranes; IIT - ion-imprinting technique; IP - interfacial polymerization; LM - liquid membrane; MEUF - micellar-enhanced ultrafiltration; MAA - methacrylic acid; MIMs - molecular imprinted membranes; ML – melamine; MOF - metal-organic framework; MSTs - membrane separation techniques; MSX – membrane-assisted solvent extraction; NF – nanofiltration; NIM - non-ion-imprinted membrane; NPOE - plasticizer 2-nitrophenyl octyl ether; NPs – nanoparticles; PAUF - polymer-assisted ultrafiltration; PA - *p*-phenylenediamine; PAN – polyacrylonitrile; PEI – polyethylenimine; PES - polyethersulfone; PIMs - polymer inclusion membranes; PPC - poly(piperazine-cresol); PSA - polysulfonamide; PTFE – polytetrafluoroethylene; PVDF - polyvinylidene fluoride; REE – rare earth element; RTIL - room temperature ionic liquids; SLM - supported liquid membrane; SWE - subcritical water extraction; TBP - tributyl phosphate; TDDA - tridodecylamine ; Tp - 2,4,6-triformylphloro-glucinol; UF – ultrafiltration;

**Declaration of Competing Interest**

The authors declare that they have no known competing financial interests or personal relationships that could have appeared to influence the work reported in this paper.